\documentclass[12pt,reqno,a4paper]{amsart}
\usepackage{amscd,amsfonts,amssymb,amsmath,amsthm}
\usepackage[arrow,matrix]{xy}
\usepackage{graphicx,epstopdf}
\usepackage{subfigure,tikz}
\usepackage{color}
\usepackage{mathrsfs,euscript,hyperref}

\newtheorem{theorem}{Theorem}[section]
\newtheorem{lemma}[theorem]{Lemma}
\newtheorem{proposition}[theorem]{Proposition}
\newtheorem{corollary}[theorem]{Corollary}
\theoremstyle{remark}
\newtheorem{remark}{Remark}[section]

\newtheorem{example}{Example}[section]
\theoremstyle{definition}
\newtheorem{definition}{Definition}[section]
\newtheorem{conjecture}{Conjecture}[section]

\newcommand{\myp}{\mbox{$\:\!$}}
\newcommand{\mypp}{\mbox{$\;\!$}}

\newcommand{\myn}{\mbox{$\;\!\!$}}
\newcommand{\mynn}{\mbox{$\:\!\!$}}

\newcommand{\rme}{\mathrm{e}}

\newcommand{\dif}{\mathrm{d}}
\newcommand{\const}{\mathrm{const}}
\DeclareMathOperator{\supp}{\rm supp} \DeclareMathOperator{\ex}{\rm
ex}

\newcommand{\xbar}[1]{%
   \kern0.15ex\hbox{%
     \vbox{%
       \hrule height 0.4pt 
       \kern0.3ex
       \hbox{%
         \kern-0.15em
         \ensuremath{#1}%
         \kern-0.1em
       }%
     }%
   \kern0.25ex}%
}

\newcommand{\xbarscr}[1]{%
   \kern0.2ex\hbox{%
     \vbox{%
       \hrule height 0.3pt 
       \kern0.2ex
       \hbox{%
         \kern-0.1em
         \ensuremath{{}_{#1}}%
         \kern-0.15em
       }%
     }%
   \kern0.25ex}%
}

\newcommand{\xbarscrscr}[1]{%
   \kern0.2ex\hbox{%
     \vbox{%
       \hrule height 0.3pt 
       \kern0.2ex
       \hbox{%
         \kern-0.1em
         \ensuremath{{}_{{}_{#1}}}%
         \kern-0.15em
       }%
     }%
   \kern0.25ex}%
}

\newcommand{\checkbxi}{\lefteqn{\boldsymbol{\xi}}\kern.13pc\check{\phantom{\xi}}\kern-.09pc}

\newcommand{\card}{\mathop{{\mathrm{card}}}\nolimits}

\def\MR#1{\href{http://www.ams.org/mathscinet-getitem?mr=#1}{MR#1}}

\numberwithin{equation}{section}

\newcommand{\PP}{\mathbb{P}}

\usepackage{cancel}
\usepackage[normalem]{ulem}

\begin{document}

\title[Potts model on a Cayley tree]{On the uniqueness of Gibbs
measure in the Potts model on a Cayley tree with external
field}

\author[L.\mypp{}V.\myp~Bogachev]
{Leonid~V.~Bogachev}
\address{L.\,V.\,Bogachev\\ Department of Statistics, School of Mathematics, University of Leeds, Leeds, LS2 9JT, UK.}
\email{L.V.Bogachev@leeds.ac.uk}

\author[U.\,A.\,Rozikov]
{Utkir~A.~Rozikov}

 \address{U.\,A.\,Rozikov\\ Institute of Mathematics,
81 Mirzo Ulug'bek str., 100170, Tashkent, Uzbekistan.}
\email{rozikovu@yandex.com}

\dedicatory{To the memory of H.-O.\ Georgii}

\begin{abstract}
The paper concerns the $q$-state Potts model (i.e., with spin values
in $\{1,\dots,q\}$) on a Cayley tree $\mathbb{T}^k$ of degree $k\geq
2$ (i.e., with $k+1$ edges emanating from each vertex) in an
external (possibly random) field. We construct the so-called
\emph{splitting Gibbs measures} (SGM) using generalized boundary
conditions on a sequence of expanding balls, subject to a suitable
compatibility criterion. Hence, the problem of existence/uniqueness
of SGM is reduced to solvability of the corresponding functional
equation on the tree. In particular, we introduce the notion of
translation-invariant SGMs and prove a novel criterion of
translation invariance. Assuming a ferromagnetic nearest-neighbour
spin-spin interaction, we obtain various sufficient conditions for
uniqueness. For a model with constant external field, we provide
in-depth analysis of uniqueness vs.\ non-uniqueness in the subclass
of completely homogeneous SGMs by identifying the phase diagrams on
the ``temperature--field'' plane for different values of the
parameters $q$ and~$k$.
In a few particular cases (e.g., $q=2$ or $k=2$), the maximal number
of completely homogeneous SGMs in this model is shown to be $2^q-1$,
and we make a conjecture (supported by computer calculations) that
this bound is valid for all $q\ge 2$ and~$k\ge2$.
\end{abstract}




\maketitle

\tableofcontents

\section{Introduction}\label{sec:1}
\subsection{Background and motivation}\label{sec:1.1}

The \emph{Potts model} was introduced by
R.\,B.~Potts \cite{Potts} as a lattice system with $q\ge2$ spin
states and nearest-neighbour interaction, aiming
to generalize the Kramers--Wannier duality \cite{Kramers-Wannier} of
the Ising model ($q=2$). Since then, it has become the darling of
statistical mechanics, both for physicists and mathematicians
\cite{Ba,Wu}, as one of few ``exactly soluble'' (or at least
tractable) models demonstrating a phase transition
\cite{Bricmont,Dembo2,Domb,Kotecky-Shlosman,Laanait,Martirosian}.
Due to its intuitive appeal to describe \emph{multistate} systems,
combined with a rich structure of inner symmetries, the Potts model
has been quickly picked up by a host of research in diverse areas,
such as probability \cite{Gr1}, algebra \cite{Martin}, graph theory
\cite{Be1}, conformally invariant scaling limits
\cite{Riva,Schramm}, computer science \cite{Galanis}, statistics
\cite{Green,Murua}, biology \cite{Graner}, medicine \cite{Sun,Sz},
sociology \cite{Sc,Schulze}, financial engineering
\cite{Reichardt,Takaishi}, computational algorithms \cite{Boy,Fr},
technological processes \cite{Sa,Ti}, and many more.

Much of this modelling has involved interacting spin system on
\emph{graphs}. In this context, tree-like graphs are especially
attractive for the analysis due to their recursive structure and the
lack of circuits. In particular, regular trees (known as
\emph{Cayley trees} or \emph{Bethe lattices} \cite{Bethe}) have
become a standard
trial template for various models of statistical physics (see, e.g.,
\cite{Abou-Chacra,Aizenman,Athreya,Martinelli,Mezard-Parisi,
Thouless,Weiss}), which are interesting in their own right but also
provide useful insights into (harder) models in more realistic
spaces (such as lattices $\mathbb{Z}^d$) as their
``infinite-dimensional'' approximation \cite[Chapter~4]{Ba}.
On the other hand, the use of Cayley trees is often motivated by the
applications, such as information flows \cite{Mos} and
reconstruction algorithms on networks \cite{Dembo2,Mezard}, DNA
strands and Holliday junctions \cite{Roz-DNA}, evolution of genetic
data and phylogenetics \cite{DMR}, bacterial growth and fire forest
models \cite{Br}, or computational complexity on graphs
\cite{Galanis}. Crucially, the criticality in such models is
governed by phase transitions in the underlying spin systems.

It should be stressed, however, that the Cayley tree is distinctly
different from finite-dimensional lattices, in that the ratio of the
number of boundary vertices to the number of interior vertices in a
large finite subset of the tree does not vanish in the thermodynamic
limit.\footnote{This is the common feature of \emph{nonamenable
graphs} (see~\cite{BRSSZ}).} For example, if $k\ge 2$ is the degree
of the tree (i.e., each vertex has $k+1$ neighbours), $V_n$ is a
``ball'' of radius $n$ (centred at some point $x_\circ$) and
$\partial V_n=V_{n+1}\setminus V_n$ is the boundary ``sphere'', then
$$
\frac{|\partial V_n|}{|V_n|}=\frac{(k+1)\myp
k^n}{1+(k+1)(k^{n}-1)/(k-1)}\to k-1\ge1,\qquad n\to\infty.
$$
Therefore, the remote boundary may be expected to have a very strong
influence on spins located deep inside the graph, which in turn
pinpoints a rich and complex picture of phase transitions, including
the number of possible pure phases of the system as a function of
temperature.

Mathematical foundations of random fields on Cayley trees were laid
by Preston \cite{Pr} and Spitzer~\cite{Sp}, followed by an extensive
analysis of Gibbs measures and phase transitions (see Georgii
\cite[Chapter 12]{Ge} and Rozikov \cite{Ro}, including historical
remarks and further bibliography). The Ising model on a Cayley tree
has been studied in most detail (see \cite[Chapter~2]{Ro} for a
review). In particular, Bleher et al.\ \cite{BRZ} described the
phase diagram of a ferromagnetic Ising model in the presence of an
external random field.\footnote{Note that perturbation caused by the
field breaks all symmetries of the model, which renders standard
arguments inapplicable (cf.\ \cite[Chapter 6]{Bov}).} Using physical
argumentation, Peruggi et al.\ \cite{Peruggi1,Peruggi2} considered
the Potts model on a Cayley tree (both ferromagnetic and
antiferromagnetic) with a (constant) external field, and discussed
the ``order/disorder'' transitions (cf.\ \cite{Dembo2,Galanis}).

In the present paper, we consider a similar (ferromagnetic) model
but we are primarily concerned with more general
``uniqueness/non-uniqueness'' transitions. We choose to work with
the so-called \emph{splitting Gibbs measures} (SGM), which are
conveniently defined in the thermodynamic limit using generalized
boundary conditions (GBC). To be consistent, permissible GBC fields
must satisfy a certain functional equation, which can then be used
as a tool to identify the number of solutions. In this approach, it
is crucial that any extremal Gibbs measure is SGM, and so the
problem of uniqueness is reduced to that in the SGM class.

K\"ulske et al.\ \cite{KRK} described the full set of completely
homogeneous SGMs for the $q$-state Potts model on a Cayley tree with
zero external field; in particular, it was shown that, at
sufficiently low temperatures, their number is $2^{q}-1$. Recently,
K\"ulske and Rozikov \cite{KR} found some regions for the
temperature parameter ensuring that a given completely homogeneous
SGM is extreme/non-extreme; in particular, there exists a
temperature interval in which there are at least $2^{q-1} + q$
extreme SGMs. In contrast, in the antiferromagnetic Potts model on a
tree, a completely homogeneous SGM is unique at all temperatures and
for any field (see \cite[Section~5.2.1]{Ro}).

\subsection{Set-up}\label{sec:1.2} We start by summarizing the basic concepts for Gibbs
measures on a Cayley tree, and also fix some notation.

\subsubsection{Cayley tree}\label{sec:1.2.1} Let $\mathbb{T}^k$ be a
(homogeneous) Cayley tree of degree $k\ge2$, that is, an infinite
connected cycle-free (undirected) regular graph with each vertex
incident to $k+1$ edges.\footnote{In the physics literature, an
infinite Cayley tree is often referred to as the \emph{Bethe
lattice}, whereas the term ``Cayley tree'' is reserved for rooted
trees truncated at a finite height \cite{Chen,Ost}.} For example,
$\mathbb{T}^1=\mathbb{Z}$. Denote by $V=\{x\}$ the set of the
vertices of the tree and by $E=\{\langle x,y\rangle\}$ the set of
its (non-oriented) edges connecting pairs of neighbouring vertices.
The natural distance $d(x,y)$ on $\mathbb{T}^k$ is defined as the
number of edges on the unique path connecting vertices $x,y\in V$.
In particular, $\langle x,y\rangle\in E$ whenever $d(x,y)=1$. A
(non-empty) set $\varLambda\subset V$ is called \emph{connected} if
for any $x,y\in\varLambda$ the path connecting $x$ and $y$ lies
in~$\varLambda$. We denote the complement of $\varLambda$ by
$\varLambda^c:=V\setminus\varLambda$ and its \emph{boundary} by
$\partial\varLambda:=\{x\in \varLambda^c\colon \exists\myp
y\in\varLambda,\ d(x,y)=1\}$, and we write
$\bar{\varLambda}=\varLambda\cup\partial\varLambda$. The subset of
edges in $\varLambda$ is denoted $E_\varLambda:=\{\langle
x,y\rangle\in E\colon x,y\in\varLambda\}$.

Fix a vertex $x_\circ\myn\in V$, interpreted as the \emph{root} of
the tree. We say that $y\in V$ is a \emph{direct successor} of $x\in
V$ if $x$ is the penultimate vertex on the unique path leading from
the root $x_\circ$ to the vertex $y$; that is,
$d(x_\circ,y)=d(x_\circ,x)+1$ and $d(x,y)=1$. The set of all direct
successors of $x\in V$ is denoted $S(x)$.  It is convenient to work
with the family of the radial subsets centred at $x_\circ$, defined
for $n\in\mathbb{N}_0:=\{0\}\cup \mathbb{N}$ by
\begin{equation*}
V_n:=\{x\in V\colon d(x_\circ,x)\leq n\},\qquad W_n:=\{x\in V\colon
d(x_\circ,x)=n\},
\end{equation*}
interpreted as the ``ball'' and ``sphere'', respectively, of radius
$n$ centred at the root $x_\circ$. Clearly, $\partial V_n=W_{n+1}$.
Note that if $x\in W_n$ then $S(x)=\{y\in W_{n+1}\colon d(x,y)=1\}$.
In the special case $x=x_\circ$ we have $S(x_\circ)=W_1$. For short,
we set $E_n:=E_{V_n}$.

\begin{remark}\label{rm:cofinal}
Note that the sequence of balls $(V_n)$ ($n\in\mathbb{N}_0$) is
\emph{cofinal} (see \cite[Section~1.2, page~17]{Ge}), that is, any
finite subset $\varLambda\subset V$ is contained in some $V_n$.
\end{remark}

\subsubsection{The Potts model and Gibbs measures}\label{sec:1.2.2}
In the $q$-state Potts model, the spin at each vertex $x\in V$ can
take values in the set $\varPhi:=\{1,\dots,q\}$. Thus, the spin
configuration on $V$ is a function $\sigma\colon V\!\to\varPhi$ and
the set of all configurations is $\varPhi^V$\mynn. For a subset
$\varLambda\subset V$, we denote by $\sigma_{\myn\varLambda}\colon
\varLambda\to\varPhi$ the restriction of configuration $\sigma$ to
$\varLambda$,
$$
\sigma_{\myn\varLambda}(x):=\sigma(x),\qquad x\in\varLambda.
$$
The Potts model with a \emph{nearest-neighbour interaction kernel}
$\{J_{xy}\}_{x,\myp y\myp\in\myp V}$ (i.e., such that
$J_{xy}=J_{yx}$ and $J_{xy}=0$ if $d(x,y)\ne1$) is defined by the
formal Hamiltonian
\begin{equation}\label{eq:H}
H(\sigma)=-\sum_{\langle x,\myp y\rangle\in E} \!J_{xy}
\mypp\delta_{\sigma(x),\myp\sigma(y)}-\sum_{x\in V}
\xi_{\sigma(x)}(x),\qquad \sigma\in\varPhi^V,
\end{equation}
where $\delta_{ij}$ is the Kronecker delta symbol (i.e.,
$\delta_{ij}=1$ if $i=j$ and $\delta_{ij}=0$ otherwise),
and $\boldsymbol{\xi}(x)=(\xi_{1}(x),\dots,\allowbreak
\xi_{q}(x))\allowbreak \in \mathbb{R}^q$ is the external (possibly
random) field. According to \eqref{eq:H}, the spin-spin interaction
is activated only when the neighbouring spins are equal, whereas the
additive contribution of the external field is provided, at each
vertex $x\in V$, by the component of the vector
$\boldsymbol{\xi}(x)$ corresponding to the spin value $\sigma(x)$.

For each finite subset $\varLambda\subset V$
($\varLambda\ne\varnothing$) and any fixed subconfiguration
$\eta\in\varPhi^{\varLambda^c}$ (called the \emph{configurational
boundary condition}), the \emph{Gibbs distribution}
$\gamma^\eta_{\myn\varLambda}$ is a probability measure in
$\varPhi^\varLambda$ defined by the formula
\begin{equation}\label{eq:GD}
\gamma^\eta_{\myn\varLambda}(\varsigma)=\frac{1}{Z^\eta_{\myn\varLambda}(\beta)}\exp\biggl\{-\beta
H_{\myn\varLambda}(\varsigma)+\beta\sum_{x\in\varLambda}\sum_{y\in
\varLambda_{\vphantom{t}}^{c}}\!J_{xy}\mypp\delta_{\varsigma(x),\myp\eta(y)}\biggr\},\qquad
\varsigma\in\varPhi^{\varLambda},
\end{equation}
where $\beta\in(0,\infty)$ is a parameter (having the meaning of
\emph{inverse temperature}), $H_{\myn\varLambda}$ is the restriction
of the Hamiltonian \eqref{eq:H} to subconfigurations in
$\varLambda$,
\begin{equation}\label{eq:H-Lambda}
H_{\myn\varLambda}(\varsigma)=-\sum_{\langle x,\myp y\rangle\in
E_\varLambda}\!\!J_{xy}
\mypp\delta_{\varsigma(x),\myp\varsigma(y)}-\sum_{x\in \varLambda}
\xi_{\varsigma(x)}(x),\qquad \varsigma\in\varPhi^{\varLambda},
\end{equation}
and $Z^\eta_{\myn\varLambda}(\beta)$ is the normalizing constant
(often called the \emph{canonical partition function}),
$$
Z^\eta_{\myn\varLambda}(\beta)=\sum_{\varsigma\in\varPhi^{\varLambda}}
\exp\myn\Biggl\{-\beta
H_{\myn\varLambda}(\varsigma)+\beta\sum_{x\in\varLambda}\sum_{y\in
\varLambda_{\vphantom{t}}^{c}}J_{xy}\mypp\delta_{\varsigma(x),\myp\eta(y)}\Biggr\}.
$$
Due to the nearest-neighbour interaction, formula \eqref{eq:GD} can
be rewritten as\footnote{It is also tacitly assumed in
\eqref{eq:GD1} that $\langle x,y\rangle\in E$, that is, $d(x,y)=1$.}
\begin{equation}\label{eq:GD1}
\gamma^\eta_{\myn\varLambda}(\varsigma)=\frac{1}{Z^\eta_{\myn\varLambda}(\beta)}\exp\myn\Biggl\{-\beta
H_{\varLambda}(\varsigma)+\beta\sum_{x\in\varLambda}\sum_{y\in\partial\varLambda}
\!J_{xy}\mypp\delta_{\varsigma(x),\myp\eta(y)}\Biggr\},\qquad
\varsigma\in\varPhi^{\varLambda}.
\end{equation}
Finally, a measure $\mu=\mu_{\beta,\myp\xi}$ on $\varPhi^V$ is
called a \emph{Gibbs measure} if, for any non-empty finite set
$\varLambda\subset V$ and any $\eta\in\varPhi^{\varLambda^c}\!$,
\begin{equation}\label{eq:Gibbs}
\mu(\sigma_{\myn\varLambda}=\varsigma\mypp|\mypp\sigma_{\myn\varLambda^{c}}\myn=\eta)\equiv
\gamma_{\myn\varLambda}^{\eta}\myn(\varsigma),\qquad
\varsigma\in\varPhi^{\varLambda}.
\end{equation}

\subsubsection{SGM construction}\label{sec:1.2.3}
It is convenient to construct Gibbs measures on the Cayley tree
$\mathbb{T}^k$ using a version of Gibbs distributions on the balls
$(V_n)$ defined via auxiliary fields encapsulating the interaction
with the exterior of the balls. More precisely, for a vector field
$V\ni x\mapsto \boldsymbol{h}(x)=(h_{1}(x),\dots, h_{q}(x))\in
\mathbb{R}^q$ and each $n\in\mathbb{N}_0$, define a probability
measure in $V_n$ by the formula
\begin{equation}\label{p*}
\mu^h_n(\sigma_n)=\frac{1}{Z_n}\exp\left\{-\beta
H_n(\sigma_n)+\beta\!\sum_{x\in W_n}
\!h_{\sigma_n(x)}(x)\right\},\qquad \sigma_n\in\varPhi^{V_n},
\end{equation}
where $Z_n=Z_n(\beta, \boldsymbol{h})$ is the normalizing factor and
$H_n:=H_{V_n}$, that is (see~\eqref{eq:H-Lambda}),
\begin{equation}\label{eq:Hn}
H_n(\sigma_n)=-\sum_{\langle x,y\rangle\in E_n} \!\!J_{xy}
\mypp\delta_{\sigma_n(x),\myp\sigma_n(y)}-\sum_{x\in
V_n}\xi_{\sigma_n(x)}(x),\qquad \sigma_n\in\varPhi^{V_n}.
\end{equation}
The vector field $\{\boldsymbol{h}(x)\}_{x\in V}$ in \eqref{p*} is
called \emph{generalized boundary conditions (GBC)}.

We say that the probability distributions \eqref{p*} are
\emph{compatible} (and the intrinsic GBC $\{\boldsymbol{h}(x)\}$ are
\emph{permissible}) if for each $n\in\mathbb{N}_0$ the following
identity holds,
\begin{equation}\label{p**}
\sum_{\omega\in
\varPhi^{W_{n+1}}}\!\mu^h_{n+1}(\sigma_n\myn\mynn\vee \omega)\equiv
\mu^h_{n}(\sigma),\qquad \sigma_n\in \varPhi^{V_{n}},
\end{equation}
where the symbol $\vee$ stands for concatenation of
subconfigurations. A criterion for permissibility of GBC is provided
by Theorem~\ref{ep} (see Section~\ref{sec:2.1} below). By
Kolmogorov's extension theorem (see, e.g., \cite[Chapter~II,
\S\mypp3, Theorem~4, page~167]{Shiryaev}), the compatibility
condition \eqref{p**} ensures that there exists a unique measure
$\mu^h=\mu^h_{\beta,\myp\xi}$ on $\varPhi^V$ such that, for all
$n\in\mathbb{N}_0$,
\begin{equation}\label{eq:mu-h}
\mu^h(\sigma_{V_n}\myn\mynn=\sigma_n)\equiv\mu^h_n(\sigma_n),\qquad
\sigma_n\in \varPhi^{V_n},
\end{equation}
or more explicitly (substituting~\eqref{p*}),
\begin{equation}\label{eq:mu-h-ex}
\mu^h(\sigma_{V_n}\!=\sigma_n)=\frac{1}{Z_n}\exp\left\{-\beta
H_n(\sigma_n)+\beta\sum_{x\in W_n}h_{\sigma_n(x)}(x)\right\},\qquad
\sigma_n\in \varPhi^{V_n}.
\end{equation}

It is easy to show that $\mu^h$ so defined is a \emph{Gibbs measure}
(see~\eqref{eq:Gibbs}); since the family $(V_n)$ is cofinal (see
Remark~\ref{rm:cofinal}), according to a standard result
\cite[Remark (1.24), page~17]{Ge} it suffices to check that, for
each $n\in\mathbb{N}_0$ and any
$\eta\in\varPhi^{V_{n\vphantom{t}}^{c}}$,
\begin{equation}\label{eq:Gibbs-n}
\mu^h(\sigma_{V_n}\myn\mynn=\sigma_n\myp|\mypp\sigma_{V_{n\vphantom{t}}^{c}}=\eta)\equiv
\gamma_{n}^{\eta}\myn(\sigma_n),\qquad \sigma_n\in\varPhi^{V_{n}},
\end{equation}
where $\gamma^{\eta}_n$ is the Gibbs distribution in $V_n$ with
configurational boundary condition $\eta$ (cf.~\eqref{eq:GD}).
Indeed, denote $\omega:=\eta_{\myp
W_{n+1}\mynn}\in\varPhi^{W_{n+1}}\mynn$, then, due to the
nearest-neighbour interaction in the Hamiltonian \eqref{eq:H} and
according to \eqref{eq:mu-h}, we have
\begin{align*}
\mu^h(\sigma_{V_n}\myn\mynn=\sigma_n\myp|\mypp\sigma_{V_{n\vphantom{t}}^{c}}=\eta)
&=\mu^h(\sigma_{V_n}\myn\mynn=\sigma_n\myp|\mypp\sigma_{W_{n+1}}\myn\mynn=\omega)\\
&=\frac{\mu^h(\sigma_{V_{n+1}\mynn}\mynn=\sigma_n\mynn\vee\omega)}{\mu^h(\sigma_{W_{n+1}}\myn\mynn=\omega)}\\
&=\frac{\mu^h_{n+1}(\sigma_n\mynn\vee\omega)}{\mu^h_{n+1}(\omega)}.
\end{align*}
Furthermore, recalling the definitions \eqref{eq:GD1} and
\eqref{p*}, and using the proportionality symbol $\propto$ to
indicate omission of factors not depending on $\sigma_{n}$, we
obtain
\begin{align}
\notag
\mu^h(\sigma_{V_n}\myn\mynn=\sigma_n\myp|\mypp\sigma_{V_{n\vphantom{t}}^{c}}=\eta)
&\propto \mu^h_{n+1}(\sigma_n\mynn\vee\omega)\\
\notag &\propto \exp\myn\Biggl\{-\beta
H_{n+1}(\sigma_n\mynn\vee\omega)+\beta\sum_{x\in
W_{n+1}}h_{\omega(x)}(x)\Biggr\}\\
\notag &\propto \exp\myn\Biggl\{-\beta
H_{n}(\sigma_n)+\beta\sum_{x\in W_n}\sum_{y\in S(x)}J_{xy}
\mypp\delta_{\sigma_n(x),\mypp\omega(y)}\Biggr\}\\
&\propto \gamma^{\eta}_{V_n}(\sigma_n), \label{eq:GD2}
\end{align}
and since both the left- and the right-hand sides of \eqref{eq:GD2}
are probability measures on $\varPhi^{V_n}\myn$, the relation
\eqref{eq:Gibbs-n} follows.

\begin{definition}
Measure $\mu^h$ satisfying \eqref{eq:mu-h} is called a
\emph{splitting Gibbs measure} (\emph{SGM}).
\end{definition}

The term \emph{splitting} was coined by Rozikov and Suhov
\cite{RoSu} to emphasize that, in addition to the Markov property
(see \cite[Section~12.1]{Ge} and also Remark~\ref{rm:Markov} below),
such measures enjoy the following factorization property:
conditioned on a fixed subconfiguration
$\sigma_n\mynn\in\varPhi^{V_n}$, the values $\{\sigma(x)\}_{x\in
W_{n+1}}$ are independent under the law~$\mu^h$. Indeed, using
\eqref{p*} and~\eqref{eq:mu-h}, it is easy to see that, for each
$n\in\mathbb{N}_0$ and any $\omega\in\varPhi^{W_{n+1}}$,
\begin{align*}
\mu^h\bigl(&\sigma_{W_{n+1}}\!=\omega\mypp |\mypp
\sigma_{V_n}\!=\sigma_n\bigr)\\
&\propto \prod_{x\in W_{n+1}} \exp\left\{\beta J_{x'x}
\mypp\delta_{\sigma_n(x'),\mypp\omega(x)}+\beta\myp
\xi_{\omega(x)}(x)+\beta\myp h_{\omega(x)}(x)\right\}\\
&\propto \prod_{x\in W_{n+1}} \mu^h\bigl(\sigma(x)=\omega(x)\myp
|\mypp \sigma_{V_n}=\sigma_n\bigr),
\end{align*}
where the proportionality symbol $\propto$ indicates omission of
factors not depending on $\omega$, and $x'\mynn=x'(x)\in W_n$ is the
unique vertex such that $x\in S(x')$.

\begin{remark}\label{rm:2.2}
Note that adding a constant $c=c(x)$ to all coordinates $h_i(x)$ of
the vector $\boldsymbol{h}(x)$ does not change the probability
measure \eqref{p*} due to the normalization $Z_n$. The same is true
for the external field $\boldsymbol{\xi}(x)$ in the Hamiltonian
\eqref{eq:Hn}. Therefore, without loss of generality we can consider
\emph{reduced GBC} $\boldsymbol{\check{h}}(x)$, for example defined
as
\begin{equation*}
\check{h}_i(x)=h_i(x)-h_q(x),\qquad i=1,\dots,q-1.
\end{equation*}
The same remark also applies to the external field
$\boldsymbol{\xi}$ and its reduced version
$\boldsymbol{\check{\xi}}(x)$, defined by
\begin{equation*}
\check{\xi}_i(x):=\xi_i(x)-\xi_q(x),\qquad i=1,\dots,q-1.
\end{equation*}
Of course, such a reduction can equally be done by subtracting any
other coordinate,
$$
{}_{\ell}\check{h}_i(x):=h_i(x)-h_\ell(x),\qquad
{}_{\ell}\check{\xi}_i(x):=\xi_i(x)-\xi_\ell(x)\qquad (i\ne \ell).
$$
\end{remark}

\begin{remark}
For $q=2$, the Potts model is equivalent to the Ising model with
redefined spins
$$
\tilde\sigma(x):=2\myp\sigma(x)-3 \in \{-1,1\},\qquad x\in V,
$$
whereby the Hamiltonians in the two models are linked through the
relations
$$
\delta_{\sigma(x),\myp\sigma(y)}=\frac{\tilde\sigma(x)\myp\tilde\sigma(y)+1}{2},\qquad
\xi_{\sigma(x)}(x)=\frac{\xi_2(x)-\xi_1(x)}{2}\,\tilde\sigma(x)+\frac{\xi_1(x)+\xi_2(x)}{2}.
$$
In turn, this leads to rescaling of the inverse temperature
$\beta=\frac12\mypp\tilde{\beta}$.
\end{remark}

\subsubsection{Boundary laws}\label{sec:1.2.4}
Let us comment on the link between the SGM construction outlined in
Section~\ref{sec:1.2.3} and an alternative (classical) approach to
defining Gibbs measures on tree-like graphs (including Cayley
trees), as presented in the book by Georgii~\cite[Chapter~12]{Ge}.
As was already mentioned in \cite[pages 641--642]{KR}, the family of
permissible GBC $\{\boldsymbol{h}(x)\}_{x\in V}$ defines a
\emph{boundary law} $\{\boldsymbol{z}(x,y)\}_{\langle x,y\rangle \in
E}$ in the sense of \cite[Definition~(12.10)]{Ge} (see also
\cite{Z0}); that is, for any $x,y\in V$ such that $\langle
x,y\rangle\in E$, and for all $i\in\varPhi$ it holds
\begin{equation}\label{eq:BL-def}
z_i(x,y)=c(x,y)\prod_{v\in\partial\{x\}\setminus\{y\}}\sum_{j\in\varPhi}
z_j(v,x)\myp\exp\bigl\{\beta
J_{xv}\myp\delta_{ij}+\beta\myp\xi_{i}(x)+\beta\myp\xi_{j}(v)\bigr\},
\end{equation}
where $c(x,y)>0$ is an arbitrary constant (not depending on
$i\in\varPhi$).

To see this, for any $y\in V$ and $x\in S(y)$ (so that
$d(x_\circ,y)=d(x_\circ,x)-1$), set
\begin{equation}\label{eq:z-in}
z_i(x,y):=\exp\{\beta\myp h_{i}(x)\},\qquad i\in\varPhi,
\end{equation}
which defines the values of the boundary law on ordered edges
$\langle x,y\rangle$ \emph{pointing to the root}~$x_\circ$. This
definition is consistent, in that the equation \eqref{eq:BL-def} is
satisfied (for such edges) due to the assumed permissibility of the
GBC $\{\boldsymbol{h}(x)\}_{x\in V}$ (see Theorem~\ref{ep}).

The values $z_i(x,y)$ on the edges $\langle x,y\rangle$
\emph{pointing away from the root} $x_\circ$ (i.e., such that
$d(x_\circ,y)=d(x_\circ,x)+1$) can be identified inductively (up to
proportionality constants) using formula~\eqref{eq:BL-def}. The base
of induction is set out by choosing $x=x_\circ$ and $y\in
\partial\{x_\circ\}=S(x_\circ)$. Then for all $v\in S(x_\circ)$
we have $z_i(v,x_\circ)=\exp\{\beta\myp h_i(v)\}$
(see~\eqref{eq:z-in}), and equation \eqref{eq:BL-def} yields
$$
z_i(x_\circ,y)=c(x_\circ, y) \prod_{v\in S(x_\circ)\setminus \{y\}}
\sum_{j\in\varPhi} \exp\bigl\{\beta\myp h_j(v)+\beta J_{x_\circ,\myp
v}
\myp\delta_{ij}+\beta\myp\xi_{i}(x_\circ)+\beta\myp\xi_{j}(v)\bigr\},
$$
which defines $z_i(x_\circ,y)$ ($i\in\varPhi$) up to an unimportant
constant factor. If $z_i(x,y)$ is already defined for all $x\in V_n$
and $y\in S(x)$, then for $x\in W_{n+1}$ and $y\in S(x)$ we have
$$
\partial\{x\}\setminus\{y\}=\bigl(S(x)\setminus\{y\}\bigr)\cup \{x'\},
$$
where $x'\in W_{n}$ is the unique vertex such that $x\in S(x')$.
Noting that the values $z_j(x'\mynn,x)$ ($j\in\varPhi$) are already
defined by the induction hypothesis and that
$z_j(v,x)=\exp\{\beta\myp h_j(v)\}$ for all $v\in S(x)$, formula
\eqref{eq:BL-def} yields $z_i(x,y)$ (again, up to a proportionality
constant), which completes the induction step.

\begin{remark}\label{rm:Markov}
As a consequence of the equivalence between (permissible) GBC
$\{\boldsymbol{h}(x)\}_{x\in V}$ and boundary laws
$\{\boldsymbol{z}(x,y)\}_{\langle x,y\rangle \in E}$, it follows
from \cite[Theorem~(12.12), page~243]{Ge} that any SGM $\mu^h$
determines a unique \emph{Markov chain} $\mu$ (see
\cite[Definition~(12.2), page~239]{Ge}), and vice versa, each Markov
chain $\mu$ defines a unique SGM $\mu^h$.
\end{remark}

\begin{remark}\label{rm:r2.1}
It is known that for each $\beta>0$ the Gibbs measures form a
non-empty convex compact set $\mathscr{G}$ in the space of all
probability measures on $\varPhi^V$ endowed with the weak topology
(see, \strut{}e.g., \cite[Chapter~7]{Ge}). A measure $\mu\in
\mathscr{G}$ is called \emph{extreme} if it cannot be expressed as
$\frac12\myp \mu_1+\frac12\myp\mu_2$ for some
$\mu_1,\mu_2\in\mathscr{G}$ with $\mu_1\ne\mu_2$. The set of all
extreme measures in \strut{}$\mathscr{G}$ \strut{}denoted by
$\ex\mathscr{G}$ is a \emph{Choquet simplex}, in the sense that any
$\mu\in\mathscr{G}$ can be represented \strut{}as $\mu =
\int_{\ex\mathscr{G}}\nu\,\rho(\dif{\nu})$, with some probability
measure $\rho$ on $\ex\mathscr{G}$. The crucial observation, which
\strut{}will be instrumental throughout the paper, is that, by
virtue of combining \cite[Theorem~(12.6)]{Ge} with
Remark~\ref{rm:Markov}, \emph{any extreme measure
$\mu\in\ex\mathscr{G}$ is SGM}; therefore, the question of
uniqueness of the Gibbs measure is reduced to that in the SGM class.
\end{remark}
Using the boundary law $\{\boldsymbol{z}(x,y)\}_{\langle
x,y\rangle\in E}$, formula \eqref{eq:mu-h-ex} can be extended to
more general subsets in $V$. Namely, according to
\strut{}\cite[formula~(12.13), page~243]{Ge} (adapted to our
notation), for any finite \emph{connected} set
$\varnothing\ne\varLambda\subset V$ (and
$\bar{\varLambda}=\varLambda\cup\partial\varLambda$),
\begin{equation}\label{eq:mu-h-ex-Lambda}
\mu^h(\sigma_{\myn\bar{\varLambda}}\myn=\varsigma)=\frac{1}{Z_{\myn\bar{\varLambda}}}\exp\myn\Biggl\{-\beta
H_{\myn\bar{\varLambda}}(\varsigma)+\beta\mynn\sum_{x\in
\partial\varLambda}h^\dag_{\varsigma(x)}(x,x_{\myn\varLambda})\Biggr\},\qquad \varsigma\in
\varPhi^{\bar{\varLambda}},
\end{equation}
where $Z_{\bar{\varLambda}}=Z_{\bar{\varLambda}}(\beta,
\boldsymbol{h})$ is the normalizing factor, $x_{\myn\varLambda}$
denotes the unique neighbour of $x\in\partial\varLambda$ belonging
to $\varLambda$, and
\begin{equation}\label{eq:h->z}
h^\dag_i(x,y):=\beta^{-1}\ln z_i(x,y),\qquad i\in\varPhi.
\end{equation}
In particular, if $x\in S(y)$ then (combining \eqref{eq:h->z}
with~\eqref{eq:z-in})
\begin{equation}\label{eq:h-inward}
h^\dag_i(x,y)=h_i(x),\qquad i\in\varPhi.
\end{equation}

To link the general expression \eqref{eq:mu-h-ex-Lambda} with
formula \eqref{eq:mu-h-ex} for balls $V_n$, consider part of the
boundary $\partial\varLambda$ defined as
\begin{equation}\label{eq:p-Lambda-circ}
\partial\varLambda^{\downarrow}:=
\{x\in \partial\varLambda\colon S(x)\cap\varLambda=\varnothing\}.
\end{equation}
In other words, $\partial\varLambda^{\downarrow}$ consists of the
points $x\in\partial\varLambda$ such that the corresponding vertex
$x_{\myn\varLambda}\in\varLambda$ is closer to the root $x_\circ$
than $x$ itself. In view of the definition \eqref{eq:z-in}, for
$x\in\partial\varLambda^{\downarrow}$ we get
$h^\dag_{i}(x,x_{\myn\varLambda})\equiv h_{i}(x)$. Clearly, if
$x_\circ\in\varLambda$ then
$\partial\varLambda^\downarrow=\partial\varLambda$, but if
$x_\circ\notin\varLambda$ then the set
$\partial\varLambda\setminus\partial\varLambda^{\downarrow}$ is
non-empty and, moreover, it contains exactly one vertex, which we
denote by~$\check{x}$. Note that $\check{x}\in\partial\varLambda$ is
closer to the root $x_\circ$ than
$\check{x}_{\myn\varLambda}\in\varLambda$, and in this case
$h^\dag_{i}(\check{x},\check{x}_{\myn\varLambda})$ is only
expressible through the GBC $\{\boldsymbol{h}(x)\}$ via a recursive
procedure, as explained above.

\begin{figure}[th]
\includegraphics
[height=6cm]{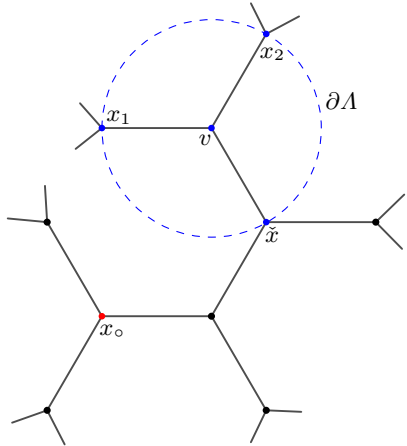}%
\put(-116.1,44.0){\mbox{\scriptsize $\textstyle x_\circ$}}%
\put(-79,115.5){\mbox{\scriptsize $v$}}%
\put(-113.6,124.5){\mbox{\scriptsize $x_1$}}%
\put(-56.1,148.6){\mbox{\scriptsize $x_2$}}%
 \put(-54.2,79.0){\mbox{\scriptsize $\check{x}$}}%
 \put(-31.5,129.4){\mbox{\scriptsize $\partial\varLambda$}}%
\put(-115.4,80.7){\lefteqn{\mbox{\begin{tikzpicture}
\draw[blue,
dashed] (0,0) circle (3.42pc);
\end{tikzpicture}}}}
\caption{Illustration of the sets in formula
\eqref{eq:mu-h-ex-Lambda-inside} relative to the root $x_\circ$:
$\varLambda=\{v\}$, $\partial\varLambda=\{x_1,x_2,\check{x}\}$ (here
$\check{x}_{\myn\varLambda}=v$),
$\bar\varLambda=\{v,x_1,x_2,\check{x}\}$, and
$\partial\varLambda^\downarrow=\{x_1,x_2\}$
(see~\eqref{eq:p-Lambda-circ}).}\label{Fig1}
\end{figure}

Thus, formula \eqref{eq:mu-h-ex-Lambda} can be represented more
explicitly as follows,
\begin{equation}\label{eq:mu-h-ex-Lambda-inside}
\mu^h(\sigma_{\myn\bar{\varLambda}}\myn=\varsigma)
=\frac{1}{Z_{\myn\bar{\varLambda}}}\exp\myn\Biggl\{-\beta
H_{\myn\bar{\varLambda}}(\varsigma)+\beta\!\sum_{x\in
\partial\varLambda^\downarrow}\!h_{\varsigma(x)}(x)+\beta\!\!\sum_{x\in
\partial\varLambda\setminus\partial\varLambda^\downarrow}
\!h^\dag_{\varsigma(x)}(x,x_{\myn\varLambda})\Biggr\}.
\end{equation}
In fact, the last sum in \eqref{eq:mu-h-ex-Lambda-inside} includes
at most one term, which corresponds to $x=\check{x}$; more
precisely, the latter sum is vacuous whenever
$x_\circ\in\varLambda$, in which case the first sum in
\eqref{eq:mu-h-ex-Lambda-inside} is reduced to the sum over all
$x\in\partial\varLambda$. In particular, the formula
\eqref{eq:mu-h-ex} is consistent with
\eqref{eq:mu-h-ex-Lambda-inside} by picking the set
$\varLambda=V_{n-1}$ ($n\ge1$), with boundary
$\partial{\varLambda}=W_n$. For a graphical illustration of the sets
involved in formula \eqref{eq:mu-h-ex-Lambda-inside}, see
Figure~\ref{Fig1} (for a single-vertex set $\varLambda=\{v\}$ with
$v\ne x_\circ$).

\subsubsection{Layout} The rest of the paper is organized as
follows (cf.\ the table of contents). We state our main results in
Section~\ref{sec:2}, starting with a general \emph{compatibility
criterion} (Theorem~\ref{ep}), which reduces the existence of SGM
$\mu^h$ to the solvability of an infinite system of non-linear
equations for permissible GBC $\{\boldsymbol{h}(x)\}$. This is
followed by various sufficient conditions for uniqueness of SGM with
\emph{uniform ferromagnetic interaction} (Theorems \ref{t2},
\ref{t3} and~\ref{t4}). As part of our general treatment of the
Potts model on the Cayley tree, in Section~\ref{sec:2.3.1} we
introduce the notion of translation-invariant SGMs (based on a
bijection between $\mathbb{T}^k$ and a free group with $k+1$
generators of period $2$ each), and state a novel criterion of
translation invariance (Proposition~\ref{pr:TI}) in terms of the
external field and the GBC.

Non-uniqueness results for a subclass of \emph{completely
homogeneous SGMs} (i.e., where the reduced fields
$\{\check{\boldsymbol{\xi}}(x)\}$ and
$\{\check{\boldsymbol{h}}(x)\}$ are constant) are summarized in
Theorems \ref{th:3.7}, \ref{th:3.8} and~\ref{th:non-U}. The number
of such measures is estimated in several special cases by $2^q-1$
(Theorem~\ref{th:k=2}), and we conjecture that this is a universal
upper bound. In Section~\ref{sec:3}, we record some auxiliary
lemmas. The proofs of the uniqueness results (Theorems \ref{t2},
\ref{t3} and~\ref{t4}) are presented in Section~\ref{sec:4}.
Section~\ref{sec:5} is devoted to the in-depth analysis of
completely homogeneous SGMs, culminating in the proof of
Theorems~\ref{th:3.7}--\ref{th:k=2} (given in
Sections~\ref{sec:5.2}--\ref{sec:5.5}, respectively). In
Section~\ref{sec:6}, we study some fine properties (such as
monotonicity, bounds and zeros) of the critical curves on the
temperature--field plane, summarized in Propositions
\ref{pr:alpha-pm-limits}, \ref{pr:alpha-pm><},
\ref{pr:theta-symmetry}--\ref{pr:K-monotone}, \ref{pr:alpha-zeros}
and~\ref{pr:alpha-pm_q=2}. Finally, Appendix~\ref{sec:A} presents
the proof of Proposition~\ref{pr:TI}, while Appendix~\ref{sec:B} is
devoted to the proof of a technical Lemma~\ref{lm:q=3} addressing
the special case $q=3$.

\section{Results}\label{sec:2}
\subsection{Compatibility criterion}\label{sec:2.1}
In view of Remark \ref{rm:2.2}, when working with vectors and
vector-valued functions and fields it will often be convenient to
pass from a generic vector $\boldsymbol{u}=(u_1,\dots,u_q)\in
\mathbb{R}^q$ to a ``reduced vector''
$\boldsymbol{\check{u}}=(\check{u}_1,\dots,\check{u}_{q-1})\in
\mathbb{R}^{q-1}$ by setting $\check{u}_i:=u_i-u_q$
($i=1,\dots,q-1$).

The following general statement describes a
criterion\footnote{Earlier versions of this theorem are found in
\cite[Proposition~1, page~375]{GaRo} or \cite[Theorem~5.1,
page~106]{Ro}.} for the GBC $\{\boldsymbol{h}(x)\}_{x\in V}$ to
guarantee compatibility of the measures
$\{\mu^h_n\}_{n\in\mathbb{N}_0}$.

\begin{theorem}\label{ep}
The probability distributions $\{\mu^h_n\}_{n\in\mathbb{N}_0}$
defined in \eqref{p*} are compatible (and the underlying GBC
$\{\boldsymbol{h}(x)\}_{x\in V}$ are permissible) if and only if the
following vector identity holds
\begin{equation}\label{p***}
\beta\myp\boldsymbol{\check{h}}(x)=\sum_{y\in
S(x)}\boldsymbol{F}\bigl(\beta\myp\boldsymbol{\check{h}}(y)+\beta\myp\boldsymbol{\check{\xi}}(y);
\myp\rme^{\beta J_{xy}}\bigr),\qquad x\in V,
\end{equation}
where
$\boldsymbol{\check{h}}(x)=(\check{h}_{i}(x),\dots,\check{h}_{q-1}(x))$,
\,$\boldsymbol{\check{\xi}}(x)=(\check{\xi}_{i}(x),\dots,\check{\xi}_{q-1}(x))$,
\begin{equation}\label{hxi}
\check{h}_{i}(x):=h_{i}(x)-h_{q}(x), \qquad \check
\xi_{i}(x):=\xi_{i}(x)-\xi_{q}(x),\qquad i=1,\dots,q-1,
\end{equation}
and the map
$\boldsymbol{F}(\boldsymbol{u};\theta)=(F_1(\boldsymbol{u};\theta),\dots,F_{q-1}(\boldsymbol{u};\theta))$
is defined for
$\boldsymbol{u}=(u_1,\dots,u_{q-1})\in\mathbb{R}^{q-1}$ and
$\theta>0$ by the formulas
\begin{equation}\label{eq:Fi}
F_i(\boldsymbol{u};\theta):=\ln\frac{(\theta-1)\mypp\rme^{u_i}+1+\sum_{j=1}^{q-1}\rme^{u_j}}{\theta+
\sum_{j=1}^{q-1}\rme^{u_j}},\qquad i=1,\dots, q-1.
\end{equation}
\end{theorem}

\begin{remark}
Likewise, Theorem \ref{ep} is true for any of the $q$ possible
reductions (see Remark~\ref{rm:2.2}).
\end{remark}

\begin{remark}\label{rm:F(0)}
Note that $\boldsymbol{F}(\boldsymbol{0};\theta)=\boldsymbol{0}$ for
any $\theta>0$.
\end{remark}

\begin{remark}
In view of the link (discussed in Section~\ref{sec:1.2.4}) between
GBC $\{\boldsymbol{h}(x)\}_{x\in V}$ and boundary laws
$\{\boldsymbol{z}(x,y)\}_{\langle x,y\rangle\in E}$, the
compatibility criterion \eqref{p***} is but a reformulation of the
consistency property \eqref{eq:BL-def} of the boundary law.
\end{remark}

By virtue of Theorem \ref{ep}, if the GBC $\{\boldsymbol{h}(x)\}$
and the external field $\{\boldsymbol{\xi}(x)\}$ satisfy the
functional equation \eqref{p***} for some $\beta>0$ then there is a
(unique) SGM $\mu^{h}_{\beta,\myp\xi}$.

\subsection{Uniqueness results}\label{sec:2.2}
From now on, we confine ourselves to the case of \emph{uniform}
(ferromagnetic) nearest-neighbour interaction by setting
$J_{xy}=J\geq 0$ if $d(x,y)=1$ (and $J_{xy}=0$ otherwise). It will also
be convenient to re-parameterize the model by introducing the new
parameter $\theta=\rme^{\beta J}\ge1$ termed \emph{activity}.

\smallskip
For $\theta\ge 1$, consider the function
\begin{equation}\label{eq:phi}
\varphi(t;\theta):=\frac{(\theta-1)\mypp t}{\bigl(\sqrt{\theta\myp
(t-1)}+\sqrt{t-\theta}\mypp\bigr)^2},\qquad t\ge \theta,
\end{equation}
which can also be written as
\begin{equation}\label{eq:phi1}
\varphi(t;\theta)= \frac{\sqrt{\theta\myp(t-1)}-\sqrt{t-\theta}}
{\sqrt{\theta\myp(t-1)}+\sqrt{t-\theta}},\qquad t\ge \theta.
\end{equation}
Noting that
\begin{equation*}
\varphi(t;\theta)=\frac{\theta-1}
{\left(\sqrt{\theta-\theta/t}+\sqrt{1-\theta/t}\mypp\right)^2},
\end{equation*}
it is evident that $t\mapsto \varphi(t;\theta)$ is a decreasing
function; in particular, for all $t\ge\theta$
\begin{equation}\label{eq:>phi>}
1=\varphi(\theta;\theta)\ge\varphi(t;\theta)\ge
\varphi(\infty;\theta)=\frac{\sqrt{\theta}-1}{\sqrt{\theta}+1}.
\end{equation}

For brevity, introduce the notation
\begin{equation}\label{eq:Q}
Q(\theta):=(q-2)\mypp\frac{\sqrt{\theta}-1}{\sqrt{\theta}+1},
\end{equation}
and for $k\ge2$, $q\ge2$ consider the equation
\begin{equation}\label{eq:theta0phi}
Q(\theta)+\varphi(\theta+1;\theta)=\frac{1}{k},\qquad \theta\ge 1,
\end{equation}
or more explicitly (noting that
$\varphi(\theta+1;\theta)=(\theta-1)/(\theta+1)$),
\begin{equation}\label{eq:theta0}
Q(\theta)+\frac{\theta-1}{\theta+1}=\frac{1}{k},\qquad \theta\ge 1.
\end{equation}
The left-hand side of \eqref{eq:theta0} is a continuous increasing
function of $\theta\in[1,\infty)$ ranging from $0$ to $q-1>k^{-1}$,
which implies that there is a unique solution of the equation
\eqref{eq:theta0}, denoted $\theta_0=\theta_0(k,q)$. In particular,
for $q=2$ we get
\begin{equation}\label{eq:theta0q=2}
\theta_0(k,2)=\frac{k+1}{k-1}.
\end{equation}

Let us also consider the equation
\begin{equation}\label{eq:eq1}
(q-1)\mypp\frac{\theta-1}{\theta+1}=\frac{1}{k},
\end{equation}
which has the unique root
\begin{equation}\label{eq:theta1}
\theta_*=\theta_*(k,q):=\frac{k(q-1)+1}{k(q-1)-1}.
\end{equation}
Noting from \eqref{eq:Q} that, for any $\theta>1$,
$$
(q-1)\mypp \frac{\sqrt{\theta}-1}{\sqrt{\theta}+1}<
Q(\theta)+\frac{\theta-1}{\theta+1}\le
(q-1)\mypp\frac{\theta-1}{\theta+1},
$$
and comparing equations \eqref{eq:theta0} and \eqref{eq:eq1}, it
follows that
\begin{equation*}
\theta_*(k,q)\le \theta_0(k,q)< \left(\theta_*(k,q)\right)^2,
\end{equation*}
where the first inequality is in fact strict unless $q=2$.

\begin{theorem}\label{t2}
Let $\theta_0=\theta_0(k,q)$ be the  unique solution of the
equation~\eqref{eq:theta0}. Then the Gibbs measure
$\mu_{\theta,\myp\xi}$ is unique for $\theta\in(1,\theta_0)$ and any
external field $\boldsymbol{\xi}$.
\end{theorem}

\begin{remark}
It is known that the Ising model on a Cayley tree with zero external
field has a unique Gibbs measure if and only if \strut{}$\theta\le
\theta_{\rm c}(k)=\sqrt{1+\vphantom{(^T}\smash{\frac{2}{k-1}}}$
\,(see \cite{BRZ}); that is to say, $\theta_{\rm c}(k)$ is the
\emph{critical activity} of the Ising model. Since
\strut{}$\theta_{\rm c}(k)=\theta_0(k,2)$, our Theorem \ref{t2} is
sharp in this case. Let $\theta_{\rm cr}(k,q)$ be the critical
activity for the Potts model; its exact value is known only for the
binary tree ($k=2$), namely $\theta_{\rm
cr}(2,q)=1+2\sqrt{q-1}\,$~\cite{KRK}. Note that $\theta_{\rm
cr}(2,2)=\theta_0(2,2)\,(= 3)$ but $\theta_{\rm
cr}(2,q)>\theta_0(2,q)$ for $q\ge 3$, so Theorem \ref{t2} is not
sharp already for $k=2$, $q\ge 3$.
\end{remark}

For $k\geq 2$, $q\geq 2$ and any $\gamma\in\mathbb{R}$, consider the
equation (cf.\ \eqref{eq:theta0phi})
\begin{equation}\label{eq:equation-alpha*}
Q(\theta)+
\varphi\bigl(t_\gamma(\theta);\theta\bigr)=\frac{1}{k},\qquad
\theta\ge1,
\end{equation}
where $\varphi(t;\theta)$ and $Q(\theta)$ are defined in
\eqref{eq:phi} and \eqref{eq:Q}, respectively, and
\begin{equation}\label{eq:theta_K}
t_\gamma(\theta):=\theta+1+(q-2)\mypp\theta^\gamma.
\end{equation}
It can be shown (see Lemma \ref{lm:unique}) that equation
\eqref{eq:equation-alpha*} has a unique root,
$\theta^*_{\myn\gamma}=\theta^*_{\myn\gamma}(k,q)$. More
specifically, if $q=2$ then $t_\gamma(\theta)=\theta+1$ and equation
\eqref{eq:equation-alpha*} is reduced to equation
\eqref{eq:theta0phi} (with $Q(\theta)\equiv 0$), so that
$\theta_{\gamma}^*(k,2)\equiv \theta_0(k,2)=(k+1)/(k-1)$
(see~\eqref{eq:theta0q=2}). However, if $q\ge 3$ then the root
$\theta^*_{\myn\gamma}$ is an increasing function of parameter
$\gamma$ with the asymptotic bounds
\begin{equation}\label{eq:bounds_for_theta*}
\theta_0(k,q)=\lim_{\gamma\to-\infty}\theta^*_{\myn\gamma}(k,q)<\theta^*_{\myn\gamma}(k,q)<
\lim_{\gamma\to+\infty}\theta^*_{\myn\gamma}(k,q)=\left(\theta_*(k,q)\right)^2.
\end{equation}

\begin{definition}\label{def:Delta}
Given the external field
$\boldsymbol{\xi}(x)=(\xi_1(x),\dots,\xi_q(x))$ ($x\in V$), define
the asymptotic ``gap'' between its coordinates as follows,
\begin{equation}\label{eq:Delta}
\varDelta^\xi:=\max_{1\le \ell\le q}\liminf_{x\in V}\,{}_\ell
\check{\xi}_{(1)}(x),
\end{equation}
where
\begin{equation}\label{eq:xi_min}
{}_\ell\check{\xi}_{(1)}(x):=\min_{i\ne\ell}
{}_\ell\check{\xi}_i(x)\equiv\min_{i\ne\ell}
(\xi_i(x)-\xi_\ell(x)),\qquad x\in V.
\end{equation}
\end{definition}
\begin{theorem}\label{t3}
The Gibbs measure $\mu_{\beta,\myp\xi}$ is unique for any
$\beta\in\bigl(0,\ln \theta^*_{\mynn\varDelta^\xi-k}\bigr)$, where
$\theta_{\gamma}^*\!$ denotes the unique  solution of the equation
\eqref{eq:equation-alpha*}.
\end{theorem}

\begin{remark}\label{rm:q=2}
As already mentioned, if $q=2$ then
$\theta_{\gamma}^*(k,2)\equiv\theta_0(k,2)$ and we recover Theorem
\ref{t2} in this case. But if $q\ge 3$ then
$\theta_{\mynn\myp\gamma}^*(k,q)>\theta_0(k,q)$ for any
$\gamma\in\mathbb{R}$ (see~\eqref{eq:bounds_for_theta*}), so that
Theorem \ref{t3} ensures the uniqueness of the SGM
$\mu_{\beta,\myp\xi}$ on a wider interval of temperatures as
compared to Theorem~\ref{t2}, for any $\varDelta^\xi$. Moreover, due
to the monotonicity of the map
$\gamma\mapsto\theta_{\mynn\myp\gamma}^*$ (Lemma \ref{lm:unique}), a
larger gap $\varDelta^\xi$ facilitates uniqueness of SGM; however,
the domain of guaranteed uniqueness (in parameter $\theta$) is
bounded in all cases (see~\eqref{eq:bounds_for_theta*}) by
$(\theta_*(k,q))^2\le (\theta_*(2,3))^2=25/9\doteq 2.7778$.
\end{remark}

\begin{remark} If the external field $\boldsymbol{\xi}$ is random then
the gap \eqref{eq:Delta} is a random variable measurable with
respect to the ``tail'' $\sigma$-algebra
$\mathscr{F}^\infty=\bigcap_{n=0}^\infty
\sigma\{\boldsymbol{\xi}(x),\,x\in V_n^c\}$. Intuitively, this means
that $\varDelta^\xi$ does not depend on the values of \strut{}the
field $\boldsymbol{\xi}(x)$ on any finite set $\varLambda\subset V$.
If the values of $\boldsymbol{\xi}(x)$ are assumed to be independent
(not necessarily identically distributed) for different $x\in V$
then, by Kolmogorov's zero--one law, $\varDelta^\xi=\const$ (and
therefore $\theta^*_{\mynn\varDelta^\xi-k}=\const$) almost surely
(a.s.).
\end{remark}

\begin{example}\label{ex:Delta}
Let us compute the asymptotic gap $\varDelta^\xi$ in a few examples.
\begin{itemize}
\item[(a)] Let the random vectors $\boldsymbol{\xi}(x)$ ($x\in V$) be mutually
independent, with independent and identically distributed (i.i.d.)\
coordinates $\xi_i(x)$ ($i=1,\dots,q)$, each taking the values
$\pm1$ with probabilities $\frac12$. Note that
${}_\ell\check{\xi}_{(1)}(x)\in\{0,\pm2\}$ \,($\ell=1,\dots,q$) and
\begin{align}
\notag \PP\bigl({}_\ell\check{\xi}_{(1)}(x)=-2\bigr)&=
\PP\Bigl(\xi_\ell(x)=1,\,\min_{i\ne\ell}
\xi_i(x)=-1\Bigr)\\
\notag
&=\PP\bigl(\xi_\ell(x)=1\bigr)\cdot\bigl(1-\PP(\xi_i(x)=1,\,i\ne\ell)\bigr)\\
&=\tfrac12\Bigl(1-\left(\tfrac12\right)^{q-1}\Bigr)>0.
\label{eq:=-1/2}
\end{align}
The Borel--Cantelli lemma then implies that $\liminf_{x\in
V}{}_\ell\check{\xi}_{(1)}(x)=-2$ a.s.\ and hence, according to
\eqref{eq:Delta}, $\varDelta^\xi=-2$ a.s.

\smallskip
\item[(b)] In the previous example, let us remove the i.i.d.\
assumption for the coordinates, and instead suppose that each of the
random vectors $\boldsymbol{\xi}(x)$ (still mutually independent for
different $x\in V$) can take two values, $\pm(1,\dots,1)$, with
probability $\frac12$ each. Then it is clear that
${}_\ell\check{\xi}_{(1)}(x)\equiv0$ ($\ell=1,\dots,q$), hence
$\varDelta^\xi=0$ a.s.

\smallskip
\item[(c)] \strut{}Extending example (b), suppose
that, with some $\alpha\in\mathbb{R}$,
\begin{equation*}
\PP\bigl(\xi_i(x)=\alpha\ \text{and}\ \xi_j(x)=0\ \text{for all}\
j\ne i\bigr)=q^{-1}\qquad (i=1,\dots,q).
\end{equation*}
Of course, for $\alpha=0$ we have $\boldsymbol{\xi}(x)\equiv
\boldsymbol{0}$ and hence $\varDelta^\xi=0$; thus, let $\alpha\ne0$.
If $q=2$ then it is straightforward to see that
$$
\PP\bigl({}_\ell\check{\xi}_{(1)}(x)=\pm\alpha\bigr)=\tfrac12,\qquad\ell=1,2.
$$
For $q\ge3$, if $\alpha>0$ then,
similarly,
\begin{equation}\label{eq:check+}
\PP\bigl({}_\ell\check{\xi}_{(1)}(x)=-\alpha\bigr)=q^{-1},\qquad
\PP\bigl({}_\ell\check{\xi}_{(1)}(x)=0\bigr)=1-q^{-1},
\end{equation}
whereas if $\alpha<0$ then
\begin{equation}\label{eq:check-}
\PP\bigl({}_\ell\check{\xi}_{(1)}(x)=-\alpha\bigr)=q^{-1},\qquad
\PP\bigl({}_\ell\check{\xi}_{(1)}(x)=\alpha\bigr)=1-q^{-1}.
\end{equation}
Thus, in all cases, the Borel--Cantelli lemma yields that
$\varDelta^\xi=-|\alpha|$ a.s.\ (which also includes the case
$\alpha=0$).

\smallskip
\item[(d)] Consider i.i.d.\ vectors $\boldsymbol{\xi}(x)$ ($x\in V$) with i.i.d.\
coordinates $\xi_i(x)$ ($i=1,\dots,q)$, each with the uniform
distribution on $[0,1]$. Note that $-1\le
{}_\ell\check{\xi}_{(1)}(x)\le 1$ ($\ell=1,\dots,q$) and, for any
$\varepsilon\in(0,1)$,
\begin{align*}
\PP\bigl({}_\ell\check{\xi}_{(1)}(x)\le -1+2\varepsilon\bigr)&\ge
\PP\Bigl(\xi_\ell(x)\ge
1-\varepsilon,\,\min_{i\ne\ell}\xi_i(x)\le\varepsilon\Bigr)\\
&=\PP\bigl(\xi_\ell(x)\ge 1-\varepsilon\bigr)\cdot\Bigl(1-
\PP\bigl\{\min_{i\ne\ell}\xi_i(x)\ge\varepsilon\bigr\}\Bigr)\\
&=\varepsilon\left(1-(1-\varepsilon)^{q-1}\right)>0.
\end{align*}
The Borel--Cantelli lemma then implies that $\liminf_{x\in
V}{}_\ell\check{\xi}_{(1)}(x)\le -1+2\varepsilon$ a.s., and since
$\varepsilon>0$ is arbitrary, it follows that $\liminf_{x\in
V}{}_\ell\check{\xi}_{(1)}(x)=-1$ a.s., for each $\ell=1,\dots,q$;
hence, according to \eqref{eq:Delta}, $\varDelta^\xi=-1$
a.s.

\smallskip
\item[(e)] For a ``non-ergodic'' type of example leading to
a random gap $\varDelta^\xi$, suppose that
$\boldsymbol{\xi}(x)\equiv \boldsymbol{\xi}(x_\circ)$ ($x\in V$),
where the distribution of $\boldsymbol{\xi}(x_\circ)$ is as in
example~(a). That is to say, the values of the field
$\boldsymbol{\xi}(x)$ are obtained by duplicating its (random) value
at the root. Then, similarly to \eqref{eq:=-1/2}, we compute
\begin{align*}
\PP(\varDelta^\xi=-2)= \tfrac12-\left(\tfrac12\right)^{q},\qquad
\PP(\varDelta^\xi=2)= \left(\tfrac12\right)^{q},\qquad
\PP(\varDelta^\xi=0)=\tfrac12.
\end{align*}

\smallskip
\item[(f)] Finally, the simplest ``coordinate-oriented'' choice
$\boldsymbol{\xi}(x)\equiv(\alpha,0,\dots,0)$, $x\in V$, with a
fixed $\alpha\in\mathbb{R}$, exemplifies translation-invariant
(non-random) external fields, including the case of zero field,
$\alpha=0$. Our results for this model will be stated in
Section~\ref{sec:2.3}; for now, let us calculate the value of the
gap $\varDelta^\xi$. Again, for $\alpha=0$ we have
$\boldsymbol{\xi}(x)\equiv \boldsymbol{0}$ and hence
$\varDelta^\xi=0$; thus, let $\alpha\ne0$. If $q=2$ then
${}_1\check{\xi}_{(1)}(x)=-\alpha$,
${}_2\check{\xi}_{(1)}(x)=\alpha$, hence it is easy to see that
$\varDelta^\xi=\max \{-\alpha,\alpha\}=|\alpha|$. For $q\ge 3$, if
$\alpha>0$ then ${}_1\check{\xi}_{(1)}(x)=-\alpha$ and
${}_\ell\check{\xi}_{(1)}(x)=0$ ($\ell\ne1$), whereas if $\alpha<0$
then still ${}_1\check{\xi}_{(1)}(x)=-\alpha$ but
${}_\ell\check{\xi}_{(1)}(x)=\alpha$ ($\ell\ne1$); as a result,
$\varDelta^\xi=0$ for $\alpha\ge0$ and $\varDelta^\xi=|\alpha|$ for
$\alpha<0$.
\end{itemize}
\end{example}

\smallskip
The following general assertion summarizes Example \ref{ex:Delta}.
Recall that random variables $X_1,\dots,X_q$ are said to be
\emph{exchangeable} if the distribution of the random vector
$(X_1,\dots,X_q)$ is invariant with respect to permutations of the
coordinates. The \emph{support} $\supp X$ of (the distribution of) a
random variable $X$ is defined as the (closed) set comprising all
points $u\in\mathbb{R}$ such that for any $\varepsilon>0$ we have
$\mathbb{P}(|X-u|\le\varepsilon)>0$.
\begin{proposition}\label{pr:Delta}
Suppose that the random vectors $\{\boldsymbol{\xi}(x)\}_{x\in V}$
are i.i.d., and for each $x\in V$ their coordinates
$\xi_1(x),\dots,\xi_q(x)$ are exchangeable. Then
$$
\varDelta^\xi=\inf\{\supp \left(\xi_{1}(x)-\xi_{q}(x)\right)\}\quad
\text{a.s.}
$$
In particular, $\varDelta^\xi<0$ a.s., unless
$\xi_1(x)=\dots=\xi_q(x)$ a.s., in which case  $\varDelta^\xi=0$
a.s.
\end{proposition}

\proof Observe that, by exchangeability of $\{\xi_i(x)\}$, the
distribution of ${}_\ell\xi_{(1)}(x)$ does not depend on
$\ell=1,\dots,q$ and, moreover,
\begin{equation}\label{eq:supp}
\supp {}_\ell\check{\xi}_{(1)}(x)=\supp
\left(\xi_{1}(x)-\xi_{q}(x)\right).
\end{equation}
Denote $u_0:=\inf\{\supp \left(\xi_{1}(x)-\xi_{q}(x)\right)\}$. From
\eqref{eq:supp}, it follows that ${}_\ell\check{\xi}_{(1)}(x)\ge
u_0$ a.s., and therefore, according to \eqref{eq:Delta},
$\varDelta^\xi\ge u_0$ a.s. On the other hand, for any
$\varepsilon>0$ we have
\begin{equation*}
\PP\bigl({}_\ell\check{\xi}_{(1)}(x)\le u_0+\varepsilon\bigr)=
\PP\bigl(\xi_{1}(x)-\xi_{q}(x)\le u_0+\varepsilon\bigr)>0,
\end{equation*}
and the Borel--Cantelli lemma implies that $\liminf_{x\in
V}{}_\ell\check{\xi}_{(1)}(x)\le u_0+\varepsilon$ a.s., so that
$\varDelta^\xi\le u_0$ a.s. As a result, $\varDelta^\xi=u_0$ a.s.,
as claimed.
\endproof

\begin{theorem}\label{t4}
Assume that the random external field
$\boldsymbol{\xi}=\{\boldsymbol{\xi}(x)\}_{x\in V}$ is as in
Proposition~\textup{\ref{pr:Delta}}. Let
$\theta^{\dag}=\theta^{\dag}(k,q)$ be the root of the
equation\footnote{The left-hand side of
\eqref{eq:equation-alpha-new} does not depend on $x\in V$ due to the
i.i.d.\ assumption on $\{\boldsymbol{\xi}(x)\}$.}
\begin{equation}\label{eq:equation-alpha-new}
Q(\theta)+\mathbb{E}\bigl\{\varphi\bigl(t_{\check{\xi}_{(1)}(x)-k}(\theta);\theta\bigr)\bigr\}
=\frac{1}{k},\qquad \theta\ge1,
\end{equation}
where
$\check{\xi}_{(1)}(x)\equiv{}_q\myp\check{\xi}_{(1)}(x)=\min_{i\ne
q} \bigl(\xi_i(x)-\xi_q(x)\bigr)$
\textup{(}cf.~\eqref{eq:xi_min}\textup{)} and the notation
$t_\gamma(\theta)$ is introduced in~\eqref{eq:theta_K}. Then, for
each $\theta\in[1,\theta^{\dag})$ and for $\mathbb{P}$-almost all
realizations of the random field $\boldsymbol{\xi}$, there is a
unique Gibbs measure $\mu_{\theta,\myp\xi}$.
\end{theorem}

\begin{remark}\label{rm:3.5}
Note that Theorem \ref{t3} guarantees uniqueness of the Gibbs
measure in the interval $1<\theta<\theta^*_{\mynn\varDelta^\xi-k}$,
where $\theta^*_{\mynn\varDelta^\xi-k}$ is the solution of the
equation
\begin{equation*}
Q(\theta)+\varphi\bigl(t_{\myn\varDelta^\xi-k}(\theta);\theta\bigr)=\frac{1}{k},\qquad
\theta\ge1.
\end{equation*}
By Proposition \ref{pr:Delta}, we have $\check{\xi}_{(1)}(x)\ge
\varDelta^\xi$ (a.s.), and moreover, $\check{\xi}_{(1)}(x)>
\varDelta^\xi$ with positive probability, unless
$\xi_1(x)=\dots=\xi_q(x)$ a.s. Thus, excluding the case $q=2$ where
$t_\gamma(\theta)=\theta+1$, by monotonicity of the function
$\gamma\mapsto \varphi(t_\gamma;\theta)$ we conclude that
$\theta^*_{\mynn\varDelta^\xi-k}(k,q)<\theta^{\dag}(k,q)$, and
therefore the domain of uniqueness in Theorem \ref{t4} is wider than
\strut{}that in Theorem~\ref{t3}.
\end{remark}

\begin{example}
Let us illustrate Theorem \ref{t4} with a simple model described in
Example \ref{ex:Delta}(c), assuming that $q\ge3$. Suppose first that $\alpha>0$.
Then, according to the distribution \eqref{eq:check+} and notation \eqref{eq:theta_K},
equation \eqref{eq:equation-alpha-new} specializes to
\begin{equation*}
Q(\theta)+\frac{1}{q}\,\varphi\left(\theta+1+
\frac{q-2}{\theta^{\alpha+k}_{\vphantom{X}}};\theta\right)+
\frac{q-1}{q}\,\varphi\left(\theta+1+\frac{q-2}{\theta^{k}};\theta\right)
=\frac{1}{k}.
\end{equation*}
By monotonicity of the function $t\mapsto \varphi(t;\theta)$, it is
clear that the root $\theta^{\dag}$ of this equation is strictly
bigger than the root
$\theta^*_{\mynn\varDelta^\xi-k}=\theta^*_{-\alpha-k}$ of the
equation
\begin{equation*}
Q(\theta)+\varphi\left(\theta+1+\frac{q-2}{\theta^{\alpha+k}_{\vphantom{X}}};\theta\right)
=\frac{1}{k},
\end{equation*}
in accordance with Remark \ref{rm:3.5}. Similarly, if $\alpha<0$
then the distribution \eqref{eq:check+} is replaced by
\eqref{eq:check-} and equation \eqref{eq:equation-alpha-new} takes
the form
\begin{equation*}
Q(\theta)+\frac{1}{q}\,\varphi\left(\theta+1+
\frac{q-2}{\theta^{\alpha+k}_{\vphantom{X}}};\theta\right)+
\frac{q-1}{q}\,\varphi\left(\theta+1+\frac{q-2}{\theta^{-\alpha+k}_{\vphantom{X}}};\theta\right)
=\frac{1}{k},
\end{equation*}
and by the monotonicity argument it is evident that its root
$\theta^{\dag}$ is strictly bigger than the root
$\theta^*_{\mynn\varDelta^\xi-k}=\theta^*_{\alpha-k}$ of the
equation
\begin{equation*}
Q(\theta)+\varphi\left(\theta+1+\frac{q-2}{\theta^{-\alpha+k}};\theta\right)
=\frac{1}{k},
\end{equation*}
again confirming the observation of Remark~\ref{rm:3.5}.
\end{example}

\smallskip
\subsection{Translation-invariant SGM and the problem of
non-uniqueness}\label{sec:2.3}

\subsubsection{Translation invariance}\label{sec:2.3.1}
\begin{figure}[th]
\vspace{-.5pc} \centering
\includegraphics[height=6cm]{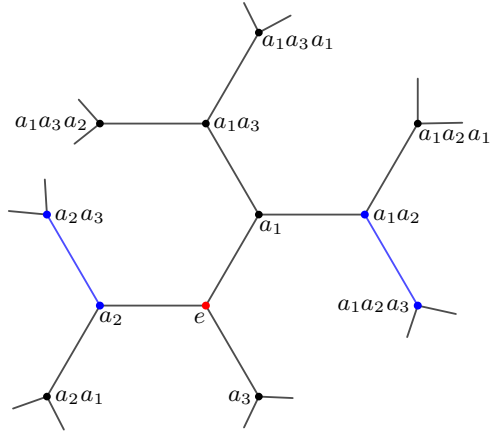}
\put(-108.0,45.5){\mbox{\scriptsize$e$}}%
\put(-95.2,16.2){\mbox{\scriptsize$a_3$}}%
\put(-143.4,45.9){\mbox{\scriptsize$a_2$}}%
\put(-83.5,80.3){\mbox{\scriptsize$a_1$}}%
\put(-40.9,84.5){\mbox{\scriptsize$a_1\kern-.02pc a_2$}}%
\put(-100.6,119.2){\mbox{\scriptsize$a_1\kern-.02pc a_3$}}%
\put(-160.0,84.5){\mbox{\scriptsize$a_2\myp a_3$}}%
\put(-159.9,16.2){\mbox{\scriptsize$a_2\myp a_1$}}%
\put(-54.3,50.7){\mbox{\scriptsize$a_1\kern-.02pc a_2\myp a_3$}}%
\put(-23.9,114.7){\mbox{\scriptsize$a_1\kern-.02pc a_2\myp a_1$}}%
\put(-82.5,150.1){\mbox{\scriptsize$a_1\kern-.02pc a_3\myp a_1$}}%
\put(-175.0,119.5){\mbox{\scriptsize$a_1\kern-.02pc a_3\myp a_2$}}%
\caption{A fragment of the Cayley tree $\mathbb{T}^2$ $(k=2$), with
vertices represented (one-to-one) by elements of the free group
$\mathscr{A}_2$ with generators $\{a_1,a_2,a_3\}$ (of order $2$
each). The identity element $e\in\mathscr{A}_2$ designates the root
$x_\circ\in V$. Starting from $e$, the elements $a\in\mathscr{A_2}$
are inductively expanded along the tree via right-multiplication by
one of the generators. Translations on $\mathscr{A}_2$ are defined
as left shifts, $T_g\colon a\to ga$ ($g,a\in\mathscr{A}_2$). For
example, under the shift $T_{a_1}\!$ the edge $\langle a_2,a_2\myp
a_3\rangle$ is mapped to the edge $\langle a_1a_2,a_1a_2\myp
a_3\rangle$.} \label{Fig2}
\end{figure}

To introduce the notion of \emph{translations} on the Cayley tree
$\mathbb{T}^k$\myn, let $\mathscr{A}_k$ be the free group with
generators $a_1, \dots,\allowbreak a_{k+1}$ of order $2$ each (i.e.,
$a_i^{-1}=a_i$). It is easy to see (cf.\ \cite{Ga0} and also
\cite[Section~2.2]{Ro}) that the Cayley tree $\mathbb{T}^k=(V,E)$ is
in a one-to-one correspondence with the group $\mathscr{A}_k$.
Namely, start by associating the root $x_\circ\in V$ with the
identity element $e\in \mathscr{A}_k$, and identify the elements
$a_1,\dots,a_{k+1}$ with the $k+1$ nearest neighbours of $x_\circ$
(i.e., comprising the set $S(x_\circ)=W_1$). Proceed inductively by
expanding the elements $a\in\mathscr{A}_k$ along the tree via
\emph{right}-multiplication by the generators $a_i$
$(i=1,\dots,k+1$), yielding $k$ new elements\footnote{Indeed, if
$a=w\myp a_j$ then $a\myp a_j=w\myp a_j^2=w$, so this particular
multiplication returns the element $w$ already obtained at the
previous step.} corresponding to the set of direct successors of $a$
(see Figure~\ref{Fig2}). This establishes a bijection
$\mathfrak{b}\colon V\! \rightarrow\mathscr{A}_k$.



Consider the family of \emph{left} shifts $T_g\colon
\mathscr{A}_k\to\mathscr{A}_k$ ($g\in \mathscr{A}_k$) defined by
\begin{equation*}
T_g(a):=g\myp a,\qquad a\in\mathscr{A}_k.
\end{equation*}
By virtue of the bijection $\mathfrak{b}$, this determines conjugate
translations on $V$,
\begin{equation}\label{eq:Tconj}
\widetilde{T}_z:= \mathfrak{b}^{-1}\mynn\circ
T_{\mathfrak{b}(z)}\myn\circ \mathfrak{b},\qquad z\in V.
\end{equation}
Clearly, $\widetilde{T}_z$ is an automorphism of $V$ preserving the
nearest-neighbour relation; indeed, if $\langle x,y\rangle\in E$ and
$y\in S(x)$ (so that $\mathfrak{b}(y)=\mathfrak{b}(x)\mypp a_j$,
with some generator $a_j$) then, according to \eqref{eq:Tconj},
$x'\mynn:=\widetilde{T}_z(x)$ and \strut{}$y'\mynn:=
\widetilde{T}_z(y)=\mathfrak{b}^{-1}(\mathfrak{b}(z)\myp
\mathfrak{b}(y))=\mathfrak{b}^{-1}(\mathfrak{b}(x')\mypp a_j)$ are
nearest \strut{}neighbours, $\langle x'\myn,y'\rangle\in E$ (see
Figure~\ref{Fig2}). For example, if $k=1$ (whereby the Cayley tree
$\mathbb{T}^1$ is reduced to the integer lattice $\mathbb{Z}^1$),
the action of the shift $\widetilde{T}_z$ ($z\in\mathbb{Z}^1$) can
be written in closed form,
$$
\widetilde{T}_z(x)=z+(-1)^z x,\qquad x\in\mathbb{Z}^1.
$$

In turn, the map \eqref{eq:Tconj} induces shifts on
\strut{}configurations \strut{}$\sigma\in \varPhi^{V}$,
\begin{equation}\label{eq:Tsigma}
(\widetilde{T}_z\myp\sigma)(x):=\sigma(\widetilde{T}^{-1}_zx),
\qquad x\in V.
\end{equation}

\begin{definition}
We say that SGM $\mu^h=\mu^h_{\theta,\myp\xi}$ is \emph{translation
invariant} (with respect to the \strut{}group of shifts
$(\widetilde{T}_z)$) if for each $z\in V$ the measures $\mu^h\circ
\widetilde{T}_z^{-1}$ and $\mu^h$ coincide; that is, for any
(finite) $\varLambda\subset V$ and any configuration $\varsigma\in
\varPhi^{\varLambda}$
\begin{equation}\label{tim}
\mu^h\bigl(\sigma_{\widetilde{T}_z(\varLambda)}=\widetilde{T}_z(\varsigma)\bigr)
=\mu^h(\sigma_{\myn\varLambda}=\varsigma).
\end{equation}
\end{definition}

Recall that the quantities $\check{h}_i^\dag(x,y)$
\textup{(}$\langle x,y\rangle\in E$\textup{)} defining the boundary
law were introduced in Section~\ref{sec:1.2.4}.
\begin{proposition}\label{pr:TI}
An SGM $\mu^h=\mu^h_{\theta,\myp \xi}$ is translation invariant
under the group of shifts $(\widetilde{T}_z)_{z\in V}$ if and only
if the following conditions are satisfied.
\begin{itemize}
\item[\rm (i)] The reduced field $\{\check{\boldsymbol{\xi}}(x)\}$
is constant over the tree,
\begin{equation}\label{eq:xih}
\check{\boldsymbol{\xi}}(x)=
\check{\boldsymbol{\xi}}(x_\circ),\qquad x\in V.
\end{equation}
\item[\rm (ii)] The reduced field $\{\check{\boldsymbol{h}}{}^\dag(x,y)\}$ is symmetric,
\begin{equation}\label{eq:h-dag}
\check{\boldsymbol{h}}{}^\dag(x,y)=
\check{\boldsymbol{h}}{}^\dag(y,x),\qquad \langle x,y\rangle\in E.
\end{equation}

\item[\rm (iii)]
For any $z\in V$,
\begin{equation}\label{eq:h-h}
\check{\boldsymbol{h}}{}^\dag(x,y)=
\check{\boldsymbol{h}}{}^\dag\myn\bigl(\widetilde{T}_{z}(x),\widetilde{T}_{z}(y)\bigr),\qquad
\langle x,y\rangle\in E.
\end{equation}
\end{itemize}
\end{proposition}

\smallskip
This result will be proved in Appendix~\ref{sec:A}.

\smallskip
For $x\ne x_\circ$, denote by $x'\mynn$ the unique vertex such that
$x\in S(x')$. Then, according to \eqref{eq:h-inward}
and~\eqref{eq:h-h},
\begin{equation}\label{eq:h-h-0}
\check{\boldsymbol{h}}{}^\dag(x,x')=\check{\boldsymbol{h}}(x)
=\check{\boldsymbol{h}}\bigl(\widetilde{T}^{-1}_{x'}(x)\bigr).
\end{equation}
Note that $\widetilde{T}^{-1}_{x'}(x)\in
\partial\{x_\circ\}=W_1$. Thus,
there are $k+1$ (vector) values $\check{\boldsymbol{h}}(x_j)$
($x_j\in W_1$) that determine a translation-invariant SGM $\mu^h$.
By translations \eqref{eq:h-h-0}, these values (which can be
pictured as $k+1$ ``colours'') are propagated to all vertices $x\in
V$ in a periodic ``chessboard'' tiling, except the root $x=x_\circ$
where the value $\check{\boldsymbol{h}}(x_\circ)$ is calculated
separately, according to the compatibility formula~\eqref{p***}.

The criterion of translation invariance given by
Proposition~\ref{pr:TI} appears to be new. In Georgii
\cite[Corollary~(12.17)]{Ge}, a version of this result is
established (in the language of boundary laws) for \emph{completely
homogeneous} SGM, that is, assuming the invariance under the group
of \emph{all automorphisms} of the tree $\mathbb{T}^k$.\footnote{It
is worth pointing out that the latter group is generated by the
group of (left) shifts $(\widetilde{T}_z)$ and pairwise inversions
between vertices $x_j,x_\ell\in
\partial\{x_\circ\}$ \cite[\S\mypp3.5]{Magnus}.} Namely, we have the following

\begin{corollary}\label{cor:CHSGM}
An SGM $\mu^h=\mu^h_{\theta,\myp \xi}$ is completely homogeneous if
and only if\myp\ $\check{\boldsymbol{\xi}}(x)\equiv
\check{\boldsymbol{\xi}}{}^0\mynn$ for all $x\in V$ and
$\check{\boldsymbol{h}}(x)\equiv \check{\boldsymbol{h}}{}^0\mynn$
for all $x\ne x_\circ$.
\end{corollary}

\begin{remark}
In the existing studies of Gibbs measures on trees (see, e.g.,
\cite{GRR,KRK}), it is common to use the term ``translation
invariant'' (and the abbreviation TISGM) having in mind just
completely homogeneous SGM. We prefer to keep the terminological
distinction between ``single-coloured'' completely homogeneous GBC
$\check{\boldsymbol{h}}(x)\equiv \check{\boldsymbol{h}}{}^0\mynn$
and ``multi-coloured'' translation-invariant GBC as characterized by
Proposition~\ref{pr:TI}. The latter case is very interesting
(especially with regard to uniqueness) but technically more
challenging, so it is not addressed here in full generality.
However, as we will see below, the subclass of completely
homogeneous SGM in the Potts model is very rich in its own right.
\end{remark}

\subsubsection{Analysis of uniqueness}\label{sec:2.3.2}
For the rest of Section~\ref{sec:2.3}, we deal with \emph{completely
homogeneous SGM} $\mu^h=\mu^h_{\theta,\myp \xi}$, that is, with the
external field $\{\boldsymbol{\xi}(x)\}_{x\in V}$ and GBC
$\{\boldsymbol{h}(x)\}_{x\in V}$ satisfying the homogeneity
conditions of Corollary~\ref{cor:CHSGM}. For \strut{}simplicity, we
confine ourselves to the case where all coordinates of the (reduced)
vector $\check{\boldsymbol{\xi}}{}^0\mynn$ are zero except one; due
to permutational symmetry, we may assume, without loss of
generality, that $\check{\xi}_1^0=\alpha\in\mathbb{R}$ and
$\check{\xi}^0_2=\dots=\check{\xi}^0_{q-1}=0$,
\begin{equation}\label{eq:xi0}
\boldsymbol{\check{\xi}}{}^0\mynn = (\alpha,0,\dots,0)\in
\mathbb{R}^{q-1}.
\end{equation}
We also write
$$
\boldsymbol{\check{h}}{}^0\mynn=(\check{h}_1^0,\dots,\check{h}_{q-1}^0)\in\mathbb{R}^{q-1}.
$$
Then, denoting $z_i:=\theta^{\mypp\check{h}_i^0/k}$
($i=1,\dots,q-1$), the compatibility equations \eqref{p***} are
equivalently rewritten in the form
\begin{equation}\label{pt1}
\left\{\begin{aligned} z_1&=1+\frac{(\theta-1)(\theta^{\alpha}
z_1^k-1)}{\theta+\theta^{\alpha}_{\vphantom{X}}
z_1^k+\sum_{j=2}^{q-1}z_j^k},\\
z_i&=1+\frac{(\theta-1)(z_i^k-1)}{\theta+\theta^{\alpha}_{\vphantom{X}}
z_1^k+\sum_{j=2}^{q-1}z_j^k},\qquad i=2,\dots,q-1.
\end{aligned}
\right.
\end{equation}

Solvability of the system \eqref{pt1} can be analysed in some
detail; in particular, we are able to characterize the uniqueness of
its solution, which in turn gives a criterion of the uniqueness of
completely homogeneous SGM in the Potts model.

The case $\theta=1$ is trivial, as the system \eqref{pt1} will then
have the unique solution $z_1=\dots=z_{q-1}=1$. The case $\theta>1$
and $\alpha=0$ has been studied in \cite{KR}; these results can be
reproduced directly by the methods developed in the present work
similarly to a more general (and difficult) case $\alpha\ne0$, and
are also obtainable in the limit as $\alpha\to0$ (see
Lemma~\ref{lm2}(b) and Remark~\ref{rm:eta->1}).

Thus, let us focus on the new case $\alpha\ne 0$. By
Lemma~\ref{lm2}(c), the system \eqref{pt1} with $\theta>1$ is
reduced either to a single equation
\begin{equation}\label{rm=1-int}
u=1+\frac{(\theta-1)(\theta^\alpha
u^k-1)}{\theta+\theta^\alpha_{\vphantom{X}}
u^k+q-2}
\end{equation}
or to the system of equations (indexed by $m=1,\dots, q-2$)
\begin{equation}\label{uv-int}
\left\{\begin{aligned} u&=1+\frac{(\theta-1)(\theta^\alpha
u^k-1)}{\theta+\theta^\alpha_{\vphantom{X}}
u^k+m v^k+q-2-m},\\
1&=\frac{(\theta-1)(1+v+\dots+v^{k-1})}{\theta+\theta^\alpha_{\vphantom{X}}
u^k+m v^k+q-2-m},
\end{aligned}
\right.
\end{equation}
subject to the condition
\begin{equation}\label{eq:vnot=1}
v\ne1.
\end{equation}

Let us first address the solvability of the
equation~\eqref{rm=1-int}. Denote
\begin{equation}\label{eq:theta'c}
\theta_{\rm c}=\theta_{\rm c}(k,q):=
\frac{1}{2}\,\Biggl(\sqrt{(q-2)^2+4\myp(q-1)\left(\frac{k+1}{k-1}\right)^2}-(q-2)\Biggr).
\end{equation}
In particular,  if $q=2$ then $\theta_{\rm
c}(k,2)=\frac{k+1}{k-1}\equiv \theta_0(k,2)$
(cf.~\eqref{eq:theta0q=2}). Let us also set
\begin{equation}\label{eq:b}
b= b(\theta):=\frac{\theta\myp(\theta+q-2)}{q-1}.
\end{equation}
Clearly, $b(1)=1$ and $b(\theta)>1$ for $\theta>1$. Furthermore,
comparing \eqref{eq:theta'c} and \eqref{eq:b} observe that
$b(\theta_{\rm c})=\left(\frac{k+1}{k-1}\right)^2$ and
$b(\theta)>\left(\frac{k+1}{k-1}\right)^2$ for $\theta>\theta_{\rm
c}$. For $\theta\ge\theta_{\rm c}$, denote by $x_{\pm}=
x_{\pm}(\theta)$ the roots of the quadratic equation
\begin{equation}\label{eq:quadratic-intro}
(b+x)(1+x)=k\myp(b-1)\myp x
\end{equation}
with discriminant
\begin{equation}\label{eq:D}
D= D(\theta):=\bigl(k\myp(b-1)-(b+1)\bigr)^2-4b=
(b-1)(k-1)^2\left(b-\left(\frac{k+1}{k-1}\right)^2\right),
\end{equation}
that is,\footnote{Here and in what follows, formulas involving the
symbols $\pm$ and/or $\mp$ combine the two cases corresponding to
the choice of either the upper or lower sign throughout.}
\begin{equation}\label{eq:x+/--intro}
x_{\pm}= x_{\pm}(\theta):=\frac{(b-1)(k-1)-2\pm\sqrt{D}}{2}.
\end{equation}
Furthermore, introduce the notation
\begin{equation}
\label{eq:a_pm} a_{\pm}=
a_{\pm}(\theta):=\frac{1}{x_\pm}\left(\frac{1+x_\pm}{b+x_\pm}\right)^k,\qquad
\theta\ge\theta_{\rm c}.
\end{equation}
Of course, $a_{-}(\theta_{\rm c})=a_{+}(\theta_{\rm c})$, and one
can also show that $a_{-}(\theta)<a_{+}(\theta)$ for all
$\theta>\theta_{\rm c}$ (see the proof of Theorem~\ref{th:3.7} in
Section~\ref{sec:5.2}). Finally, denote
\begin{equation}\label{eq:alpha_pm}
\alpha_{\pm}=\alpha_{\pm}(\theta):=-(k+1)+\frac{1}{\ln\theta}\ln\frac{q-1}{a_{\mp}},\qquad
\theta\ge\theta_{\rm c},
\end{equation}
so that $\alpha_{-}(\theta_{\rm c})=\alpha_{+}(\theta_{\rm c})$ and
$\alpha_{-}(\theta)<\alpha_{+}(\theta)$ for $\theta>\theta_{\rm c}$.

\begin{theorem}\label{th:3.7}
Let $\nu_0(\theta,\alpha)$ denote the number of solutions $u>0$ of
the equation~\eqref{rm=1-int}. Then
\begin{equation*}
\nu_0(\theta,\alpha)=\begin{cases} 1&\text{if}\ \, \theta\leq
\theta_{\rm c} \ \,\text{or} \ \,\theta>\theta_{\rm c} \ \,
\text{and} \ \, \alpha\notin [\alpha_-,\alpha_+],\\
2&\text{if}\ \, \theta>\theta_{\rm c} \ \, \text{and} \ \,
\alpha\in\{\alpha_{-},\alpha_+\},\\
3&\text{if}\ \,
\theta>\theta_{\rm c} \ \, \text{and} \ \, \alpha\in
(\alpha_-,\alpha_+),
\end{cases}
\end{equation*}
where $\theta_{\rm c}$ is given in \eqref{eq:theta'c} and
$\alpha_\pm=\alpha_\pm(\theta)$ are defined by~\eqref{eq:alpha_pm}.
\end{theorem}

Let us now state our results on the solvability of the set of
equations \eqref{uv-int}. For each $m\in \{1,\dots,q-2\}$, consider
the functions
\begin{align}
\label{eq:L-intr} L_m(v;\theta):={}&(\theta-1)\mypp
\bigl(v^{k-1}+\dots+v\bigr)-m v^{k}
-(q-1-m),\\
\label{eq:K-intr} K_m(v;\theta):={}&
\frac{\bigl(v^{k-1}+\dots+v+1\bigr)^k
L_m(v;\theta)}{\bigl(v^{k-1}+\dots+v+L_m(v;\theta)\bigr)^k}.
\end{align}
It can be checked (see Lemma~\ref{lm:vmtheta}) that for any
$\theta>1$ there is a unique value $v_m= v_m(\theta)>0$ such that
$$
L_m^*(\theta):=L_m(v_m;\theta)=\max_{v>0}L_m(v;\theta),
$$
and moreover, the function $\theta\mapsto L_m^*(\theta)$ is strictly
increasing. Denote by $\theta_m$ the (unique) value of $\theta>1$
such that
\begin{equation}\label{eq:Lmthetam}
L_m^*(\theta_m)=0.
\end{equation}
Thus, for any $\theta>\theta_m$ the range of the functions $v\mapsto
L_m(v;\theta)$ and $v\mapsto K_m(v;\theta)$ includes positive
values,
\begin{equation}\label{eq:V+}
\mathscr{V}_m^+(\theta):=\{v>0\colon L_m(v;\theta)>0\}\equiv
\{v>0\colon K_m(v;\theta)>0\}\ne \varnothing,
\end{equation}
and, therefore, the function
\begin{equation}\label{eq:alpham}
\alpha_m(\theta):=\frac{1}{\ln\theta} \max_{v\in
\mathscr{V}_m^+(\theta)}\ln K_m(v;\theta)=\frac{\ln
K_m^*(\theta)}{\ln\theta},\qquad \theta>\theta_m,
\end{equation}
is well defined, where
$$
K_m^*(\theta):=\max_{v\in \mathscr{V}_m^+(\theta)} K_m(v;\theta).
$$

\begin{example}\label{ex:k=2_m}
In the case $k=2$, from \eqref{eq:L-intr} and \eqref{eq:Lmthetam} we
obtain explicitly
\begin{equation}\label{eq:theta|k=2}
\theta_m=1+2\sqrt{m\myp(q-m-1)}, \qquad m=1,\dots,q-2.
\end{equation}
In particular, $\theta_1=1$ for $q=2$, and $\theta_1\ge 3$ whenever
$q\ge3$. Comparing \eqref{eq:theta'c} and \eqref{eq:theta|k=2}, we
also \strut{}find that $\theta_{\rm c}\le \theta_1$ if and only if
$q\ge 6$. For example, for $q=5$ we have $\theta_1=1+2\sqrt{3}\doteq
4.4641$, $\theta_2=5$, $\theta_{\rm
c}=\frac{1}{2}\myp(\sqrt{153}-3)\doteq 4.6847$, that is,
$\theta_1<\theta_{\rm c}<\theta_2$, whereas for \strut{}$q=6$ and
$q=7$ we compute $\theta_1=\theta_{\rm c}=5$ and
$\theta_1=1+2\sqrt{5}\doteq5.4721>\theta_{\rm
c}=\frac12\myp(\sqrt{241}-5)\doteq5.2621$, respectively. Another
simple case is $q=3$ (with $m=1$ and any $k\ge2$); indeed, it is
easy to see that the condition \eqref{eq:Lmthetam} is satisfied with
$v_1^*=v_1(\theta_1)=1$ (cf.\ Lemma~\ref{lm:vmtheta}(c)), whence we
readily find $\theta_1=1+\frac{2}{k-1}$.
\end{example}

\begin{theorem}\label{th:3.8} For each $m\in \{1,\dots,q-2\}$,
let\/ $\nu_{m}(\theta,\alpha)$ denote the number of positive
solutions $(u,v)$ of the system~\eqref{uv-int}. Then
$\nu_{m}(\theta,\alpha)\ge 1$ if and only if\/ $\theta>\theta_m$ and
$\alpha\le \alpha_m(\theta)$.
\end{theorem}

As will be shown in Proposition~\ref{pr:alpha-zeros}(b),
$\alpha_{+}(\theta)$ has a unique zero given by
\strut{}$\theta^{+}_{0}=1+\frac{q}{k-1}$, whereas
$\alpha_{-}(\theta)$ has a unique zero $\theta^{-}_{0}$, which
coincides with the \strut{}zero $\theta_1^{\myp0}$
of~$\alpha_1(\theta)$. \strut{}It also holds that $\alpha_1(\theta)$
is a majorant of the family of functions $\{\alpha_m(\theta)\}$ (see
Propositions~\ref{pr:theta-symmetry} and~\ref{pr:theta-monotone}).

\subsubsection{Uniqueness of completely homogeneous SGM}
\strut{}In the case $q=3$, there appears to be an additional
critical value (see Lemma~\ref{lm:q=3})
\begin{equation}\label{eq:theta-cr-q=3}
\tilde{\theta}_1=\tilde{\theta}_1(k):=\frac{5-k+\sqrt{49k^2+62k+49}}{6\myp(k-1)}.
\end{equation}
For example,
\begin{equation}\label{eq:theta-cr-q=3_k=2}
\tilde{\theta}_1(k)=\begin{cases}\displaystyle
\frac{1+\sqrt{41}}{2}\doteq 3.7016, &k=2,\\[.5pc]
\displaystyle\ \frac{7}{3}\doteq 2.3333,&k=3,\\[.5pc]
\displaystyle\frac{1+\sqrt{1081}}{18}\doteq 1.8821,\ \ &k=4.
\end{cases}
\end{equation}

For $q\ge2$, consider the following subsets of the half-plane
$\{\theta\ge1\}=\{(\theta,\alpha)\colon
\theta\ge1,\,\alpha\in\mathbb{R}\}$,
\begin{align}
\notag A_q:={}&\{\theta>\theta_{\rm c},\,\alpha_{-}(\theta)\le
\alpha\le \alpha_{+}(\theta)\},\\[.2pc]
\label{eq:Bq} B_q:={}& \left\{
\begin{alignedat}{3}
& \,\varnothing     &&\text{if\ \,}q=2,\\
&\{\theta>\theta_1,\,\alpha<\alpha_1(\theta)\}
\cup\{\theta>\tilde{\theta}_1,
\,\alpha=\alpha_1(\theta)\}\quad&&\text{if\ \,}q=3,\\
&\{\theta>\theta_1,\,\alpha\le \alpha_1(\theta)\}\quad&&\text{if\
\,}q\ge4.
\end{alignedat}
\right.
\end{align}
Denote the total number of positive solutions
$\boldsymbol{z}=(z_1,\dots,z_{q-1})$ of the system of
equations~\eqref{pt1} by $\nu(\theta,\alpha)$ ($\theta\ge1$,
$\alpha\in\mathbb{R}$); of course, this number also depends on $k$
and~$q$. Theorems \ref{th:3.7} and \ref{th:3.8} can now be
summarized as follows.

\begin{theorem}[Non-uniqueness]\label{th:non-U}\mbox{}
\begin{itemize}
\item[\rm(a)]
If\/ $q=2$ then $\nu(\theta,\alpha)\ge2$ if and only if
$(\theta,\alpha)\in A_2\cup B_2=A_2$.

\vspace{.3pc}
\item[\rm (b)] If\/ $q=3$ then
$\nu(\theta,\alpha)\ge 2$ if $(\theta,\alpha)\in A_3\cup B_3$. The\
``only if''\ statement holds true at least for $k=2,3,4$.

\vspace{.3pc}
\item[\rm (c)] If\/ $q\ge 4$ then
$\nu(\theta,\alpha)\ge 2$ if and only if $(\theta,\alpha)\in A_q\cup
B_q$.
\end{itemize}
\end{theorem}

\begin{remark}
The special case $q=3$ in the definition \eqref{eq:Bq} and in
Theorem~\ref{th:non-U} emerges because for $\theta\le
\tilde{\theta}_1$ and $\alpha=\alpha_1(\theta)$, there is a
(hypothetically unique) solution
$(u,v)=\bigl(\theta-\frac{\theta+2}{k},1\bigr)$ of the
system~\eqref{uv-int}, which is, however, not admissible due to the
\strut{}constraint~\eqref{eq:vnot=1} and, therefore, does not
destroy the uniqueness of solution to~\eqref{pt1}. This hypothesis
is conjectured below; if it is true then ``if'' \strut{}in
Theorem~\ref{th:non-U}(b) (i.e., $q=3$) can be enhanced to ``if and
only if'' for all $k\ge2$.
\end{remark}

\begin{conjecture}\label{conj:v=1}
If $q=3$, $\theta\le \tilde{\theta}_1$ and $\alpha=\alpha_1(\theta)$ then
$(u,v)=\bigl(\theta-\frac{\theta+2}{k},1\bigr)$ is the sole solution
of the system~\eqref{uv-int}.
\end{conjecture}

\begin{remark}\label{rm:q=3-critical}
Regardless of Conjecture~\ref{conj:v=1}, the inclusion of a proper
part of the curve $\alpha=\alpha_1(\theta)$ in the uniqueness region
in the case $q=3$ is indeed necessary. Namely, it can be proved (see
Proposition~\ref{pr:theta+-}) that if $\varepsilon>0$ is small
enough then $\nu(\theta,\alpha_1(\theta))=1$ for $\theta_1<\theta<
\theta_1+\varepsilon$ but $\nu(\theta,\alpha_1(\theta))\ge 2$ for
$\theta>\theta^0_1-\varepsilon$, where $\theta_1^0$ is the zero of
$\alpha_1(\theta)$.
\end{remark}

\begin{figure} \centering
\subfigure[\,$q=5$ \,($k=2$)]
{\hspace{-1pc}\includegraphics[width=6.7cm]{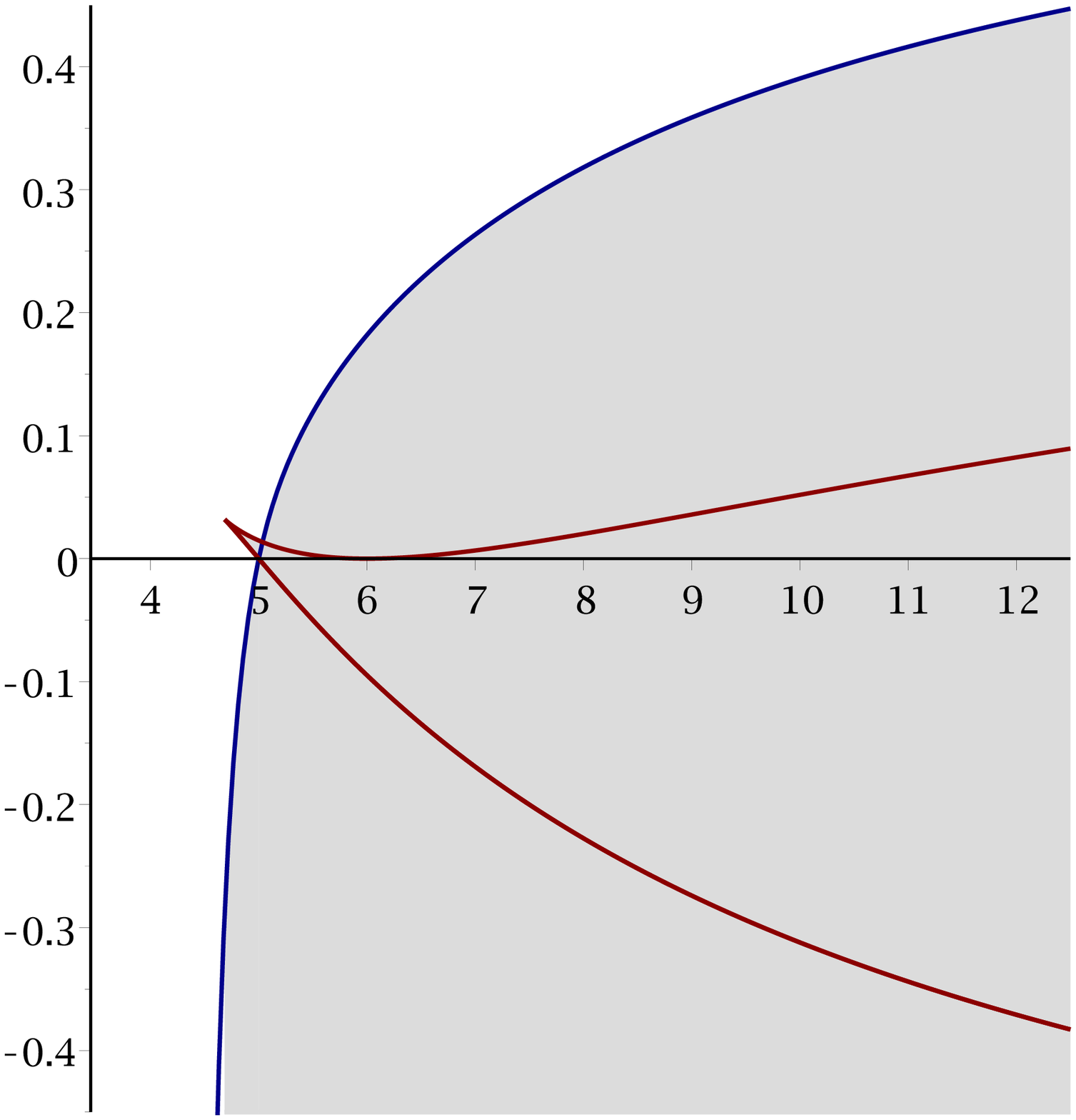}
\put(2,86){\mbox{\scriptsize$\theta$}}
\put(-174,188){\mbox{\scriptsize$\alpha$}}
\put(-106,162){\mbox{\scriptsize$\alpha_1(\theta)$}}
\put(-41,32){\mbox{\scriptsize$\alpha_{-}(\theta)$}}
\put(-74,108){\mbox{\scriptsize$\alpha_{+}(\theta)$}}
\label{Fig3a} } \hspace{1.5pc} \subfigure[\,$q=3$ \,($k=2$)]
{\raisebox{-.3pc}{\includegraphics[width=6.7cm]{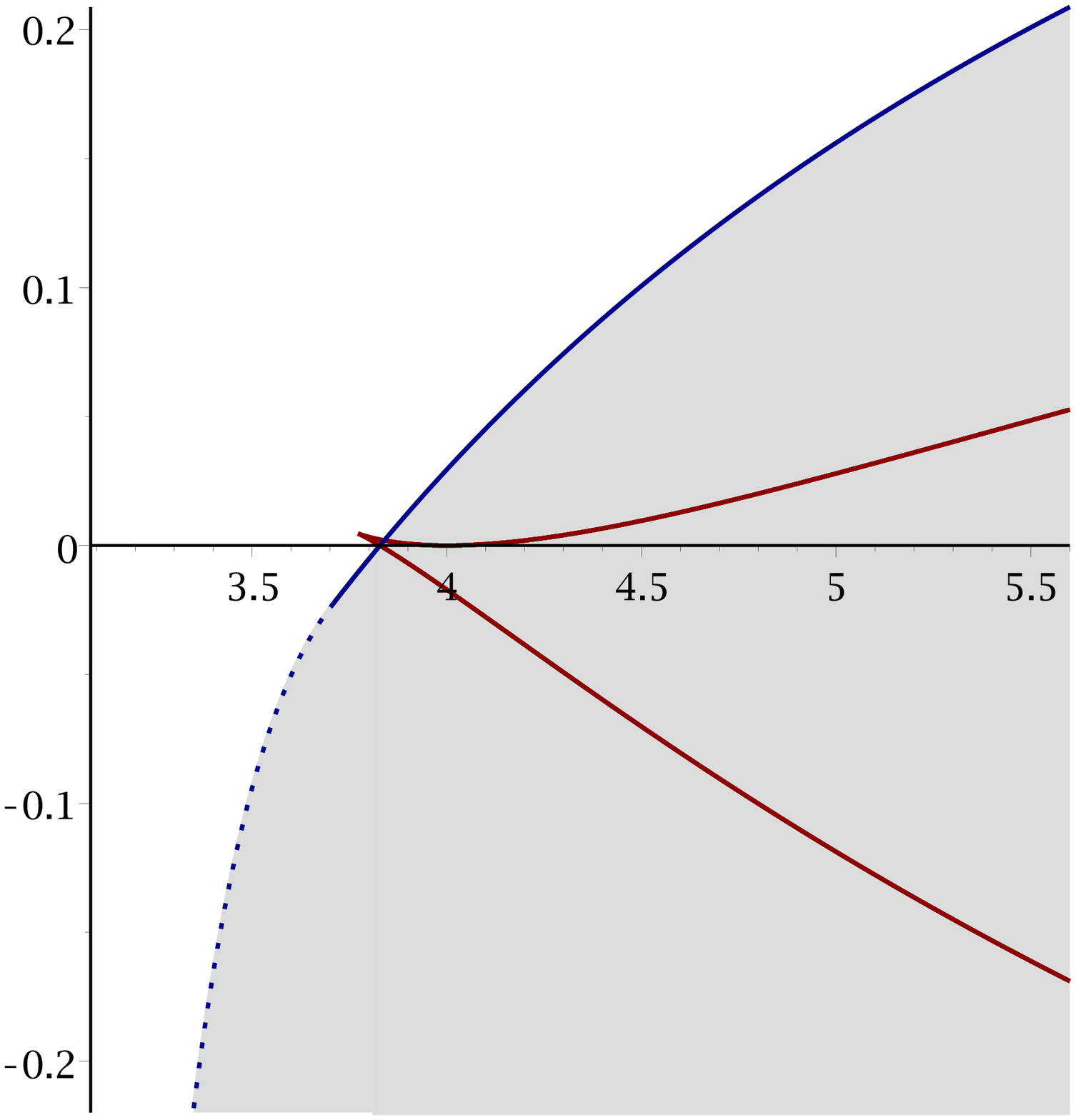}}\label{Fig3b}
\put(2,84.8){\mbox{\scriptsize$\theta$}}
\put(-173,187){\mbox{\scriptsize$\alpha$}}
\put(-84,149){\mbox{\scriptsize$\alpha_1(\theta)$}}
\put(-48,50){\mbox{\scriptsize$\alpha_{-}(\theta)$}}
\put(-68,108){\mbox{\scriptsize$\alpha_{+}(\theta)$}}}

\caption{The phase diagram for the Potts model~\eqref{eq:xi0}
showing the non-uniqueness region (shaded in grey) according to
Theorem~\ref{th:non-U}: (a) regular case $q\ge 4$ (shown here for
$q=5$); (b) special case $q=3$, both with $k=2$. The critical
boundaries are determined by (parts of) the graphs of the functions
$\alpha_\pm(\theta)$ and $\alpha_1(\theta)$ defined in
\eqref{eq:alpha_pm} and \eqref{eq:alpham}, respectively. The dotted
part of the boundary on panel (b), given by
$\alpha=\alpha_1(\theta)$, $\theta\in(\theta_1, \tilde{\theta}_1]$
(see \strut{}formula~\eqref{eq:Bq} with $q=3$), is excluded from the
shaded region (see Theorem~\ref{th:non-U}(b) and
Conjecture~\ref{conj:v=1}, proved for $2\le k\le 4$); here,
$\theta_1=3$ and
$\tilde{\theta}_1=\frac12\bigl(1+\sqrt{41}\myp\bigr)\doteq 3.7016$.
}\label{Fig3}
\end{figure}

Theorem \ref{th:non-U} provides a sufficient and (almost) necessary
condition for the uniqueness of solution of~\eqref{pt1}, illustrated
in Figure~\ref{Fig3} for $q=5$ and $q=3$, both with $k=2$.

\smallskip
To conclude this subsection, the following result describes a few
cases where it is possible to estimate the maximal number of
solutions of the system~\eqref{pt1}.
\begin{theorem}\label{th:k=2}\mbox{}
\begin{itemize}
\item[\rm (a)]
If $q=2$ then $\nu(\theta,\alpha)\le3$\textup{;} moreover,
$\nu(\theta,\alpha)=3$ for all $\theta\ge1$ large enough.

\smallskip
\item[\rm (b)] Let $\alpha=0$ and $k\ge2$. Then $\nu(\theta,0)\le 2^q-1$ for all
$\theta\ge1$\textup{;} moreover, $\nu(\theta,0)=2^q-1$ for all
$\theta$ large enough.

\smallskip
\item[\rm (c)] If $k=2$ then $\nu(\theta,\alpha)\le 2^q-1$ for all
$\theta\ge1$ and $\alpha\in\mathbb{R}$.
\end{itemize}
\end{theorem}

\begin{conjecture}\label{conj:max=4fork>2}
The upper bound $2^q-1$ in Theorem~\ref{th:k=2} appears to be
universal. There is empirical evidence from exploration of many
specific cases (using the computing package\footnote{Throughout this
paper, we used {\sf Maple 18} (Build ID 922027) licensed to the
University of Leeds.} {\sf Maple}) to conjecture that
Theorem~\textup{\ref{th:k=2}(c)} holds true for all $k\ge2$.
\end{conjecture}

\subsubsection{Some comments on earlier work}

For $q=2$, when the Potts model is reduced to the Ising model, the
result of Theorem~\ref{th:k=2}(a) is well known
(see~\cite[Section~12.2]{Ge} or \cite[Chapter~2]{Ro}).
The case $\alpha=0$ (i.e., with zero field) is also well studied
(see, e.g., \cite[Section~5.2.2.2, Proposition~5.4, pages
114--115]{Ro} and \cite[Theorem~1, page~192]{KRK}), the result of
Theorem~\ref{th:k=2}(b) can be considered as a corollary of
\cite[Theorem~1, page~192]{KRK}); in particular, it is known that
there are $\bigl\lfloor\frac12(q+1)\bigr\rfloor$ critical values of
$\theta$, including $\theta_0^-=\theta_1^{\myp0}$ and
$\theta_0^+=1+q/(k-1)$.

The general case $q\ge3$ with $\alpha\in\mathbb{R}$ was first
addressed by Peruggi et al.~\cite{Peruggi1} (and continued
in~\cite{Peruggi2}) using physical argumentation. In particular,
they correctly identified the critical point $\theta_{\rm c}$
\cite[equation~(22), page~160]{Peruggi2} (cf.~\eqref{eq:theta'c})
and also suggested an explicit critical boundary in the phase
diagram for $\alpha\ge0$, defined by the expression
\cite[equation~(21), page~160]{Peruggi2} (adapted to our notation)
$$
\tilde{\alpha}_{-}(\theta)=\frac{(k+1)\ln\bigl(1+(q-2)/\theta\bigr)-(k-1)\ln(q-1)}{2\ln\theta}.
$$
Note that this function enjoys a correct value at
$\theta=\theta_{\rm c}$ (i.e., $\tilde{\alpha}_{-}(\theta_{\rm
c})=\alpha_{\pm}(\theta_{\rm c})$, see formula~\eqref{eq:alphapm-c}
below), but $\tilde{\alpha}_{-}(\theta)>\alpha_{-}(\theta)$ for all
$\theta>\theta_{\rm c}$. The corresponding critical value of
activity $\theta$, emerging as zero of $\tilde{\alpha}_{-}(\theta)$,
is reported in \cite[equation~(20), page~158]{Peruggi2} as
$$
\theta_{\rm cr}=\theta_{\rm
cr}(k,q):=\frac{q-2}{(q-1)^{(k-1)/(k+1)}-1}.
$$
In particular, $\theta_{\rm cr}$ is bigger than the exact critical
value $\theta_0^{-}=\theta_1^{\myp0}$, where the uniqueness breaks
down at $\alpha=0$ (see Proposition~\ref{pr:alpha-zeros}). For
example, the corresponding numerical values (for $k=5$ and $q=3$ or
$q=8$) are given by (cf.\ \cite[figure~1, page~159]{Peruggi2})
$$
\theta_{\rm cr}\doteq\begin{cases}
1.7024
,&q=3,\\
2.2562
,&q=8,
\end{cases}
\qquad \theta_0^{-}\doteq\begin{cases}1.6966
,&q=3,\\
2.1803
,&q=8. \end{cases}
$$

The critical boundary in the phase diagram for $\alpha\le0$ was
described in \cite[page~160]{Peruggi2} only heuristically, as a line
``joining'' the points $\theta=\theta_{\rm cr}(k,q)$, $\alpha=0$ and
$\theta=\theta_{\rm cr}(k,q-1)$, $\alpha=-\infty$, and illustrated
by a sketch graph in the vicinity of $\theta_{\rm cr}(k,q)$ (for
$k=5$ and $q=3$ or $q=8$).

It should be stressed that the phase transition occurring at these
critical boundaries is not of type ``uniqueness/non-uniqueness'',
with which we are primarily concerned in the present paper, but in
fact the so-called ``order/disorder'' phase transition. The latter
was studied rigorously in a recent paper by Galanis et al.\
\cite{Galanis} in connection with the computational complexity of
approximating the partition function of the Potts model. The useful
classification of critical points deployed in \cite{Galanis} is
based on the notion of \emph{dominant phase}; in particular, the
critical point $\theta_0^{+}=1+q/(k-1)$ (conjectured earlier by
H\"aggstr\"om \cite{Hagg} in a more general context of random
cluster measures on trees) can be explained from this point of view
as a threshold beyond which only ordered phases are dominant. Note
that the paper \cite{Galanis} studies the Potts model primarily with
zero external field ($\alpha=0$); the authors claim that their
methods should also work in a more general ferromagnetic framework
including a non-zero field, but no details are spelled out clearly.

In the present paper, we do not investigate the thermodynamical
nature of phase transitions, instead focussing on the number of
completely homogeneous SGMs, especially on the uniqueness issue. In
particular, the order/disorder critical point $\theta_{\rm cr}$ is
not immediately detectable by our methods. It would be interesting
to look into these issues for the Potts model with external field,
thus extending the results of~\cite{Galanis}. More specifically, our
analysis (see Proposition~\ref{pr:alpha-zeros}) shows that the
critical point $\theta_0^{+}$ is the signature of the upper critical
function $\alpha_+(\theta)$ at $\alpha=0$, which has a minimum at
$\theta=\theta_0^{+}$. Therefore, it is reasonable to
\emph{conjecture} that the $\alpha>0$ analogue of the interval of
activities $\theta$ between the critical points $\theta_0^-$ and
$\theta_{\rm cr}$ is the interval $[\theta_\alpha^{-},
\theta_\alpha^{+}]$, where $\theta_\alpha^{-}$ is the (sole) root of
the equation $\alpha_{-}(\theta)=\alpha$ and $\theta_\alpha^{+}$ is
the smaller root of the similar equation
$\alpha_{+}(\theta)=\alpha$. However, it is not clear as to what
happens in the interval \emph{between} the roots of the latter
equation. The counterpart of this picture for $\alpha<0$ is likely
to be simpler, as only the equation $\alpha_{-}(\theta)=\alpha$ is
involved. We intend to address these issues in our forthcoming work.

\section{Auxiliary lemmas}\label{sec:3}
In this section, we collect a few technical results that will be
instrumental in the proofs of the main theorems. We start with an
elementary lemma.

\begin{lemma}\label{lm:f}
For $a,b,c,d>0$, consider the function
\begin{equation}\label{eq:f}
f(t):=\ln\frac{a\myp\rme^t+b}{c\myp\rme^t+d},\qquad t\in\mathbb{R}.
\end{equation}
\begin{itemize}
\item[\rm (a)] If $ad>bc$ then $f(t)$ is monotone increasing on $\mathbb{R}$ and
\begin{equation}\label{eq:f1}
\ln\frac{b}{d}\le f(t)\le \ln\frac{a}{c},\qquad t\in\mathbb{R}.
\end{equation}
Similarly, if  $ad<bc$ then $f(t)$ is monotone decreasing on
$\mathbb{R}$ and
\begin{equation*}
\ln\frac{a}{c}\le f(t)\le \ln\frac{b}{d},\qquad t\in\mathbb{R}.
\end{equation*}

\item[\rm(b)] Furthermore,
\begin{equation}\label{eq:f'}
|f'(t)|\le
\frac{|ad-bc|}{\bigl(\sqrt{ad}+\sqrt{bc}\myp\bigr)^2},\qquad t\in
\mathbb{R}.
\end{equation}
\end{itemize}
\end{lemma}
\proof (a) Differentiating equation \eqref{eq:f}, we get
\begin{equation}\label{eq:f'a}
f'(t)=\frac{a\myp\rme^t}{a\myp\rme^t+b}-\frac{c\myp\rme^t}{c\myp\rme^t+d}=\frac{ad-bc}{ac\mypp
\rme^{t}+bd\mypp\rme^{-t}+ad+bc}.
\end{equation}

If $ad>bc$ then, according to \eqref{eq:f'a}, the function $f(t)$ is
monotone increasing, and the bounds \eqref{eq:f1} follow by taking
the limit as $t\to\pm\infty$. The case $ad<bc$ is similar.

\smallskip
(b) By the inequality between the arithmetic and geometric means,
the denominator on the right-hand side of \eqref{eq:f'a} is bounded
below by
$$
2\sqrt{abcd}+ad+bc=\bigl(\sqrt{ad}+\sqrt{bc}\myp\bigr)^2,
$$
and the result \eqref{eq:f'} follows.
\endproof

Let us define two norms for vector
$u=(u_1,\dots,u_{q-1})\in\mathbb{R}^{q-1}$,
\begin{equation}\label{eq:norm}
\|u\|_\infty:=\max_{1\leq i\leq q-1} |u_i|,\qquad
\|u\|_1:=\sum_{i=1}^{q-1} |u_i|.
\end{equation}
The next two lemmas give useful estimates for the function
$\boldsymbol{F}=(F_1,\dots,F_{q-1})$ defined in \eqref{eq:Fi} and
for its partial derivatives.
\begin{lemma}\label{l1}
For any $\theta\ge1$, the following uniform estimate holds,
\begin{equation}\label{fo}
\sup_{u\in\smash{\mathbb{R}^{q-1}}}\|\boldsymbol{F}(u;\theta)\|_\infty\le
\ln\theta.
\end{equation}
\end{lemma}

\proof Note that $F_i$, as a function of $u_i$, may be represented
by formula \eqref{eq:f} with the coefficients
\begin{equation}\label{eq:abcd}
a=\theta,\quad b=1+s,\quad c=1,\quad d=\theta+s,\quad s:=\sum_{j\ne
i} \rme^{u_j}\ge0,
\end{equation}
where
$$
ad-bc=\theta\myp(\theta+s)-(1+s)=(\theta-1)(\theta+1+s)>0.
$$
Therefore, by the estimates \eqref{eq:f1} we have
\begin{equation}\label{eq:<Fi<}
\ln\frac{1+s}{\theta+s}\le F_i(u;\theta)\le \ln \theta.
\end{equation}
Furthermore, noting that $(1+s)/(\theta+s)\ge 1/\theta$, the
two-sided bound \eqref{eq:<Fi<} implies the inequality
$|F_i(u;\theta)|\le \ln\theta$, and the bound \eqref{fo} follows by
taking the maximum over $i=1,\dots,q$.
\endproof

Recall that the function $\varphi(t;\theta)$ is defined by
\eqref{eq:phi}. Denote by $\nabla F _i$ the gradient of the map
$\boldsymbol{u}\mapsto F_i(\boldsymbol{u};\theta)$,
$$
\nabla F_i(\boldsymbol{u};\theta):=\left(\frac{\partial
F_i(\boldsymbol{u};\theta)}{\partial u_1},\dots, \frac{\partial
F_i(\boldsymbol{u};\theta)}{\partial u_{q-1}}\right),\qquad
\boldsymbol{u}\in\mathbb{R}^{q-1}.
$$
Recall that the norm $\|\myp{\cdot}\myp\|_1$ is defined in
\eqref{eq:norm}.
\begin{lemma}\label{l2}
For $\gamma\in\mathbb{R}$ and any
$\boldsymbol{u}=(u_1,\dots,u_{q-1})\in \mathbb{R}^{q-1}$ such that
$\min_{1\le i\le q-1}u_i\ge \gamma \ln\theta$, it holds
\begin{equation}\label{ho1}
\max_{1\le i\le q-1}\|\nabla F_i(\boldsymbol{u};\theta)\|_1\le
Q(\theta)+\varphi(t_\gamma(\theta);\theta),
\end{equation}
where $Q(\theta)$ is defined in \eqref{eq:Q} and
$t_\gamma(\theta)=\theta+1+(q-2)\mypp\theta^\gamma$ \textup{(}see
\eqref{eq:theta_K}\textup{)}. Moreover, the following uniform
estimate holds,
\begin{equation}\label{ho1*}
\max_{1\le i\le q-1}\sup_{\boldsymbol{u}\in\mathbb{R}^{q-1}}
\|\nabla F_i(\boldsymbol{u};\theta)\|_1\le Q(\theta)+
\frac{\theta-1}{\theta+1}.
\end{equation}

\end{lemma}
\proof Like in the proof of Lemma \ref{l1}, let us represent
$F_i(\boldsymbol{u};\theta)$ by formula \eqref{eq:f} with the
coefficients \eqref{eq:abcd}. Then by Lemma \ref{lm:f}
\begin{equation}\label{eq:b1}
\left|\frac{\partial F_i(\boldsymbol{u};\theta)}{\partial
u_i}\right|\leq
\frac{(\theta-1)(\theta+1+s)}{\bigl(\sqrt{\theta\myp(\theta+s)}+\sqrt{1+s}\myp\bigr)^2}=
\varphi(\theta+1+s;\theta),
\end{equation}
where $s=\sum_{j\ne i}\rme^{u_j}\ge
(q-2)\mypp\theta^{\gamma}=t_\gamma(\theta)-\theta-1$. Hence, by
monotonicity of the map $t\mapsto \varphi(t;\theta)$, from
\eqref{eq:b1} it follows that
\begin{equation}\label{eq:b1i=j}
\left|\frac{\partial F_i(\boldsymbol{u};\theta)}{\partial
u_i}\right|\le \varphi(t_\gamma(\theta);\theta).
\end{equation}

On the other hand, expressing $F_i(\boldsymbol{u};\theta)$ by
formula \eqref{eq:f} with
\begin{equation*}
a=1,\quad b=\theta\mypp  \rme^{u_i}+1+s',\quad c=1,\quad
d=\theta+\rme^{u_i}+s',\quad s':=\sum_{\ell\ne i,j}
\rme^{u_\ell}\ge0,
\end{equation*}
by Lemma \ref{lm:f} we obtain
\begin{align} \left|\frac{\partial
F_i(\boldsymbol{u};\theta)}{\partial u_j}\right|&\le
\frac{(\theta-1)\myp|\rme^{u_i}-1|}{\bigl(\sqrt{\theta+\rme^{u_i}+s'}+\sqrt{\theta\myp\rme^{u_i}+1+s'}\myp\bigr)^2}.
\label{eq:Fij1}
\end{align}
If $u_i>0$ then the estimate \eqref{eq:Fij1} specializes to
\begin{align}
\notag \left|\frac{\partial F_i(\boldsymbol{u};\theta)}{\partial
u_j}\right|&\le
\frac{(\theta-1)(\rme^{u_i}-1)}{\bigl(\sqrt{\theta+\rme^{u_i}+s'}+
\sqrt{\theta\myp\rme^{u_i}+1+s'}\myp\bigr)^2}\\
\notag &\le\frac{\theta-1}{\bigl(\sqrt{1+(\theta+1)/(\rme^{u_i}-1)}
+\sqrt{\theta+(\theta+1)/(\rme^{u_i}-1)}\myp\bigr)^2}\\
\label{eq:ij2} &\le
\frac{\theta-1}{(\sqrt{1}+\sqrt{\theta}\myp)^2}=\frac{\sqrt{\theta}-1}{\sqrt{\theta}+1}.
\end{align}
Similarly, if $u_i\le 0$ then $1\ge \rme^{u_i}>0$ and from
\eqref{eq:Fij1} we obtain
\begin{align}
\notag \left|\frac{\partial F_i(\boldsymbol{u};\theta)}{\partial
u_j}\right|&\le \frac{(\theta-1)\myp
(1-\rme^{u_i})}{\bigl(\sqrt{\theta+\rme^{u_i}+s'}
+\sqrt{\theta\myp\rme^{u_i}+1+s'}\myp\bigr)^2}\\
\label{eq:ij3} &\le
\frac{\theta-1}{(\sqrt{\theta}+\sqrt{1}\myp)^2}=\frac{\sqrt{\theta}-1}{\sqrt{\theta}+1}.
\end{align}
Thus, combining \eqref{eq:ij2} and \eqref{eq:ij3}, we have
\begin{equation}\label{eq:b1i<>j}
\left|\frac{\partial F_i(\boldsymbol{u};\theta)}{\partial
u_j}\right|\le \frac{\sqrt{\theta}-1}{\sqrt{\theta}+1},\qquad i\ne
j.
\end{equation}
As a result, according to the definition \eqref{eq:norm} of the norm
$\|\cdot\|_1$, the bounds \eqref{eq:b1i=j} and \eqref{eq:b1i<>j}
imply the estimate~\eqref{ho1}.

Finally, since
\begin{equation}\label{eq:gamma-infty}
\lim_{\gamma\to-\infty}\varphi(t_\gamma(\theta);\theta)=\varphi(\theta+1;\theta)=\frac{\theta-1}{\theta+1},
\end{equation}
from \eqref{ho1} we obtain \eqref{ho1*}, and the proof of
Lemma~\ref{l2} is complete.
\endproof

\begin{remark} Estimates similar to \eqref{ho1}
were proved in \cite{RS}.
\end{remark}


\begin{lemma}\label{lm:mono} For integer $q\ge2$ and any
$\gamma\in\mathbb{R}$, the map $[1,\infty)\ni\theta\mapsto
\varphi(t_\gamma(\theta);\theta)$ is a continuous, strictly
increasing function with the range $[0,1)$.
\end{lemma}
\proof Denoting $\tilde{q}:=q-2\in\mathbb{N}_0$, by formula
\eqref{eq:phi1} we have
\begin{align}
\notag
\varphi(t_\gamma(\theta);\theta)=\varphi(\theta+1+\tilde{q}\mypp\theta^\gamma;\theta)&=
\frac{\sqrt{\theta\myp(\theta+\tilde{q}\mypp\theta^\gamma)}-\sqrt{1+\tilde{q}\mypp\theta^\gamma}}
{\sqrt{\theta\myp(\theta+\tilde{q}\mypp\theta^\gamma)}+\sqrt{1+\tilde{q}\mypp\theta^\gamma}}\\
&=1-2\left(1+\sqrt{\frac{\theta^2+\tilde{q}\mypp\theta^{\gamma+1}}{1+\tilde{q}\mypp\theta^\gamma}}\,\right)^{\!-1}.
\label{eq:phi11}
\end{align}
Clearly, the function \eqref{eq:phi11} is continuous, so we only
need to show that
\begin{equation}\label{eq:A}
A(\theta):=\frac{\theta^2+\tilde{q}\mypp\theta^{\gamma+1}}{1+\tilde{q}\mypp\theta^\gamma},\qquad
\theta\ge 1,
\end{equation}
is an increasing function.

First of all, if $\tilde{q}=0$ then \eqref{eq:A} is reduced to
$A(\theta)=\theta^2$, so there is nothing to prove. Suppose that
$\tilde{q}\ge 1$. Differentiating \eqref{eq:A}, it is easy to see
that
\begin{equation}\label{eq:A'}
A'(\theta)=\frac{2\myp\theta+\tilde{q}\mypp\theta^\gamma\bigl[\theta+2+\tilde{q}\mypp\theta^\gamma+(1-\gamma)(\theta-1)\bigr]}
{\bigl(1+\tilde{q}\mypp\theta^\gamma\bigr)^2},
\end{equation}
and it is evident that the right-hand side of \eqref{eq:A'} is
positive for all $\theta\ge1$ as long as $\gamma\le1$. On the other
hand, for $\gamma\ge1$ by Bernoulli's inequality we have
$$
\tilde{q}\mypp\theta^\gamma\ge
\theta^\gamma=\bigl(1+(\theta-1)\bigr)^\gamma\ge 1+\gamma\mypp
(\theta-1),
$$
and the expression in square brackets in \eqref{eq:A'} is estimated
from below by $(\theta+2)+1+(\theta-1)=2\myp\theta+2>0$. Thus, in
all cases $A'(\theta)>0$ for $\theta\ge1$, as required.

Finally, from \eqref{eq:A} we see that $A(1)=1$ and
$\lim_{\theta\to\infty}A(\theta)=\infty$, and it follows that the
range of the function \eqref{eq:phi11} is $[0,1)$, which completes
the proof of the lemma.
\endproof

\begin{lemma}\label{lm:unique}
For $k\ge 2$, $q\ge 2$ and any $\gamma\in\mathbb{R}$, the
equation~\eqref{eq:equation-alpha*} has a unique root
$\theta^*_{\myn\gamma}=\theta^*_{\myn\gamma}(k,q)$. If $q=2$ then
$\theta^*_{\myn\gamma}(k,2)\equiv \theta_0(k,2)=(k+1)/(k-1)$, where
$\theta_0(k,q)$ is the root of the equation \eqref{eq:theta0}. For
$q\ge 3$, the root $\theta^*_{\myn\gamma}$ is a continuous monotone
increasing function of\/ $\gamma$, such that
\begin{equation}\label{eq:gamma-limits}
\lim_{\gamma\to-\infty}\theta^*_{\myn\gamma}(k,q)=\theta_0(k,q),\qquad
\lim_{\gamma\to+\infty}\theta^*_{\myn\gamma}(k,q)=\left(\theta_*(k,q)\right)^2,
\end{equation}
where $\theta_*(k,q)$ is the root of the equation
\eqref{eq:theta1}\textup{;} in particular,
\begin{equation}\label{eq:gamma-bounds}
\theta_0(k,q)<\theta^*_{\myn\gamma}(k,q)<\left(\theta_*(k,q)\right)^2.
\end{equation}
\end{lemma}
\proof The case $q=2$ is straightforward, so assume that $q\ge3$.
Due to continuity and monotonicity of the function $Q(\theta)$ (see
\eqref{eq:Q}) and by virtue of Lemma \ref{lm:mono}, the left-hand
side of equation \eqref{eq:equation-alpha*} is a continuous
increasing function of $\theta\in[1,\infty)$, with the range
$[0,q-1)$ because
$$
\lim_{\theta\downarrow 1} \bigl(
Q(\theta)+\varphi(t_\gamma(\theta);\theta)\bigr)=0,\qquad
\lim_{\theta\to\infty}
\bigl(Q(\theta)+\varphi(t_\gamma(\theta);\theta)\bigr)=q-1.
$$
Hence, the equation \eqref{eq:equation-alpha*} always has a unique
solution, $\theta^*_{\myn\gamma}=\theta^*_{\myn\gamma}(k,q)$. Since
$t_\gamma(\theta)$ is a continuous increasing function of $\gamma$,
while the map $t\mapsto\varphi(t;\theta)$ is continuous and
decreasing, it follows that the root $\theta^*_{\myn\gamma}$ is
continuous and increasing in $\gamma$.

Finally, observing that (see \eqref{eq:>phi>}
and~\eqref{eq:gamma-infty})
$$
\lim_{\gamma\to-\infty}
\varphi(t_\gamma(\theta);\theta)=\varphi(\theta+1;\theta)=\frac{\theta-1}{\theta+1},\qquad
\lim_{\gamma\to\infty}
\varphi(t_\gamma(\theta);\theta)=\varphi(\infty;\theta)=\frac{\sqrt{\theta}-1}{\sqrt{\theta}+1},
$$
and comparing equation \eqref{eq:equation-alpha*} with the limiting
equations as $\gamma\to\pm\infty$ (which have the roots $\theta_0$
and $\theta_*^2$, respectively), we obtain the required limits
\eqref{eq:gamma-limits}, and hence the asymptotic bounds
\eqref{eq:gamma-bounds} for $\theta^*_{\myn\gamma}$.
\endproof

\section{Proofs of the main results related to uniqueness}\label{sec:4}

\subsection{Proof of Theorem \textup{\ref{ep}} (criterion of compatibility)}\label{sec:4.1}

For shorthand, denote temporarily
$\boldsymbol{\zeta}(x):=\boldsymbol{h}(x)+\boldsymbol{\xi}(x)$.
Suppose that the compatibility condition \eqref{p**} holds. On
substituting \eqref{p*},
it is easy to see that \eqref{p**} simplifies to
\begin{multline}\label{puu}
\prod_{x\in W_{n}}\prod_{y\in S(x)}
\sum_{\omega(y)\in\varPhi}\exp\bigl\{\beta \bigl(J_{xy}
\mypp\delta_{\sigma_n(x),\mypp\omega(y)}+\zeta_{\omega(y)}(y)\bigr)\bigr\}=\frac{Z_{n+1}}{Z_{n}}
\prod_{x\in W_{n}}\exp\bigl\{\beta h_{\sigma_n(x)}(x)\bigr\},
\end{multline}
for any $\sigma_{n} \in\varPhi^{V_n}$. Consider the equality
\eqref{puu} on configurations
$\sigma_n^1,\sigma_n^2\in\varPhi^{V_n}$ that coincide everywhere in
$V_n$ except at vertex $x\in W_{n}$, where $\sigma_n^1(x)=i\le q-1$
and $\sigma_n^2(x)=q$. Taking the log-ratio of the two resulting
relations, we obtain
\begin{equation*}
\sum_{y\in
S(x)}\ln\frac{\exp\bigl\{\beta\bigl(J_{xy}+\zeta_{i}(y)\bigr)\bigr\}+\sum_{j\ne
i}\exp\bigl\{\beta\myp\zeta_{j}(y)\bigr\}}
{\exp\bigl\{\beta\bigl(J_{xy}+\zeta_{q}(y)\bigr)\bigr\}+\sum_{j=1}^{q-1}\exp\bigl\{\beta\myp\zeta_{j}(y)\bigr\}}=
\beta\bigl(h_{i}(x)-h_{q}(x)\bigr),
\end{equation*}
which is readily reduced to \eqref{p***} in view of the notation
\eqref{hxi} and~\eqref{eq:Fi}.

Conversely, again using \eqref{hxi} and~\eqref{eq:Fi}, equation
\eqref{p***} can be rewritten in the coordinate form as follows,
\begin{equation}\label{pru}
\prod_{y\in S(x)} \sum_{j=1}^q\exp\bigl\{\beta\bigl(J_{xy}
\mypp\delta_{ij}+\zeta_{j}(y)\bigr)\bigr\}= a(x)\exp\{\beta
h_{i}(x)\},\qquad i=1,\dots,q-1,
\end{equation}
where (omitting the immaterial dependence on $\beta$,
$\boldsymbol{h}$ and $\boldsymbol{\xi}$) we denote
$$
a(x):= \exp\{\beta h_{q}(x)\} \prod_{y\in S(x)}
\sum_{j=1}^q\exp\bigl\{\beta\bigl(J_{xy}
\mypp\delta_{qj}+\zeta_{j}(y)\bigr)\bigr\},\qquad x\in V.
$$
Hence, using \eqref{pru} and setting $A_n:=\prod_{x \in W_{n}}
a(x)$, we get
\begin{align}
\notag \sum_{\omega\in
\varPhi^{W_{n+1}}}\mu^h_{n+1}(\sigma_{n}\mynn\vee
\omega)&=\frac{\exp\{-\beta H_{n}(\sigma_{n})\}}{Z_{n+1}}
\prod_{x\in W_{n}} \prod_{y\in
S(x)}\sum_{j=1}^q\exp\bigl\{\beta\bigl(J_{xy}\mypp
\delta_{\sigma_{n}(x),\myp j}+\zeta_{j}(y)\bigr)\bigr\}\\
\label{pru2} &= \frac{A_{n}}{Z_{n+1}}\exp\left\{-\beta
H_{n}(\sigma_{n})+\beta\sum_{x\in
W_{n}}h_{\sigma_{n}(x)}(x)\right\}=\frac{A_{n}Z_n}{Z_{n+1}}\mu_n^h(\sigma_n).
\end{align}
Finally, observe that
$$
\sum_{\sigma_{n}\in\varPhi^{V_{n}}} \sum_{\omega\in
\varPhi^{W_{n+1}}}\mu_{n+1}^h(\sigma_{n}\mynn\vee \omega)
=\sum_{\sigma_{n+1}\in\varPhi^{V_{n+1}}}
\mu_{n+1}^h(\sigma_{n+1})=1,
$$
whereas from the right-hand side of \eqref{pru2} the same sum is
given by
$$
\sum_{\sigma_{n}\in\varPhi^{V_{n}}}\frac{A_{n}Z_n}{Z_{n+1}}\mu_n^h(\sigma_n)=\frac{A_{n}Z_n}{Z_{n+1}}.
$$
Hence, $A_{n}Z_{n}/Z_{n+1}=1$ and formula \eqref{pru2} yields
\eqref{p**}, as required. This completes the proof of
Theorem~\ref{ep}.

\subsection{Preparatory results for the uniqueness of SGM}\label{sec:4.2}
First, let us rewrite the functional equation \eqref{p***} in a form
more convenient for iterations. Recall that we assume $J_{xy}\equiv
J>0$ ($d(x,y)=1$) and use the notation $\theta=\rme^{\beta J}$.
\begin{lemma}\label{lm:h->g}
Via the substitutions
\begin{equation}\label{eq:g}
\boldsymbol{g}(x)
=\boldsymbol{F}\bigl(\beta\myp\boldsymbol{\check{h}}(x)+\beta\myp\boldsymbol{\check{\xi}}(x);\theta\bigr)\in\mathbb{R}^{q-1},
\qquad x\in V,
\end{equation}
and
\begin{equation}\label{eq:h->g}
\boldsymbol{\check{h}}(x)=\beta^{-1}\!\sum_{y\in
S(x)}\boldsymbol{g}(y),\qquad x\in V,
\end{equation}
equation \eqref{p***} is equivalent to the fixed-point equation
\begin{equation}\label{rz3.2}
\boldsymbol{g}(x)=\boldsymbol{\varPsi} \boldsymbol{g}(x),\qquad x\in
V,
\end{equation}
where the mapping\/ $\boldsymbol{\varPsi}\colon
(\mathbb{R}^{q-1})^{V}\to(\mathbb{R}^{q-1})^{V}$ is defined by
\begin{equation}\label{eq:Psi}
\boldsymbol{\varPsi}
\boldsymbol{g}(x):=\boldsymbol{F}\!\left(\beta\myp\boldsymbol{\check{\xi}}(x)+\sum_{\smash{y\in
S(x)}}\boldsymbol{g}(y);\theta\right),\qquad x\in V.
\end{equation}
\end{lemma}
\proof By means of \eqref{eq:g}, the recursive equation \eqref{p***}
for $\boldsymbol{\check{h}}$ can be written as \eqref{eq:h->g}.
Substituting this into \eqref{eq:g} and using the notation
\eqref{eq:Psi}, we see that $\boldsymbol{g}$ solves the functional
equation~\eqref{rz3.2}. Conversely, if $\boldsymbol{g}$ satisfies
the equation~\eqref{rz3.2} then for $\boldsymbol{\check{h}}$ defined
by \eqref{eq:h->g} we have, using \eqref{eq:Psi},
\begin{align*}
\beta\myp\boldsymbol{\check{h}}(x)=\sum_{y\in
S(x)}\boldsymbol{g}(y)&=\sum_{y\in S(x)}\boldsymbol{\varPsi}\boldsymbol{g}(y)\\
&=\sum_{y\in
S(x)}\boldsymbol{F}\!\left(\beta\myp\boldsymbol{\check{\xi}}(y)+\sum_{\smash{z\in
S(y)}}\boldsymbol{g}(z);\theta\right)\\
&=\sum_{y\in
S(x)}\boldsymbol{F}\!\left(\beta\myp\boldsymbol{\check{\xi}}(y)+\beta\myp
\boldsymbol{\check{h}}(y);\theta\right),
\end{align*}
so that $\boldsymbol{\check{h}}$ solves the equation~\eqref{p***}.
Thus, Lemma~\ref{lm:h->g} is proved.
\endproof
In particular, Lemma~\ref{lm:h->g} implies that for the proof of
uniqueness of SGM it suffices to show that the equation
\eqref{rz3.2} has a unique solution $\boldsymbol{g}(x)$ ($x\in V$).

Let us state and prove one general result in the contraction case.
On the vector space  $(\mathbb{R}^{q-1})^V\!$ of
$\mathbb{R}^{q-1}$-valued functions on the vertex set $V$ of the
Cayley tree $\mathbb{T}^k$, introduce the $\sup$-norm
\begin{equation*}
\|\boldsymbol{g}\|_V:=\sup_{x\in V}\|\boldsymbol{g}(x)\|_\infty=
\sup_{x\in V}\max_{1\le i\le q-1}|g_i(x)|,\qquad
\boldsymbol{g}(x)=(g_1(x),\dots,g_{q-1}(x)).
\end{equation*}
Sometimes, we need the similar norm for functions restricted to
subsets $\varLambda\subseteq V$,
\begin{equation}\label{eq:norm*}
\|\boldsymbol{g}\|_{\varLambda}:=\sup_{x\in
\varLambda}\|\boldsymbol{g}(x)\|_\infty,\qquad \boldsymbol{g}\in
(\mathbb{R}^{q-1})^{\varLambda}\!.
\end{equation}

The next lemma and its proof are an adaptation of a standard result
for $\ell^\infty(\mathbb{R})$.

\begin{lemma}\label{lm:complete}
For any subset $\varLambda\subseteq V$, the space
$(\mathbb{R}^{q-1})^{\varLambda}$ is complete with respect to the
sup-norm \eqref{eq:norm*}.
\end{lemma}
\proof Let $\{\boldsymbol{g}^n\}$ be a Cauchy sequence in
$(\mathbb{R}^{q-1})^{\varLambda}\!$, that is, for any
$\varepsilon>0$ there is $N\in\mathbb{N}$ such that for any $n,m\ge
N$ we have
$\|\boldsymbol{g}^n-\boldsymbol{g}^m\|_{\varLambda}<\varepsilon$. In
particular, $\{\boldsymbol{g}^n\}$ is bounded,
$\|\boldsymbol{g}^n\|_{\varLambda}\!\le M<\infty$ for some $M>0$ and
all $n\in\mathbb{N}$. Note that every coordinate sequence
$\{g^n_i(x)\}$ ($i=1,\dots,q-1$, $x\in \varLambda$) is also a Cauchy
sequence (in $\mathbb{R}$) because, according to \eqref{eq:norm},
$|g^n_i(x)-g^m_i(x)|\le
\|\boldsymbol{g}^n-\boldsymbol{g}^m\|_{\varLambda}<\varepsilon$;
hence, it converges to a limit which we denote $g_i(x)$. Clearly,
$|g_i(x)|\le M$ and $\|\boldsymbol{g}\|_{\varLambda}\!=\sup_{x\in
\varLambda}\max_i |g_i(x)|\le M<\infty$.

Now, passing to the limit as $m\to\infty$ in each inequality
$|g^n_i(x)-g^m_i(x)|<\varepsilon$, we obtain $|g^n_i(x)-g_i(x)|\le
\varepsilon$, which implies that
$\|\boldsymbol{g}^n-\boldsymbol{g}\|_{\varLambda}\!\le \varepsilon$,
for all $n\ge N$. Since $\varepsilon>0$ is arbitrary, we conclude
that $\|\boldsymbol{g}^n-\boldsymbol{g}\|_{\varLambda}\!\to0$ as
$n\to\infty$, and the lemma is proved.
\endproof

We also require the following simple estimate.

\begin{lemma}\label{lm:grad<} Let $f(\boldsymbol{u})\colon \mathbb{R}^{q-1}\!\to\mathbb{R}$ be a
$C^1$-function and $\nabla f(\boldsymbol{u})=\bigl(\frac{\partial
f(\boldsymbol{u})}{\partial u_1},\dots,\frac{\partial
f(\boldsymbol{u})}{\partial u_{q-1}}\bigr)$ its gradient. Then, for
any $\boldsymbol{v},\boldsymbol{w}\in \mathbb{R}^{q-1}$,
\begin{equation}\label{eq:<norm}
|f(\boldsymbol{w})-f(\boldsymbol{v})|\le
\|\boldsymbol{w}-\boldsymbol{v}\|_\infty\, \sup_{\boldsymbol{u}\in
\smash{\mathbb{R}^{q-1}}}\|\nabla f(\boldsymbol{u})\|_1.
\end{equation}
\end{lemma}
\proof Define the function
$\psi(t):=f(\boldsymbol{v}+t\myp(\boldsymbol{w}-\boldsymbol{v}))$,
$t\in[0,1]$, then
$$
f(\boldsymbol{w})-f(\boldsymbol{v})=\psi(1)-\psi(0)=\int_0^1
\psi'(t)\,\dif{t}=\int_0^1\sum_{i=1}^{q-1}\frac{\partial
f(\boldsymbol{v}+t\myp(\boldsymbol{w}-\boldsymbol{v}))}{\partial
u_i}\mypp(w_i-v_i)\,\dif{t},
$$
whence the estimate \eqref{eq:<norm} readily follows.
\endproof

\begin{theorem}\label{th:FP}
Suppose that, for some $\theta>1$,
\begin{equation}\label{eq:contr}
\lambda(\theta):=k\max _{1\le i\le q-1}\sup _{\boldsymbol{u}\in
\smash{\mathbb{R}^{q-1}}} \|\nabla F_i(\boldsymbol{u};\theta)\|_1<1.
\end{equation}
Then, for every realization of the field
$\boldsymbol{\xi}=\{\boldsymbol{\xi}(x)\}_{x\in V}$, the equation
\eqref{rz3.2} has a unique solution.
\end{theorem}
\proof Consider a mapping
$\boldsymbol{\varPsi}=(\varPsi_1,\dots,\varPsi_{q-1})$ of the space
$(\mathbb{R}^{q-1})^V$ to itself defined by formula \eqref{eq:Psi}.
Solving the functional equation \eqref{rz3.2} is then equivalent to
finding a fixed point of $\boldsymbol{\varPsi}$, that is,
$\boldsymbol{\varPsi} \boldsymbol{g}^*=\boldsymbol{g}^*$. The
lemma's hypothesis implies that $\boldsymbol{\varPsi}$ is a
\emph{contraction} on $(\mathbb{R}^{q-1})^{V_0^c}$; indeed, for any
functions $\boldsymbol{g},\bar{\boldsymbol{g}}\in
(\mathbb{R}^{q-1})^V$ and each $i=1,\dots,q-1$, we obtain, using
\eqref{eq:Psi} and Lemma~\ref{lm:grad<},
\begin{align*}
|\varPsi_i\myp
\boldsymbol{g}(x)-\varPsi_i\myp\bar{\boldsymbol{g}}(x)|&\le
\sup_{\boldsymbol{u}\in\smash{\mathbb{R}^{q-1}}}\|\nabla
F_i(\boldsymbol{u};\theta)\|_1\sum_{y\in
S(x)}\|\boldsymbol{g}(y)-\bar{\boldsymbol{g}}(y)\|_\infty.
\end{align*}
Noting that for $x\ne x_\circ$ the set $S(x)$ contains exactly $k$
vertices, and recalling condition \eqref{eq:contr} with
$\lambda(\theta)\in[0,1)$, it follows that
\begin{align*}
\|\boldsymbol{\varPsi}\boldsymbol{g}-\boldsymbol{\varPsi}\myp\bar{\boldsymbol{g}}\|_{V_0^c}&=\sup_{x\in
V_0^c}\|\boldsymbol{\varPsi} \boldsymbol{g}(x)-\boldsymbol{\varPsi}\myp\bar{\boldsymbol{g}}(x)\|_\infty\\
&\le\max_{1\le i\le q-1}
\sup_{\boldsymbol{u}\in\smash{\mathbb{R}^{q-1}}}\|\nabla
F_i(\boldsymbol{u};\theta)\|_1\cdot
k\mypp\|\boldsymbol{g}-\bar{\boldsymbol{g}}\|_{V_0^c}=\lambda(\theta)\,\|\boldsymbol{g}-\bar{\boldsymbol{g}}\|_{V_0^c}.
\end{align*}
Thus, because $(\mathbb{R}^{q-1})^{V_0^c}$ is a Banach space
(Lemma~\ref{lm:complete}), the well-known Banach contraction
principle (e.g., \cite[Theorem~9.23, page 220]{Rudin}) implies that
$\|\boldsymbol{g}-\bar{\boldsymbol{g}}\|_{V_0^c}=0$, that is,
$\boldsymbol{g}(x)=\bar{\boldsymbol{g}}(x)$ for all $x\in V_0^c$. It
remains to notice that the value of the solution $\boldsymbol{g}(x)$
at $x=x_\circ$ is uniquely determined from formulas \eqref{rz3.2}
and \eqref{eq:Psi} using the (unique) values outside
$V_0=\{x_\circ\}$. This completes the proof of Theorem~\ref{th:FP}.
\endproof

\begin{remark}\label{rm:5.1}
The unique solution $\boldsymbol{g}^*$ can be obtained by iterations
\cite{Rudin}; for example, put $\boldsymbol{g}^{0}\equiv
\boldsymbol{0}$ and define
$\boldsymbol{g}^{n}\!:=\boldsymbol{\varPsi} \boldsymbol{g}^{n-1}$
($n\in\mathbb{N}$), then $\boldsymbol{g}^n\to \boldsymbol{g}^*$ as
$n\to\infty$ (i.e.,
$\lim_{n\to\infty}{\|\boldsymbol{g}^n-\boldsymbol{g}^*\|_V=0}$).
\end{remark}

\begin{remark}\label{rm:extension}
It is straightforward to generalize Theorem \ref{th:FP} to the case
where the vector $\beta\myp\boldsymbol{\check{\xi}}(x)+\sum_{y\in
S(x)}\boldsymbol{g}(y)$ (see~\eqref{eq:Psi}) is guaranteed to be in
a convex domain $B(x)\subseteq\mathbb{R}^{q-1}$ for any function
$\boldsymbol{g}\colon V\to\mathbb{R}^{q-1}$ from a suitable subspace
$\mathscr{D}\subseteq (\mathbb{R}^{q-1})^V$, such that $\mathscr{D}$
is closed with respect to the norm $\|\mypp{\cdot}\mypp\|_V$ and
$\boldsymbol{\varPsi}(\mathscr{D})\subseteq \mathscr{D}$\mynn. In
that case, the supremum in \eqref{eq:contr} should be taken over all
$\boldsymbol{u}\in B(x)$,
\begin{equation*}
\lambda(\theta):=k\sup_{x\in V}\max _{1\le i\le
q-1}\sup_{\boldsymbol{u}\in B(x)} \|\nabla
F_i(\boldsymbol{u};\theta)\|_1<1,
\end{equation*}
and the unique solution $\boldsymbol{g}^*$ automatically
\strut{}belongs to~$\mathscr{D}\mynn$. For our purposes, it will
suffice to consider the balls
$B(x)=\{\boldsymbol{u}\in\mathbb{R}^{q-1}\colon
\|\boldsymbol{u}-\beta\myp\boldsymbol{\check{\xi}}(x)\|_\infty\le
k\ln\theta\}$ and the corresponding subspace
$\mathscr{D}=\{\boldsymbol{g}\in(\mathbb{R}^{q-1})^V\colon\,\|\boldsymbol{g}\|_V\le
\ln\theta\}$ \strut{}(see Lemma~\ref{l1}).
\end{remark}

\subsection{Proofs of Theorems \ref{t2}, \ref{t3} and \ref{t4} (uniqueness)}\label{sec:4.3}

By virtue of Remark \ref{rm:r2.1}, for the uniqueness in the class
of all Gibbs measure it suffices to prove it for SGMs.

\subsubsection{Proof of Theorem \textup{\ref{t2}}}
By virtue of the uniform bound \eqref{ho1*} of Lemma \ref{l2}, for
every $\theta\in[1,\theta_0)$ we have
$$
\lambda(\theta)=k\max_{1\le i\le q-1}
\sup_{\boldsymbol{u}\in\smash{\mathbb{R}^{q-1}}}\|\nabla
F_i(\boldsymbol{u};\theta)\|_1\le
k\left(Q(\theta)+\frac{\theta-1}{\theta+1}\right)<1,
$$
and the required result follows by Theorem~\ref{th:FP}.

\subsubsection{Proof of Theorem \textup{\ref{t3}}} In
view of equation \eqref{eq:equation-alpha*} with
$\gamma=\varDelta^\xi-k$, we have
\begin{equation}\label{eq:<1/k}
Q(\theta)+\varphi\bigl(t_{\myn\varDelta^\xi-k}(\theta);\theta\bigr)<\frac{1}{k},\qquad
\theta\in[1,\theta^*_{\mynn\varDelta^\xi-k}).
\end{equation}
By continuity of the map
$\gamma\mapsto\varphi(t_\gamma(\theta);\theta)$, inequality
\eqref{eq:<1/k} extends to
\begin{equation}\label{eq:<1/k-delta}
Q(\theta)+\varphi\bigl(t_{\myn\varDelta^\xi-\delta-k}(\theta);\theta\bigr)<\frac{1}{k},
\end{equation}
for some $\delta>0$ small enough.

According to the definition \eqref{eq:Delta}, there exists an
integer $N$ such that
\begin{equation}\label{eq:eps}
\max_{1\le \ell\le q}\inf_{x\in V_{\myn
N}^c}{}_\ell\check{\xi}_{(1)}(x)\ge \varDelta^{\xi}-\delta.
\end{equation}
For a specific reduction
${}_{\ell\myp}\boldsymbol{\check{\xi}}(x)\!\in\!
{}_\ell\check{\mathbb{R}}^q$, with components
${}_{\ell}\check{\xi}_j(x)=\xi_j(x)-\xi_\ell(x)$ ($j\ne\ell$),
denote by ${}_{\ell}F_i(\boldsymbol{u};\theta)$ ($i\ne \ell$) the
corresponding functions analogous to  $F_i(\boldsymbol{u};\theta)$
that were defined in \eqref{eq:Fi} under the standard reduction
(i.e., with $\ell=q$). Lemma \ref{l2} (modified to the case of
reduction via the $\ell$-th coordinate) implies that
\begin{equation}
\label{eq:F<K} \max_{i\ne\ell}\sup_{\boldsymbol{u}\in
B_{\ell}(x)}\|\nabla{}_{\ell}F_i(\boldsymbol{u};\theta)\|_1\le
Q(\theta)+\varphi\bigl(t_{\myn{}_\ell\check{\xi}_{(1)}(x)-k}(\theta);\theta\bigr),
\end{equation}
where
\begin{align*}
B_\ell(x):={}&\!\Bigl\{\boldsymbol{u}\in{}_\ell\check{\mathbb{R}}^{q}\!:
\,\min_{j\ne\ell} u_j\ge
\bigl({}_\ell\myp\check{\xi}_{(1)}(x)-k\bigr) \ln\theta\Bigr\}.
\end{align*}
Furthermore, exploiting monotonicity and continuity of the function
$t\mapsto\varphi(t;\theta)$, we obtain from \eqref{eq:F<K}
\begin{align}
\min_{1\le \ell\le q}\sup_{x\in V_{\myn N}^c}\max_{i\ne \ell}
\sup_{\boldsymbol{u}\in
B_\ell(x)}\|\nabla{}_{\ell}F_i(\boldsymbol{u};\theta)\|_1
\label{eq:F<K*} &\le
Q(\theta)+\varphi\bigl(t^*_{N}(\theta);\theta\bigr),
\end{align}
with
\begin{equation*}
t^*_{N}(\theta):=\max_{1\le \ell\le q}\,\inf_{x\in V_{\myn
N}^c}t_{\myn{}_\ell\check{\xi}_{(1)}(x)-k}(\theta).
\end{equation*}
Due to the bound \eqref{eq:eps}, we have $ t^*_{N}(\theta)\ge
t_{\myn\varDelta^\xi-\delta-k}(\theta)$, and by monotonicity of
$t\mapsto \varphi(t;\theta)$ it follows that
$$
Q(\theta)+\varphi\bigl(t^*_{N}(\theta);\theta\bigr)\le
Q(\theta)+\varphi\bigl(t_{\myn\varDelta^\xi-\delta-k}(\theta);\theta\bigr)<\frac{1}{k},
$$
according to the estimate \eqref{eq:<1/k-delta}. Together with
\eqref{eq:F<K*}, this implies that (cf.\ condition \eqref{eq:contr})
$$ \lambda_N(\theta):=k\min_{1\le \ell\le
q}\sup_{x\in V_{\myn N}^c}\max_{i\ne \ell} \sup_{\boldsymbol{u}\in
B_\ell(x)}\|\nabla_{\ell}F_i(\boldsymbol{u};\theta)\|_1 <1.
$$
Hence, by an extended version of Theorem~\ref{th:FP} (see
Remark~\ref{rm:extension}), it follows that the solution
$\boldsymbol{g}(x)$ to the functional equation \eqref{rz3.2} is
unique on the subset $\{x\in V_{\myn N}^c\}$. Finally, the values of
the solution $\boldsymbol{g}(x)$ for $x\in V_N$ are retrieved
uniquely by the ``backward'' recursion \eqref{rz3.2}
using~\eqref{eq:Psi}. Thus, the proof of Theorem \ref{t3} is
complete.

\subsubsection{Proof of Theorem \textup{\ref{t4}}} Let
$\mu_{\beta,\myp\xi}$ and $\bar{\mu}_{\beta,\myp\xi}$ be two SGMs
determined by the functions $\boldsymbol{g}(x)$ and
$\bar{\boldsymbol{g}}(x)$, respectively, each satisfying the
functional equation~\eqref{rz3.2}. Our aim is to show that, under
the theorem's hypotheses,
$\boldsymbol{g}(x)\equiv\bar{\boldsymbol{g}}(x)$, which would imply
that $\mu_{\beta,\myp\xi}=\bar{\mu}_{\beta,\myp\xi}$. The idea of
the proof is to obtain a suitable upper bound on the norm
$\|\boldsymbol{g}(x)-\bar{\boldsymbol{g}}(x)\|_\infty$ for $x\in
W_n$ in terms of
$\|\boldsymbol{g}(y)-\bar{\boldsymbol{g}}(y)\|_\infty$ for $y\in
W_{n+1}$, and to propagate this estimate along the tree. To
circumvent cumbersome notation arising from the direct iterations,
we will use mathematical induction. Consider the filtration
$\mathscr{F}_{0}\subset \mathscr{F}_{1}\subset \dots
\mathscr{F}_{n}\subset \cdots$ consisting of the sigma-algebras
$\mathscr{F}_{n}$ generated by the values of the random field
$\boldsymbol{\xi}$ in the sequence of expanding balls $V_n=\{x\in
V\colon d(x_\circ,x)\le n\}$,
$$
\mathscr{F}_{n}:=\sigma\{\boldsymbol{\xi}(x)\colon x\in
V_{n}\},\qquad n\in\mathbb{N}_0.
$$
Put
\begin{equation}\label{eq:lambda-T3.5}
\lambda(\theta):=k\,\bigl(Q(\theta)
+\mathbb{E}\bigl[\varphi\bigl(t_{\check{\xi}_{(1)}(x)-k}(\theta);\theta\bigr)\bigr]\bigr),
\end{equation}
where the expectation does not depend on $x\in V$ due to the i.i.d.\
assumption on the field $\{\boldsymbol{\xi}(x)\}$. Let us first show
that for each $x\in W_n$ ($n\ge 1$) we have the upper bound
\begin{equation}\label{eq:|g|}
\mathbb{E}\bigl(\|\boldsymbol{g}(x)-\bar{\boldsymbol{g}}(x)\|_\infty\myp|\mypp
\mathscr{F}_{n-1}\bigr)\le 2\ln\theta
\left(\lambda(\theta)\right)^m,\qquad m\in\mathbb{N}_0,
\end{equation}
where $\mathbb{E}(\cdot\mypp|\mypp\mathscr{F}_{n-1})$ stands for the
conditional expectation.

Fix $x\in W_n$. The base of induction ($m=0$) is obvious, noting
that, due to \eqref{rz3.2}, \eqref{eq:Psi} and Lemma~\ref{l1},
$$
\|\boldsymbol{g}(x)-\bar{\boldsymbol{g}}(x)\|_\infty\le
\|\boldsymbol{g}(x)\|_\infty+\|\bar{\boldsymbol{g}}(x)\|_\infty\le
2\ln\theta.
$$
Suppose now that the bound \eqref{eq:|g|} is true for some
$m\in\mathbb{N}_0$, and show that it holds for $m+1$ as well. By
Lemma \ref{lm:grad<} we have
\begin{align}
\notag
\|\boldsymbol{g}(x)-\bar{\boldsymbol{g}}(x)\|_\infty &=
\|\boldsymbol{\varPsi}\myp
\boldsymbol{g}(x)-\boldsymbol{\varPsi}\myp\bar{\boldsymbol{g}}(x)\|_\infty\\
&\le \max_{1\le i\le q-1}\sup_{\boldsymbol{u}\in B(x)}\|\nabla
F_i(\boldsymbol{u};\theta)\|_1\sum_{y\in
S(x)}\|\boldsymbol{g}(y)-\bar{\boldsymbol{g}}(y)\|_\infty,
\label{eq:ind_m}
\end{align}
where $B(x)\subset \mathbb{R}^{q-1}$ is the ball of radius
$k\ln\theta$ centred at $\beta\myp\boldsymbol{\check{\xi}}(x)$,
$$
B(x):=\bigl\{\boldsymbol{u}\in\mathbb{R}^{q-1}\colon
\|\boldsymbol{u}-\beta\myp\boldsymbol{\check{\xi}}(x) \|_\infty\le
k\ln\theta\bigr\}.
$$
Recalling that $\beta=\ln\theta$, observe that if
$\boldsymbol{u}=(u_1,\dots,u_{q-1})\in B(x)$ then, for each
$i=1,\dots,q-1$,
$$
u_i\ge
\beta\myp\check{\xi}_i(x)-k\ln\theta=\ln\theta\mypp(\check{\xi}_i(x)-k),
$$
and hence
$$
\min_{1\le i\le q-1} u_i\ge\ln\theta \min_{1\le i\le q-1}
\bigl(\check{\xi}_i(x)-k\bigr)=\ln\theta
\mypp\bigl(\check{\xi}_{(1)}(x)-k\bigr),
$$
with $\check{\xi}_{(1)}(x)=\min_{1\le i\le q-1} \check{\xi}_i(x)$.
Therefore, on applying Lemma \ref{l2} we have
\begin{equation*}
\max_{1\le i\le q-1}\sup_{\boldsymbol{u}\in B(x)}\|\nabla
F_i(\boldsymbol{u};\theta)\|_1\le Q(\theta) +
\varphi\bigl(t_{\check{\xi}_{(1)}(x)-k}(\theta);\theta\bigr).
\end{equation*}
Thus, returning to \eqref{eq:ind_m} we get
\begin{equation}\label{eq:max<}
\|\boldsymbol{g}(x)-\bar{\boldsymbol{g}}(x)\|_\infty \le
\bigl[Q(\theta) +
\varphi\bigl(t_{\check{\xi}_{(1)}(x)-k}(\theta);\theta\bigr)\bigr]\sum_{y\in
S(x)}\|\boldsymbol{g}(y)-\bar{\boldsymbol{g}}(y)\|_\infty.
\end{equation}

Now, take the conditional expectation
$\mathbb{E}(\cdot\mypp|\mypp\mathscr{F}_{n})$ on both sides of
\eqref{eq:max<}, noting that the factor in front of the sum is a
random variable measurable with respect to the sigma-algebra
$\mathscr{F}_n$, so it can be pulled out of the expectation (see
\cite[Property~{\bf K*}, page~216]{Shiryaev}). This yields
\begin{align}
\notag
\mathbb{E}\bigl[\|\boldsymbol{g}(x)-\bar{\boldsymbol{g}}(x)\|_\infty\myp|\mypp\mathscr{F}_n\bigr]
&\le \bigl[Q(\theta) +
\varphi\bigl(t_{\check{\xi}_{(1)}(x)-k}(\theta);\theta\bigr)\bigr]\sum_{y\in
S(x)}
\mathbb{E}\bigl[\|\boldsymbol{g}(y)-\bar{\boldsymbol{g}}(y)\|_\infty\myp|\mypp\mathscr{F}_n\bigr]\\
&\le k\bigl[Q(\theta) +
\varphi\bigl(t_{\check{\xi}_{(1)}(x)-k}(\theta);\theta\bigr)\bigr]
\cdot 2\ln\theta \left(\lambda(\theta)\right)^m,
\label{eq:g<*}
\end{align}
where in the last inequality we used that $\card S(x)=k$ and also
applied the induction hypothesis to each $y\in S(x)$
(see~\eqref{eq:|g|}). In turn, using the tower property of
conditional expectation (see \cite[Property~{\bf H*},
page~216]{Shiryaev}) with $\mathscr{F}_{n-1}\subset\mathscr{F}_{n}$,
from \eqref{eq:g<*} we obtain
\begin{align*}
\mathbb{E}\bigl(\|\boldsymbol{g}(x)-\bar{\boldsymbol{g}}(x)\|_\infty\myp\bigl|\mypp\mathscr{F}_{n-1}\bigr)
&=\mathbb{E}\bigl[\mathbb{E}\bigl[\|\boldsymbol{g}(x)-\bar{\boldsymbol{g}}(x)\|_\infty\myp|\mypp\mathscr{F}_n\bigr]
\bigl|\mypp\mathscr{F}_{n-1}\bigr]\\
&\le k\, \mathbb{E}\bigl[Q(\theta) +
\varphi\bigl(t_{\check{\xi}_{(1)}(x)-k}(\theta);\theta\bigr)\myp
\bigl|\mypp\mathscr{F}_{n-1}\bigr] \cdot 2\ln\theta
\left(\lambda(\theta)\right)^m\\
&=k\, \mathbb{E}\bigl[Q(\theta) +
\varphi\bigl(t_{\check{\xi}_{(1)}(x)-k}(\theta);\theta\bigr)\bigr]
\cdot 2\ln\theta \left(\lambda(\theta)\right)^m\\
&= 2\ln\theta \left(\lambda(\theta)\right)^{m+1}
\end{align*}
(see \eqref{eq:lambda-T3.5}). Thus, the induction step is completed
and, therefore, the claim \eqref{eq:|g|} is true for all $m\ge0$. In
particular, again using the tower property of conditional
expectation, from \eqref{eq:|g|} we readily get
\begin{align}
\notag
\mathbb{E}\bigl(\|\boldsymbol{g}(x)-\bar{\boldsymbol{g}}(x)\|_\infty\bigr)
&=\mathbb{E}\bigl[\mathbb{E}\bigl(\|\boldsymbol{g}(x)-\bar{\boldsymbol{g}}(x)\|_\infty\myp|\mypp
\mathscr{F}_{n-1}\bigr)\bigr]\\
&\le 2\ln\theta \left(\lambda(\theta)\right)^m. \label{eq:g<*m}
\end{align}

Now, if $\theta^\dag>1$ is the (unique) solution of the
equation~\eqref{eq:equation-alpha-new}, then $\lambda(\theta)<1$ for
all $\theta\in[1,\theta^\dag)$. Hence, taking the limit of
\eqref{eq:g<*m} as $m\to\infty$ gives
$$
\mathbb{E}\bigl(\|\boldsymbol{g}(x)-\bar{\boldsymbol{g}}(x)\|_\infty\bigr)=0,\qquad
x\ne x_\circ,
$$
and therefore $\boldsymbol{g}(x)=\bar{\boldsymbol{g}}(x)$ (a.s.)\
for any $x\ne x_\circ$. It remains to notice that this equality
extends to $x=x_\circ$  by the recursion~\eqref{rz3.2}.

\section{Analysis of the model with constant field}\label{sec:5}
\subsection{Classification of positive solutions to the
system~\eqref{pt1}}\label{sec:5.1}

Denote
\begin{equation}\label{eq:p(z)}
p_{k}(z):=z^{k-1}+\dots+z,
\end{equation}
so that
\begin{equation}\label{eq:=p(z)}
z^k-1=(z-1)\bigl(p_k(z)+1\bigr).
\end{equation}

\begin{lemma}\label{lm2} Let $(z_1,\dots,z_{q-1})$ be a solution to
\eqref{pt1}, with $z_i>0$ \textup{(}$i=1,\dots,q-1$\textup{)}.
\begin{itemize}
\item[\rm (a)]
If $\theta=1$ then $z_1=\dots=z_{q-1}=1$ is the unique solution.

\smallskip
\item[\rm (b)] If $\theta>1$ and $\alpha=0$ then
either $z_1=\dots=z_{q-1}=1$ or there is a non-empty subset
$\mathcal{I}_0\subseteq \{1,\dots,q-1\}$, with
$m:=\card\mathcal{I}_0$ ranging from $1$ to $q-1$, such that
$$
z_i=\begin{cases} u&\text{if \,}i\in \mathcal{I}_0,\\
1&\text{otherwise},
\end{cases}
$$
where $u=u(\theta,m)\ne1$ satisfies the equation
\begin{equation}\label{eq:1=}
1=\frac{(\theta-1)(p_k(u)+1)}{\theta+m u^k+q-1-m}.
\end{equation}

\smallskip
\item[\rm (c)] If $\theta>1$ and $\alpha\ne 0$ then\textup{:}

\smallskip
\begin{itemize}
\item[\rm(i)]
either $z_1=u$ and $z_2=\dots=z_{q-1}=1$, where $u=u(\theta,\alpha)$
satisfies the equation
\begin{equation}\label{rm=1}
u=1+\frac{(\theta-1)(\theta^\alpha u^k-1)}{\theta+\theta^\alpha
u^k+q-2}
\end{equation}
and, in particular, $u\ne1$\myp\textup{;}

\smallskip
\item[\rm(ii)] or, provided that $q\ge3$, there is a non-empty subset $\mathcal{I}_1\subseteq
\{2,\dots,q-1\}$, with $m:=\card\mathcal{I}_1$ ranging from $1$ to
$q-2$, such that
$$
z_i=\begin{cases} u &\text{if \,} i=1,\\
v&\text{if \,}i\in \mathcal{I}_1,\\
1\,&\text{otherwise},
\end{cases}
$$
where $u= u(\theta,\alpha,m)$ and $v= v(\theta,\alpha,m)\ne1$
satisfy the set of equations
\begin{equation}\label{uv}
\left\{\begin{aligned} u&=1+\frac{(\theta-1)(\theta^\alpha
u^k-1)}{\theta+\theta^\alpha u^k+m v^k+q-2-m},\\
1&=\frac{(\theta-1)(p_k(v)+1)}{\theta+\theta^\alpha u^k+m
v^k+q-2-m},
\end{aligned} \right.
\end{equation}
and, in particular, $u\ne1$ and $u\ne v$.
\end{itemize}
\end{itemize}
\end{lemma}

\smallskip
\proof As a general remark, observe that $z_i=1$ solves the $i$-th
equation of the system \eqref{pt1} regardless of all other $z_j$
with $j\ne i$.
\begin{itemize}
\item[\rm (a)] Obvious.

\smallskip
\item[(b)] In this case, the system \eqref{pt1} takes the form
\begin{equation}\label{eq:pt1*}
z_i=1+\frac{(\theta-1)(z_i^k-1)}{\theta+\sum_{j=1}^{q-1}z_j^k},\qquad
i=1,\dots,q-1.
\end{equation}
Suppose that the set $\mathcal{I}_0:=\{i\ge1\colon z_i\ne1\}$ is
non-empty. By virtue of the identity \eqref{eq:=p(z)}, for any
$i\in\mathcal{I}_0$ equation \eqref{eq:pt1*} is reduced to
\begin{equation}\label{eq:z-=}
(\theta-1)\mypp p_{k}(z_i) =1+\sum_{j=1}^{q-1}z_j^k.
\end{equation}
Because the right-hand side of \eqref{eq:z-=} does not depend on
$i\in\mathcal{I}_0$ and the function $p_{k}(z)$ is strictly
increasing for $z>0$, it follows that $z_i=: u=\const$
($i\in\mathcal{I}_0$). Specifically, if $\card\mathcal{I}_0=m\ge1$
then equation \eqref{eq:pt1*} specializes to~\eqref{eq:1=}.

\smallskip
\item[(c)] The proof is similar to part~(b). First of all, note that $u:=z_1\ne1$, for otherwise the first
equation in \eqref{pt1}  is not satisfied unless $\theta = 1$ or
$\alpha = 0$, either of which is ruled out. Next, if
$z_2=\dots=z_{q-1}=1$ then the first equation for $z_1=u$ in
\eqref{pt1} specializes to~\eqref{rm=1}, as stated.

Suppose now that $\mathcal{I}_1:=\{i\ge 2 \colon z_i\ne1\}\ne
\varnothing$, then similarly as above we show that $z_i=\const$
($i\in\mathcal{I}_1$), and the system \eqref{pt1} specializes to
equations~\eqref{uv} with $z_1=u$ and $z_i=v$ ($i\in\mathcal{I}_1$).

Finally, assuming to the contrary that $u=v$ and comparing the
equations in~\eqref{uv}, we would conclude that $\theta^\alpha
u^k=u^k$, that is, $\alpha=0$, which is ruled out. Hence, $u\ne v$
as claimed.
\end{itemize}

\smallskip
Thus, the proof of Lemma \ref{lm2} is complete.
\endproof

\begin{remark}\label{rm:eta->1}
It is not hard to check that, in the limit as $\alpha\to0$, case (c)
of Lemma~\ref{lm2} transforms into case~(b).
\end{remark}

\subsection{Proof of Theorem
\textup{\ref{th:3.7}}}\label{sec:5.2} By the substitution
\begin{equation}\label{eq:x}
u^{k}=\frac{(q-1)\mypp x}{\theta^{\alpha+1}_{\vphantom{X}}}, \qquad
x>0,
\end{equation}
equation \eqref{rm=1-int} can be represented in
the form
\begin{equation}\label{bu1}
ax=f(x),\qquad f(x):=\left(\frac{1+x}{b+x}\right)^{k},
\end{equation}
with the coefficients (cf.~\eqref{eq:b})
\begin{equation}\label{eq:ab}
a= a(\theta):=\frac{q-1}{\theta^{k+1+\alpha}}>0,\qquad b=
b(\theta):=\frac{\theta\myp(\theta+q-2)}{q-1}\ge 1.
\end{equation}
Equation \eqref{bu1} is well known in the theory of Markov chains on
the Cayley tree (see, e.g., \cite[Proposition~10.7]{Pr} or
\cite[page~389]{Sp}), and it is easy to analyse the number of its
positive solutions. The case $b=1$ is obvious. Assuming $b>1$, it is
straightforward to check that $f(x)$ is an increasing function, with
$f(0)=b^{-k}<1$ and $\lim_{x\to\infty}f(x)=1$; also, it has one
inflection point $x_0=\frac12\left(b\mypp(k-1)-(k+1)\right)$, such
that $f(x)$ is convex for $x<x_0$ and concave for $x>x_0$ (note that
$x_0>0$ only when $b>\frac{k+1}{k-1}$). Therefore, the equation
\eqref{bu1} has at least one and at most three positive solutions.
In fact, by fixing $b>0$ and gradually increasing the slope $a>0$ of
the ray $y=ax$ ($x\ge0$), it is evident that there are more than one
solutions (i.e., intersections with the graph $y=f(x)$) if and only
if the equation $x\myp f'(x)=f(x)$ has at least one solution, each
such solution $x=x_*$ corresponding to the line $y=ax$, with
$a=f'(x_*)$, serving as a tangent to the graph $y=f(x)$ at point
$x=x_*$. In turn, from \eqref{bu1} we compute
\begin{equation}\label{eq:f'b}
f'(x)=k\left(\frac{1+x}{b+x}\right)^{k-1}\frac{b-1}{(b+x)^2}=f(x)\,\frac{k\myp(b-1)}{(b+x)(1+x)},
\end{equation}
and it readily follows that the condition $x\myp f'(x)=f(x)$
transcribes as the quadratic equation \eqref{eq:quadratic-intro},
with discriminant $D$ given by~\eqref{eq:D}. Thus, if $D>0$, that
is, $b>\left(\frac{k+1}{k-1}\right)^2$, then the equation
\eqref{eq:quadratic-intro} has two distinct roots
$0<x_{-}\myn<x_{+}$, corresponding to the ``critical'' values
$a_{\pm}=f(x_\pm)/x_\pm$ (see \eqref{eq:x+/--intro}
and~\eqref{eq:a_pm}). Furthermore, using \eqref{eq:f'b} it is easy
to see that the function $x\mapsto f(x)/x$ is increasing on the
interval $x\in[x_{-},x_{+}]$; hence, $a_{-}\myn<a_{+}$.

To summarize, if $b\le \left(\frac{k+1}{k-1}\right)^2$ then the
equation \eqref{bu1} has a unique solution, whereas if
$b>\left(\frac{k+1}{k-1}\right)^2$ then there are one, two or three
solutions according as $a\notin [a_{-},a_+]$, $a\in\{a_{-},a_+\}$ or
$a\in (a_{-},a_+)$, respectively. Adapting these results to equation
\eqref{rm=1-int}, in view of the second formula in \eqref{eq:ab} the
condition $b(\theta)>\left(\frac{k+1}{k-1}\right)^2$ is equivalent
to $\theta>\theta_{\rm c}$, with $\theta_{\rm c}=\theta_{\rm
c}(k,q)$ defined in \eqref{eq:theta'c}. The corresponding critical
values $\alpha_\pm$ of the field parameter $\alpha$ are determined
by the first formula in \eqref{eq:ab}, that is,
\begin{equation}\label{eq:alpha=a+/-}
\theta^{\myp k+1+\alpha_\pm}=\frac{q-1}{a_\mp},
\end{equation}
leading to formula~\eqref{eq:alpha_pm}. This completes the proof of
Theorem~\ref{th:3.7}.

\subsection{Proof of Theorem \textup{\ref{th:3.8}}}\label{sec:5.3}
For $m\in\{1,\dots,q-2\}$, denote by $m'\myn:=q-1-m$ the
``conjugate'' index, $m'\in\{1,\dots,q-2\}$. Recall the
notation~\eqref{eq:L-intr},
\begin{equation}\label{eq:L}
L_m(v;\theta):=(\theta-1)\mypp p_{k}(v)-m\myp v^{k} -m',\qquad
\theta\ge 1, \ v\ge0,
\end{equation}
where the polynomial $p_k(v)$ is defined in~\eqref{eq:p(z)}.

\begin{lemma}\label{lm:vmtheta}\mbox{}
\begin{itemize}\item[\rm (a)]
For every $\theta>1$, there is $v_m= v_m(\theta)>0$ such that
the function $v\mapsto L_m(v;\theta)$ is increasing for $0<v<v_m$
and decreasing for $v> v_m$, thus attaining its unique maximum value
at $v=v_m$,
\begin{equation}\label{eq:chi}
L_m^*(\theta):=
L_m(v_m(\theta);\theta)=\max_{v>0}L_m(v;\theta),\qquad \theta>1.
\end{equation}

\item[\rm (b)] For each $m\ge 1$, the function $\theta\mapsto L_m^*(\theta)$
defined in \eqref{eq:chi} is continuous and monotone increasing,
with $\lim_{\theta\to\infty}L_m^*(\theta)=\infty$. Furthermore,
$L_m^*(\theta)$ has a unique zero $\theta_m>1$, that is,
\begin{equation}\label{eq:theta*}
L_m^*(\theta_m)=L_m(v_m(\theta_m);\theta_m)=0.
\end{equation}
\item[\rm (c)] The value $v_m^*:=v_m(\theta_m)$ is the unique positive root of the
equation
\begin{equation}\label{vy}
m\sum_{i=1}^{k-1}i\myp v^{k-i}-m'\sum_{i=1}^{k-1}i\myp v^{i-k}=0.
\end{equation}
In particular, $v_m^*=1$ if\/ $m=\frac12(q-1)$ and $v_m^*>1$ if\/
$m<\frac12(q-1)$.
\end{itemize}
\end{lemma}
\proof (a) Differentiating \eqref{eq:L} with respect to $v$, we get
\begin{align}
\notag
\frac{\partial L_m(v;\theta)}{\partial v}&=(\theta-1)\mypp
p'_k(v)-km\myp v^{k-1}\\
&=v^{k-1}\left((\theta-1)\sum_{i=1}^{k-1}\frac{k-i}{v^i}-km\right).
\label{eq:R'v}
\end{align}
It is evident that the function in the parentheses in \eqref{eq:R'v}
is continuous and monotone decreasing in $v>0$, with the limiting
values $+\infty$ as $v\downarrow0$ and $-km<0$ as $v\to\infty$.
Hence, there is a unique root $v_m=v_m(\theta)$ of the equation
$\partial L_m(v;\theta)/\partial v=0$, that is,
\begin{equation}\label{eq:R'v*}
(\theta-1)\mypp p'_k(v_m)-km\myp v_m^{k-1} = 0,
\end{equation}
and, moreover, $\partial L_m/\partial v>0$ for $0<v<v_m$ and
$\partial L_m/\partial v<0$ for $v>v_m$. Thus, claim (a) is proved.

\smallskip
(b) Note that the derivative $v'_m(\theta)$ exists by the inverse
function theorem applied to equation~\eqref{eq:R'v}.
Differentiating~\eqref{eq:chi} and using~\eqref{eq:R'v*}, we get
\begin{align*}
\frac{\dif L_m^*(\theta)}{\dif\theta}&=\frac{\partial
L_m(v;\theta)}{\partial v}\biggr|_{v=v_m(\theta)}\!\times \frac{\dif
v_m(\theta)}{\dif\theta}+\frac{\partial L_m(v;\theta)}{\partial
\theta}\biggr|_{v=v_m(\theta)}\\
&=p_k(v_m(\theta))>0.
\end{align*}
Thus, $L_m^*(\theta)$ is continuously differentiable and (strictly)
increasing.\footnote{The monotonicity of $L_m^*(\theta)$ is obvious
without proof, because the function $\theta\mapsto L_m(v;\theta)$ is
monotone increasing for each $v>0$, since $\partial
L_m/\partial\theta=p_k(v)>0$.} Observe from \eqref{eq:chi} that
\begin{align}
\notag
L_m^*(\theta)
&\ge L_m(v;\theta)|_{v=1}\\
\notag
&=(\theta-1)\mypp p_{k}(1)-m-m'\\
&=(\theta-1)(k-1)-(q-1)\to+\infty,\qquad \theta\to\infty.
\label{eq:L*+}
\end{align}
On the  other hand, from \eqref{eq:R'v} we see that if
$1<\theta<\frac{k}{k-1}$ then, for all $m\ge1$,
\begin{align*}
\frac{\partial L_m(v;\theta)}{\partial
v}\biggr|_{v=1}\!&=(\theta-1)\sum_{i=1}^{k-1}(k-i)-km\\
&=(\theta-1)\,\frac{(k-1)\myp k}{2}-km\\
&<k\left(\frac12-m\right)<0.
\end{align*}
Therefore, by part~(a), for such $\theta$ we have $0<v_m(\theta)<1$,
hence, for all $m\le q-2$,
\begin{align}
\notag
L_m^*(\theta)&=(\theta-1)\mypp p_k(v_m)-m\myp v_m^k-m'\\
\notag
&<\left(\frac{k}{k-1}-1\right) p_k(1)-m'\\
&=1-m'\le 0. \label{eq:L*-}
\end{align}
Thus, combining \eqref{eq:L*+} and \eqref{eq:L*-}, it follows that
there is a unique root $\theta=\theta_m$ of the equation
$L_m^*(\theta)=0$, which proves claim~(b).

\smallskip
(c) Elimination of $\theta=\theta_m$ from the system of equations
\eqref{eq:theta*} and \eqref{eq:R'v*} gives for
$v_m^*=v_m(\theta_m)$ a closed equation,
\begin{equation}\label{eq:L(v)=0}
mk\myp v^{k-1}p_k(v)-\bigl(m\myp v^{k}+m'\bigr)\mypp p'_k(v)=0,
\end{equation}
which can be rearranged to a more symmetric form~\eqref{vy}. The
uniqueness of the root $v^*_m$ is obvious, because the left-hand
side of~\eqref{vy} is a continuous, increasing function in $v>0$,
with the range from $-\infty$ to $+\infty$. Finally, observe that
for $v=1$ the left-hand side of~\eqref{vy} is reduced to
$(m-m')\cdot k\myp(k-1)/2$, which vanishes if $m=m'$ and is negative
if $m<m'$, so that, respectively, $v_m^*=1$ or $v_m^*>1$, as
claimed.

\smallskip
Thus, the proof of Lemma \ref{lm:vmtheta} is complete.
\endproof

\begin{remark}\label{rm:m_cont0}
The statements of Lemma \ref{lm:vmtheta} including the identity
\eqref{eq:theta*} are valid with a \emph{continuous} parameter~$m$.
\end{remark}

We can now proceed to the proof of Theorem~\ref{th:3.8}. Assume that
$\alpha\ne0$. The second equation in the system \eqref{uv} is
reduced to
\begin{equation}\label{v2}
\theta+\theta^{\alpha} u^{k}+m v^{k} +m'-1=(\theta-1)\mypp
\bigl(p_{k}(v)+1\bigr),
\end{equation}
which can be rewritten, using  the notation \eqref{eq:L}, in the
form
\begin{equation}\label{v3}
\theta^{\alpha} u^{k}=L_m(v;\theta).
\end{equation}
Furthermore, substituting \eqref{v2} and \eqref{v3} into the
denominator and numerator, respectively, of the ratio in the first
equation of \eqref{uv}, we get
\begin{equation}\label{eq:u=v^k}
u=\frac{p_k(v)+L_m(v;\theta)}{p_k(v)+1}.
\end{equation}
Finally, substituting \eqref{eq:u=v^k} back into \eqref{v3}, we
obtain the equation
\begin{equation}\label{eta1}
\theta^\alpha=K_m(v;\theta),
\end{equation}
where (cf.\ \eqref{eq:K-intr})
\begin{equation}\label{eq:K}
K_m(v;\theta):=\frac{L_m(v;\theta)(p_{k}(v)+1)^k}{(p_k(v)+L_m(v;\theta))^k}.
\end{equation}
Conversely, all steps above are reversible, so equations
\eqref{eq:u=v^k} and \eqref{eta1} imply the system~\eqref{uv}.

Note from \eqref{eta1} that $v>0$ must satisfy the condition
$K_m(v;\theta)>0$, that is, $v\in\mathscr{V}_m^+(\theta)$
(see~\eqref{eq:V+}); by Lemma \ref{lm:vmtheta}(b), this is possible
if and only if $\theta>\theta_m$. Moreover, the
equation~\eqref{eta1} has a solution $v>0$ if and only if $\alpha\le
\alpha_m(\theta)$, with the critical threshold $\alpha_m(\theta)$
defined in~\eqref{eq:alpham}. This completes the proof of
Theorem~\ref{th:3.8}.

\subsection{Proof of Theorem~\ref{th:non-U}}\label{sec:5.4}
Recall that the critical point $\tilde{\theta}_1$ was defined
in~\eqref{eq:theta-cr-q=3}. If $q=2$ then the only solutions of the
compatibility system \eqref{pt1} are provided by equation
\eqref{rm=1-int}; therefore, Theorem~\ref{th:non-U}(a) readily
follows from Theorem~\ref{th:3.7}.

More generally (i.e., for $q\ge 3$), in order that
$\nu(\theta,\alpha)\ge2$, either there must be at least two
solutions of equation~\eqref{rm=1-int}, that is, $(\theta,\alpha)\in
A_q$ (see Theorem~\ref{th:3.7}), or, since we always have
$\nu_0(\theta,\alpha)\ge1$, there should exist at least one solution
$(u,v)$ of the system~\eqref{uv-int}. By Theorem~\ref{th:3.8}, such
solutions exist if $\alpha\le \alpha_m(\theta)$ for some $m$; since
$\alpha_1(\theta)$ is a majorant of the family
$\{\alpha_m(\theta)\}$ (see Proposition~\ref{pr:theta-monotone}),
the latter condition is reduced to $\alpha\le \alpha_1(\theta)$,
which leads to the inclusion $(\theta,\alpha)\in B_q$. However, we
must ensure that this solution also satisfies the constraint $v\ne1$
(see~\eqref{eq:vnot=1}). By Lemma~\ref{lm:w>1}, this is certainly
true if $m=1<\frac12(q-1)$, that is, $q>3$, which proves
Theorem~\ref{th:non-U}(c).

Finally, Theorem~\ref{th:non-U}(b) (for $q=3$) readily follows by
the next lemma about the maximum of the function $v\mapsto
K_1(v;\theta)$ over the domain $v\in\mathscr{V}_1^+(\theta)$
(see~\eqref{eq:V+}).

\begin{lemma}\label{lm:q=3}
Let $q=3$ and $k\ge2$. \begin{itemize} \item[(a)] For all\/
$\theta>\tilde{\theta}_1$, we have $
K_1(v;\theta)|_{v=1}<\max_{v\in\mathscr{V}_1^+(\theta)}
K_1(v;\theta)$.

\item[(b)]
Let $k\in\{2,3,4\}$. If $1<\theta\le\tilde{\theta}_1$ then the
function $v\mapsto K_1(v;\theta)$ has the unique maximum at $v=1$,
that is, $K_1(v;\theta)<K_1(1;\theta)$ for any $v\ne1$.
\end{itemize}
\end{lemma}
The proof of the lemma is elementary but tedious, so it is deferred
to Appendix~\ref{sec:B}.

\begin{figure}
\centering
\subfigure[$\theta=5.4$]{\raisebox{-.35pc}{\includegraphics[width=7.2cm]{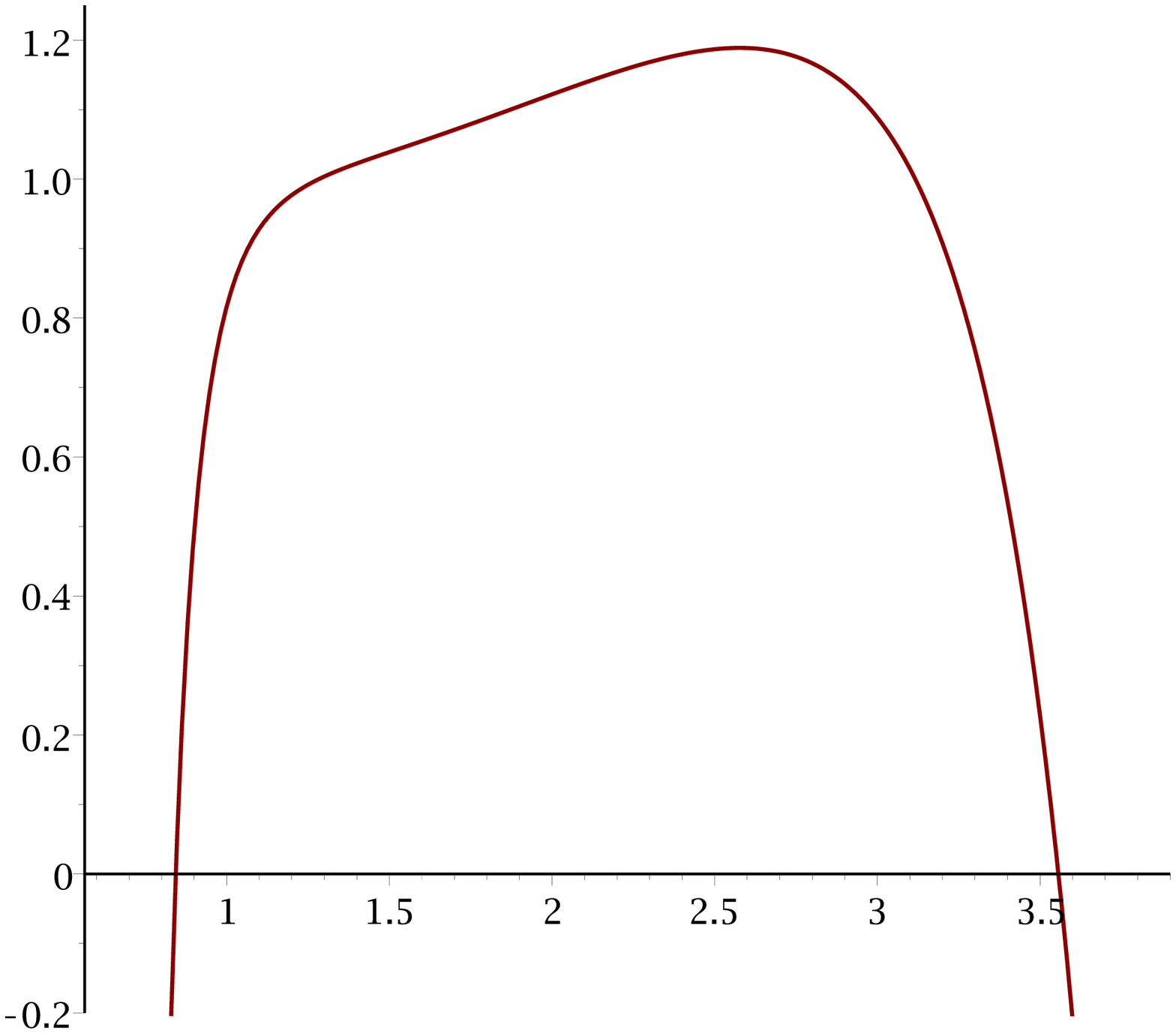}}\label{Fig4a}}
\ \ \
\subfigure[$\theta=6.9$]{\includegraphics[width=7.0cm]{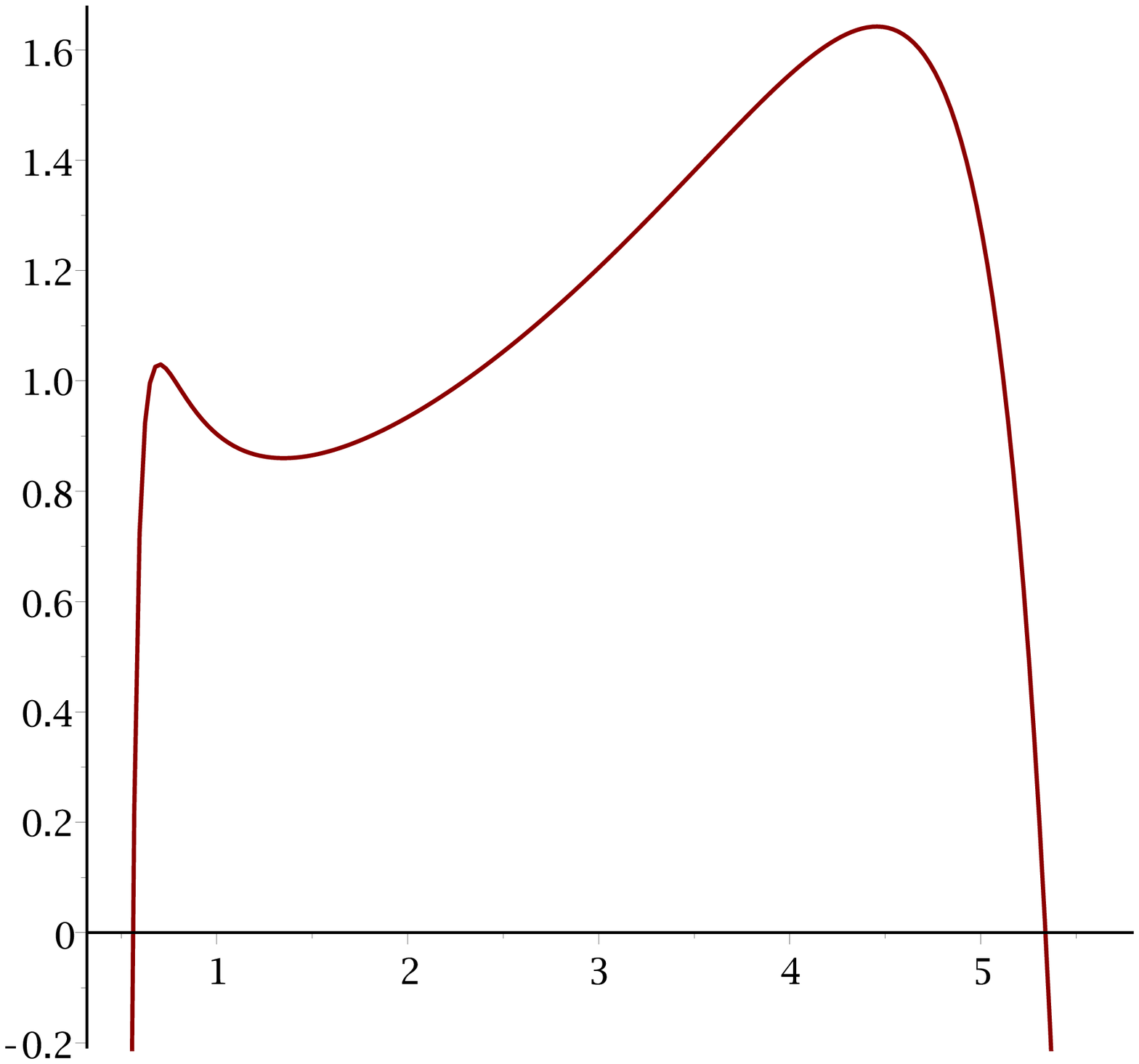}
\label{Fig4b} \put(-219,16){\mbox{\scriptsize$v$}}
\put(-4,16){\mbox{\scriptsize$v$}}}
\put(-333,130){\mbox{\scriptsize$K_1(v;\theta)$}}
\put(-132,130){\mbox{\scriptsize$K_1(v;\theta)$}} \caption{The graph
of the function $v\mapsto K_1(v;\theta)$ (i.e., with $m=1$) for
$k=2$, $q=5$ and various values of
$\theta>\theta_1=1+2\mypp\sqrt{3}\doteq 4.4641$,
illustrating different possible numbers of solutions
$\nu_1=\nu_1(\theta,\alpha)$ of the equation~\eqref{eta1}: (a)
\mypp$\theta=5.4$, \mypp$0\le\nu_1\le 2$; (b) \mypp$\theta=6.9$,
\mypp$0\le\nu_1\le4$.\label{Fig4}}
\end{figure}

\begin{figure}
\centering
\subfigure[$\theta=3.4$]{\hspace{-.1pc}\raisebox{.06pc}{\includegraphics[width=6.5cm]{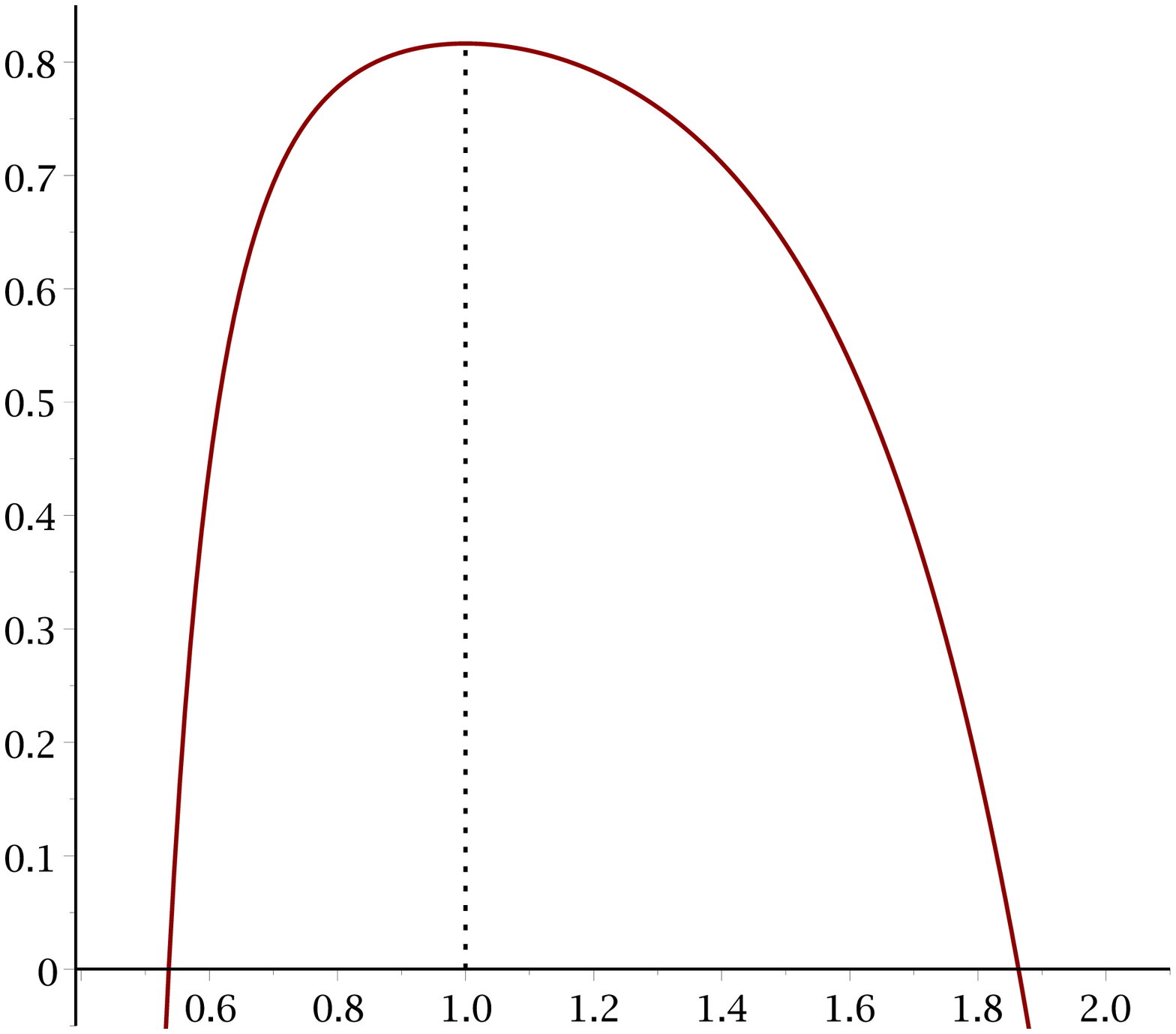}}\label{Fig5a}}
\ \subfigure[$\theta=4.1$]{\includegraphics[width=6.1cm]{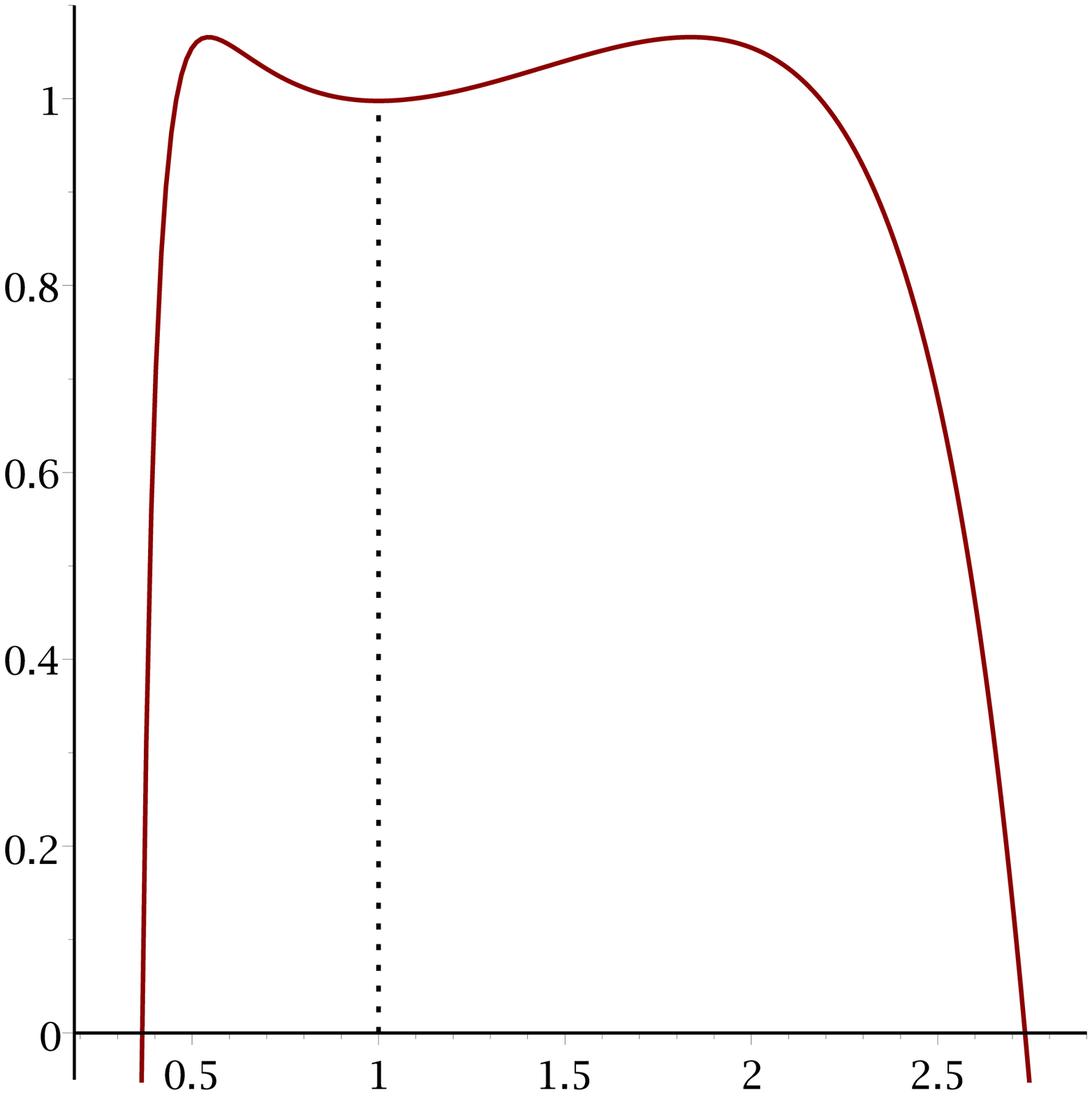}
\label{Fig5b} \put(-187,11.5){\mbox{\scriptsize$v$}}
\put(-4,11.5){\mbox{\scriptsize$v$}}}
\put(-243,135){\mbox{\scriptsize$K_1(v;\theta)$}}
\put(-92,135){\mbox{\scriptsize$K_1(v;\theta)$}} \caption{The graph
of the function $v\mapsto K_1(v;\theta)$ for $q=3$ and $k=2$,
illustrating the \strut{}location of its maximum point depending on
whether $\theta_{\rm c}<\theta<\tilde{\theta}_1$ or
$\theta>\tilde{\theta}_1$; here, $\theta_{\rm c}=3$ and
$\tilde{\theta}_1=\frac12\bigl(1+\sqrt{41}\myp\bigr)\doteq 3.7016$.
\label{Fig5}}
\end{figure}

\begin{remark}
The maximum of the function $v\mapsto K_m(v;\theta)$, as well as the
number of solutions $v>0$ of the equation \eqref{eta1} for various
values of parameters are illustrated in Figure~\ref{Fig4} (for the
regular case $q\ge4$) and in Figure~\ref{Fig5} (for the special case
$q=3$).
\end{remark}

As mentioned in Remark~\ref{rm:q=3-critical}, the case $q=3$ is
truly critical with regard to the uniqueness. Recall that
$\nu(\theta,\alpha)$ denotes the number of positive solutions of the
system \eqref{pt1}; the function $\alpha_1(\theta)$ is defined
in~\eqref{eq:alpham} and $\theta_1^0$ is its zero. The proof of the
next proposition relies on some lemmas that will be proved later, in
Sections \ref{sec:7.2} and~\ref{sec:7.3}.

\begin{proposition}\label{pr:theta+-}Let $q=3$ and $k\ge2$. There exists $\varepsilon>0$ small
enough such that
\begin{itemize}
\item[\rm (a)] $\nu(\theta,\alpha_1(\theta))=1$ if\/ $1\le \theta<\theta_1+\varepsilon$,

\smallskip
\item[\rm (b)] $\nu(\theta,\alpha_1(\theta))\ge 2$ for
all\/ $\theta>\theta^0_1-\varepsilon$.
\end{itemize}
\end{proposition}
\proof (a) As shown in the proof of Lemma~\ref{lm:q=3} (see
Appendix~\ref{sec:B}), $\partial^2 K_1/\partial v^2|_{v=1}\myn<0$
for all $\theta\in[\myp\theta_1,\tilde{\theta}_1)$; in particular,
$\partial^2 K_1(\theta_1,v)/\partial v^2|_{v=1}\myn<0$. By
Lemma~\ref{lm:vmtheta}(c), the set $\mathscr{V}_1^+(\theta)$
(see~\eqref{eq:V+}) is reduced for $\theta=\theta_1$ to the single
point $v=1$, and
$K_1^*(\theta_1)=\max_{v\in\mathscr{V}^+_1(\theta_1)}K_1(v;\theta_1)=K_1(1;\theta_1)=0$.
By continuity, it follows that  $\partial^2 K_1(\theta;v)/\partial
v^2<0$ for each $\theta\in[\myp\theta_1,\theta_1+\varepsilon)$ and
all $v\in \mathscr{V}^+_1(\theta)$; that is, the function $v\mapsto
K_1(v;\theta)$ is concave on $\mathscr{V}^+_1(\theta)$ and therefore
has a unique maximum located at $v=1$ (remembering that $\partial
K_1/\partial v|_{v=1}\myn=0$). But the solution $(u,v)$ with $v=1$
is not admissible (see~\eqref{eq:vnot=1}), hence
$\nu(\theta,\alpha_1(\theta))=1$, with the only solution of
\eqref{pt1} coming from equation~\eqref{rm=1-int}.

\smallskip
(b) We have $\alpha_1(\theta_1^0)=0$, hence $K_1^*(\theta_1^0)=1$
(see~\eqref{eq:alpham}) and, by Lemma~\ref{lm:K=L=1}(b), also
$L_1^*(\theta^0_1)=1$. If $v_1=v_1(\theta_1^0)$ is the point where
the latter maximum is attained, that is,
$L_1(v_1;\theta_1^0)=L_1^*(\theta_1^0)=1$, then it holds that
$v_1>1$ (see Lemma~\ref{lm:v>1} below). Furthermore, according to
Lemma~\ref{lm:vmtheta}(a), $v_1$ is the unique maximum of the
function $v\mapsto L_1(v;\theta_1^0)$, and in particular
$L_1(1;\theta_1^0)<L_1^*(\theta_1^0)=1$.

From the definition \eqref{eq:K}, it follows that also
$K_1(v_1;\theta_1^0)=1=K_1^*(\theta_1^0)$,\footnote{By the scaling
property \eqref{eq:K-scaled}, we also have
$K_1(v_1^{-1};\theta_1^0)=K_1(v_1;\theta_1^0)=1$.} and since
$L_1(1;\theta_1^0)<1$, Lemma~\ref{lm:K=L=1}(a) implies that
$K_1(1;\theta_1^0)<1=K_1^*(\theta_1^0)$. Thus, the corresponding
solution of the system~\eqref{uv-int} satisfies the
condition~\eqref{eq:vnot=1}, and therefore
$\nu(\theta_1^0,\alpha_1(\theta_1^0))\ge 2$.

By continuity of $K_1(1;\theta)$ and $K_1^*(\theta)$, for all
$\theta\in(\theta_1^0-\varepsilon,\theta_1^0\myp]$ (with
$\varepsilon>0$ small enough) we still have that
$K_1(1;\theta)<K_1^*(\theta)$, so the maximum is attained outside
$v=1$. Thus, by the same argument as before, the claim follows.
\endproof

\subsection{Proof of Theorem \ref{th:k=2}}\label{sec:5.5}
(a) First, let $q=2$. According to Lemma~\ref{lm2}, the system
\eqref{pt1} is reduced to the single equation \eqref{rm=1-int}, and
by Theorem~\ref{th:3.7} the number of its solutions is not more than
$3=2^{2}-1$; furthermore, for $\theta>\theta_{\rm
c}=\frac{k+1}{k-1}$ and
$\alpha_{-}(\theta)<\alpha<\alpha_{+}(\theta)$, there are exactly
three solutions, so the upper bound is attained.

\smallskip
(b) Let now $\alpha=0$ (and $q\ge 3$). Due to Lemma~\ref{lm2}(b),
either $z_i\equiv 1$ or the system \eqref{rm=1-int} is reduced to
the equation~\eqref{eq:1=} indexed by $m=1,\dots,q-1$, which can be
rewritten as $L_m(u;\theta)=1$ (see~\eqref{eq:L}). By
Lemma~\ref{lm:vmtheta}(a), the latter equation has no more than two
roots. Hence, considering permutations of the values $u\ne1$ over
the $q-1$ places, it is clear that the total number of solutions to
\eqref{rm=1-int} is bounded by
\begin{equation}\label{eq:2q}
1+2 \sum_{m=1}^{q-1}\binom{q-1}{m}=1+2\mypp(2^{q-1}-1)=2^q-1.
\end{equation}
Moreover, for $\theta>1$ large enough, there will be exactly two
roots of each of the equations $L_m(u;\theta)=1$, because
$L_m^*(\theta)=\max_{u>0}L_m(u;\theta)\to \infty$ as
$\theta\to\infty$ (see Lemma~\ref{lm:vmtheta}(b)). Therefore, the
upper bound \eqref{eq:2q} is attained.

\smallskip
(c) Finally, let $k=2$ and $\alpha\ne0$. First of all, up to three
solutions of the system \eqref{rm=1-int} arising from the
equation~\eqref{rm=1} are ensured by Lemma~\ref{lm2} (see also
Theorem~\ref{th:3.7}). Other solutions are determined by the system
\eqref{uv} indexed by $m=1,\dots,q-2$, which in turn depends on the
solvability of the equation~\eqref{eta1}. In the case $k=2$, the
latter is a polynomial equation of degree~$4$, and therefore has at
most four roots $v>0$, for each~$m$. The value $u>0$ is then
determined uniquely by formula \eqref{eq:u=v^k}, and it occupies the
first place in the vector $\boldsymbol{z}=(z_1,\dots,z_{q-1})$. As
for the root $v>0$, it occupies $m$ out of the $q-2$ remaining
places. Counting the total number of such permutations, we get the
upper bound
$$
3+4 \sum_{m=1}^{q-2}\binom{q-1}{m}=3+4\mypp(2^{q-2}-1)=2^q-1,
$$
as required. This completes the proof of Theorem~\ref{th:k=2}.

\section{Further properties of the critical curves
$\alpha_\pm(\theta)$ and $\alpha_m(\theta)$}\label{sec:6}

\subsection{Properties of $\alpha_\pm(\theta)$}

\begin{lemma}\label{lm:a+a-}
The quantities $a_\pm$, defined in \eqref{eq:a_pm} for
$b\ge\left(\frac{k+1}{k-1}\right)^2$, satisfy the identity
\begin{equation}\label{eq:a+xa-}
a_+ \times a_{-}=b^{-k-1}.
\end{equation}
\end{lemma}
\proof Using \eqref{eq:a_pm} and noting that $x_{-}=b/x_{+}$ (see
equation~\eqref{eq:quadratic-intro}), we find
\begin{align*}
a_{-}&=\frac{1}{x_{-}}\left(\frac{1+x_{-}}{b+x_{-}}\right)^k\\
&=\frac{x_{+}}{b}\left(\frac{1+b/x_{+}}{b+b/x_{+}}\right)^k\\
&=\frac{x_{+}}{b^{k+1}}\left(\frac{b+x_{+}}{1+x_{+}}\right)^k\\
&=\frac{b^{-k-1}}{a_{+}},
\end{align*}
and formula \eqref{eq:a+xa-} follows.
\endproof

\begin{lemma}\label{lm:a+<1/b}
Suppose that $b\ge\left(\frac{k+1}{k-1}\right)^2$. Then the
following inequalities hold,
\begin{equation}\label{eq:a+<}
a_{+}<b^{-1},\qquad a_{-}>b^{-k}.
\end{equation}
\end{lemma}
\proof From \eqref{eq:f'b}, for all $x\ge0$ we get the upper bound
\begin{align*}
f'(x)
&=\left(1+\frac{b-1}{1+x}\right)^{-k}\frac{k\myp(b-1)}{(b+x)(1+x)}<
\frac{1}{b+x}\le\frac{1}{b},
\end{align*}
noting that, by Bernoulli's inequality,
$$
\left(1+\frac{b-1}{1+x}\right)^{k}
>\frac{k\myp(b-1)}{1+x}.
$$
Hence, $a_{+}=f'(x_{+})<b^{-1}$, and the first inequality in
\eqref{eq:a+<} is proved. The second inequality then readily follows
from the identity \eqref{eq:a+xa-}.
\endproof

\begin{lemma}\label{lm:alpha+alpha}
The functions $\alpha_\pm(\theta)$ satisfy the following identity,
\begin{equation}\label{eq:alpha+alpha}
\alpha_{-}(\theta)+\alpha_{+}(\theta)=\frac{2\myp\ln\myp\mynn(q-1)+(k+1)\bigl(\ln
b(\theta)-2\bigr)}{\ln\theta},\qquad \theta\ge \theta_{\rm c},
\end{equation}
where $b(\theta)$ is defined in~\eqref{eq:b}. In particular, if
$q=2$ then $\alpha_{-}(\theta)+\alpha_{+}(\theta)\equiv 0$ for all
$\theta\ge\theta_{\rm c}=\frac{k+1}{k-1}$.
\end{lemma}
\proof Using \eqref{eq:alpha=a+/-}, we obtain
$$
\theta^{\myp2\myp(k+1)+\alpha_{-}(\theta)+\alpha_{+}(\theta)}=\frac{(q-1)^2}{a_{-}(\theta)\mypp
a_{+}(\theta)},\qquad \theta\ge \theta_{\rm c},
$$
and the identity \eqref{eq:alpha+alpha} follows upon substituting
formula~\eqref{eq:a+xa-}.
\endproof

\begin{proposition}\label{pr:alpha-pm-limits} The
functions $\alpha_\pm(\theta)\colon [\theta_{\rm
c},\infty)\to\mathbb{R}$ defined in \eqref{eq:alpha_pm} have the
following ``boundary''\ values,
\begin{align}\label{eq:alphapm-c}
\alpha_\pm(\theta_{\rm c})&=-(k+1)+\frac{1}{\ln\theta_{\rm c}}\left(\ln\myp\mynn(q-1)+(k+1)\ln\frac{k+1}{k-1}\right),\\
\label{ali} \alpha_\pm(\theta)&\to\pm (k-1)\qquad (\theta\to\infty).
\end{align}
In particular, $\alpha_\pm(\theta_{\rm c})=0$ if\/ $q=2$ and
$\alpha_\pm(\theta_{\rm c})>0$ if\/ $q>2$.

\end{proposition} \proof Recall from the proof of
Theorem~\ref{th:3.7} (see Section~\ref{sec:5.2}) that the critical
value $\theta=\theta_{\rm c}$ corresponds to $b(\theta_{\rm
c})=\bigl(\frac{k+1}{k-1}\bigr)^2$, whereby the quadratic equation
\eqref{eq:quadratic-intro} has the double root
$$
x_\pm(\theta_{\rm c})=\sqrt{b(\theta_{\rm c})}=\frac{k+1}{k-1}.
$$
Hence, using \eqref{eq:a_pm}, we find
$$
a_\pm(\theta_{\rm c})=\left(\frac{k-1}{k+1}\right)^{k+1},
$$
which, together with \eqref{eq:alpha=a+/-}, yields
formula~\eqref{eq:alphapm-c}.

If $q=2$ then formula \eqref{eq:theta'c} gives $\theta_{\rm
c}=\frac{k+1}{k-1}$, and it readily follows from
\eqref{eq:alphapm-c} that $\alpha(\theta_{\rm c})=0$. For $q>2$,
using the relation \eqref{eq:alpha=a+/-} observe that the required
inequality $\alpha_\pm(\theta_{\rm c})>0$ is reduced to
\begin{equation*}
\theta_{\rm c}^{k+1}<\frac{q-1}{a_\pm(\theta_{\rm
c})}=\left(\frac{k+1}{k-1}\right)^{k+1}(q-1),
\end{equation*}
that is,
\begin{equation}\label{eq:alpha>0a}
\theta_{\rm c}<\left(\frac{k+1}{k-1}\right)(q-1)^{1/(k+1)}.
\end{equation}
Denote
$$
\rho_k:=\frac{k+1}{k-1}>1,\qquad s:=(q-1)^{1/(k+1)}>1,
$$
then \eqref{eq:alpha>0a} takes the form
\begin{equation}\label{eq:alpha>0b}
\theta_{\rm c}<\rho_k\myp s.
\end{equation}
Furthermore, recalling that $b(\theta_{\rm
c})=\left(\frac{k+1}{k-1}\right)^2=\rho_k^2$ and $b(\theta)$ is
monotone increasing for $\theta>1$ (see \eqref{eq:theta'c}
and~\eqref{eq:b}), the inequality \eqref{eq:alpha>0b} is equivalent
to
$$
\rho_k^2=b(\theta_{\rm c})<b(\rho_k\myp s)=\frac{\rho_k\myp
s\bigl(\rho_k\myp s+s^{k+1}-1\bigr)}{s^{k+1}},
$$
that is,
$$
s^{k+1}-1>\rho_k\myp (s^{k}-s),
$$
which is reduced, upon dividing by $s-1>0$ and substituting
$\rho_k-1=\frac{2}{k-1}$, to
\begin{equation}\label{eq:alpha>0c}
\phi_k(s):=s^k+1-\frac{2\mypp p_k(s)}{k-1}>0.
\end{equation}
In fact, it is easy to show that $\phi_k(s)>0$ for any $s>1$.
Indeed, since $p_k(1)=k-1$, we have $\phi_k(1)=0$, while
\begin{align*}
\phi_k'(s)&=k\myp s^{k-1}-\frac{2}{k-1}\sum_{i=1}^{k-1}i\myp
s^{i-1}\\
&=s^{k-1}\left(k-\frac{2}{k-1}\sum_{i=1}^{k-1}i\myp s^{-(k-i)}\right)\\
&>s^{k-1}\left(k-\frac{2}{k-1}\sum_{i=1}^{k-1}i\right)\\
&=s^{k-1}\left(k-\frac{2}{k-1}\cdot \frac{k\mypp(k-1}{2}\right)=0.
\end{align*}
Thus, inequality \eqref{eq:alpha>0c} is verified, which implies that
$\alpha(\theta_{\rm c})>0$, as argued above.

Let us now prove~\eqref{ali}. Using the definition of $b=b(\theta)$
and $D=D(\theta)$ (see \eqref{eq:b} and \eqref{eq:D}, respectively),
we obtain the following asymptotics as $\theta\to\infty$,
$$
b=\frac{\theta^2}{q-1}+O(\theta), \qquad
\sqrt{D}=\frac{\theta^2(k-1)}{q-1}+O(\theta),
$$
and
\begin{align*}
x_+=\frac{\theta^2(k-1)}{q-1}+O(\theta),\qquad x_-=\frac{b}{x_+}
=\frac{1}{k-1} +O(\theta^{-1}).
\end{align*}
Hence,
$$
\ln x_\pm=(1\pm1)\ln \theta +O(1),\qquad
\ln\frac{b+x_\pm}{1+x_\pm}=(1\mp1)\ln \theta+O(1).
$$
Using formula \eqref{eq:a_pm}, this yields
\begin{align*}
-\ln a_\pm(\theta)&=\ln
x_\pm+k\ln\frac{b+x_{\pm}}{1+x_{\pm}}\\
&=\bigl((k+1)\mp (k-1)\bigr)\ln\theta+O(1).
\end{align*}
Therefore, from \eqref{eq:alpha=a+/-} we get
$$
k+1+\alpha_\pm=-\frac{\ln a_\mp(\theta)}{\ln\theta}+o(1)=(k+1)\pm
(k-1)+o(1),
$$
and the limit \eqref{ali} follows.
\endproof

\begin{proposition}\label{pr:alpha-pm><}
The functions $\alpha_{\pm}(\theta)$ satisfy the following bounds,
\begin{alignat}{2}\label{eq:alpha->}
\alpha_{-}(\theta)&>-(k-1),\qquad&\theta&\ge
\theta_{\rm c},\\
\label{eq:alpha+<} \alpha_{+}(\theta)&<k-1,\qquad
&\theta&\ge\max\{\theta_{\rm c}, \bar{\theta}\myp\},
\end{alignat}
where
\begin{equation}\label{eq:bar-theta}
\bar{\theta}=\bar{\theta}(k,q):=\begin{cases} \displaystyle
\frac{q-2}{(q-1)^{(k-1)/k}-1},\quad&q>2,\\[.6pc]
\ 1,&q=2.
\end{cases}
\end{equation}
\end{proposition}
\proof Using the relation~\eqref{eq:alpha=a+/-}, the first
inequality in~\eqref{eq:a+<} and the definition of $b$
in~\eqref{eq:ab}, we get
\begin{align*}
\theta^{\myp k+1+\alpha_{-}}=
\frac{q-1}{a_{+}}&>(q-1)\mypp b\\
&=\theta\myp(\theta+q-2)\ge \theta^{2},\qquad q\ge2,
\end{align*}
which proves the bound~\eqref{eq:alpha->}.

Similarly, using the second inequality in \eqref{eq:a+<} we have
\begin{align}
\notag
\theta^{\myp k+1+\alpha_{+}}=\frac{q-1}{a_{-}}&<(q-1)\mypp
b^k\\[-.4pc]
&=(q-1)^{1-k}\mypp\theta^{2k}\!\left(1+\frac{q-2}{\theta}\right)^k.
\label{eq:alpha+<-proof}
\end{align}
Noting from \eqref{eq:bar-theta} that, for $\theta\ge\bar{\theta}$,
$$
1+\frac{q-2}{\theta}\le
1+\frac{q-2}{\bar{\theta}}=(q-1)^{(k-1)/k},\qquad q\ge 2,
$$
it is easy to see that the right-hand side of
\eqref{eq:alpha+<-proof} is bounded above by $\theta^{2k}$; hence,
the inequality \eqref{eq:alpha+<} follows.
\endproof

\begin{remark}
Note that $\bar{\theta}>1$ for $q\ge 3$. Of course, if $\theta_{\rm
c}\ge \bar{\theta}$ then the upper bound \eqref{eq:alpha+<} holds
for all $\theta\ge\theta_{\rm c}$; however, the ordering between
$\theta_{\rm c}$ and $\bar{\theta}$ depends on $k$ and~$q$. For
example, for $k=2$ and $q=5$
$$
\bar{\theta}=3<\theta_{\rm c}\doteq 4.6847,
$$
whereas for $k=2$ and $q=50$
$$
\bar{\theta}=8>\theta_{\rm c}\doteq 7.8904.
$$
In fact, for large $q$ the upper bound \eqref{eq:alpha+<} fails near
$\theta_{\rm c}$; indeed, using \eqref{eq:theta'c} we get
$$
\theta_{\rm
c}=\frac{q-2}{2}\left(\sqrt{1+\frac{4\myp(q-1)}{(q-2)^2}\left(\frac{k+1}{k-1}\right)^2}-1\right)\sim
\left(\frac{k+1}{k-1}\right)^2,\qquad q\to\infty,
$$
and, according to \eqref{eq:alphapm-c},
$$
\alpha_\pm(\theta_{\rm c})\sim\frac{\ln q}{\ln\theta_{\rm
c}}\to\infty,\qquad q\to\infty.
$$
\end{remark}

\begin{figure}
\centering \subfigure[$q=2$
($k=3$)]{\hspace{-.1pc}\raisebox{.06pc}{\includegraphics[width=6.1cm]{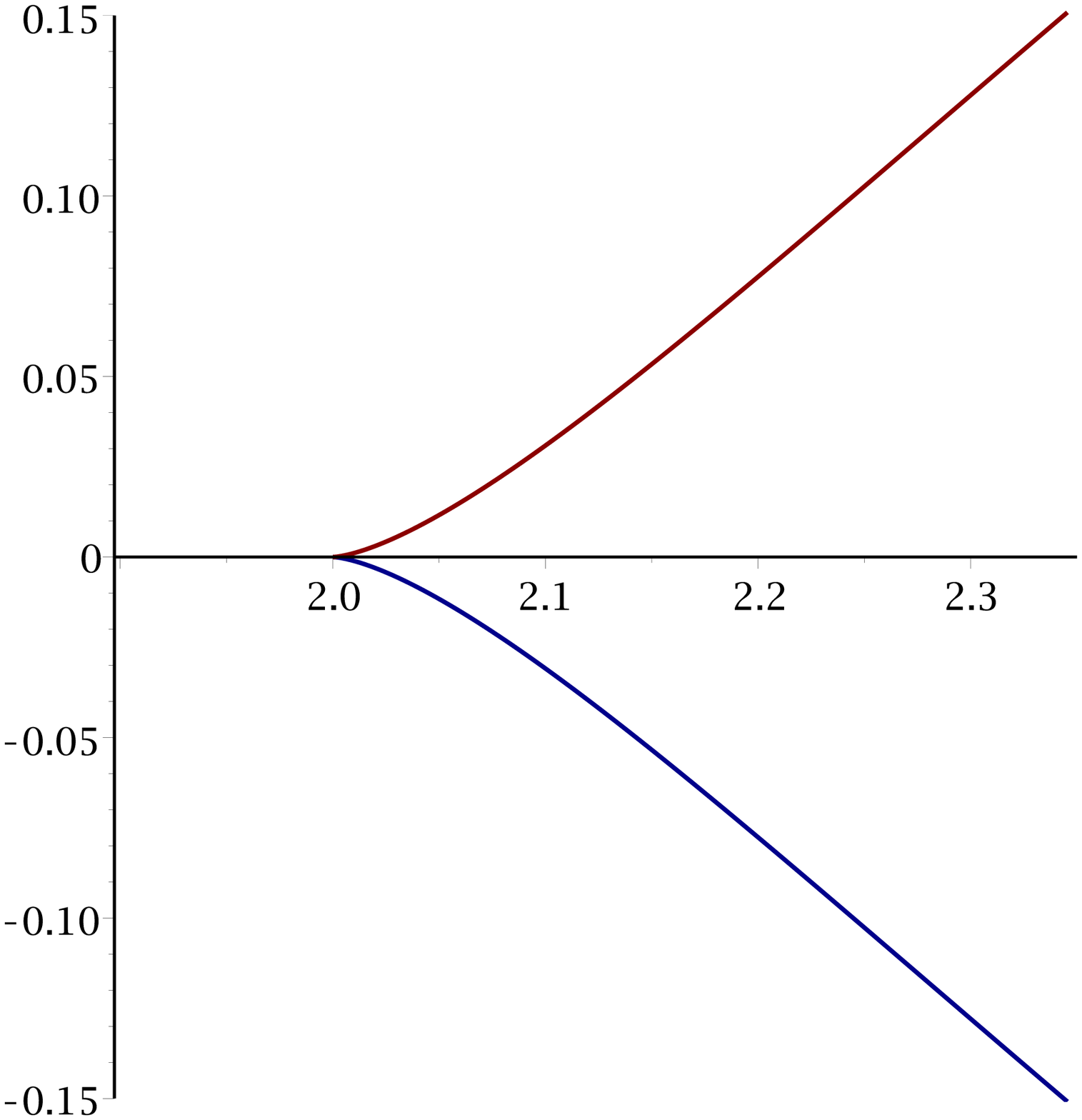}}\label{Fig6a}}
\ \ \ \ \subfigure[$q=5$
($k=2$)]{\includegraphics[width=6.1cm]{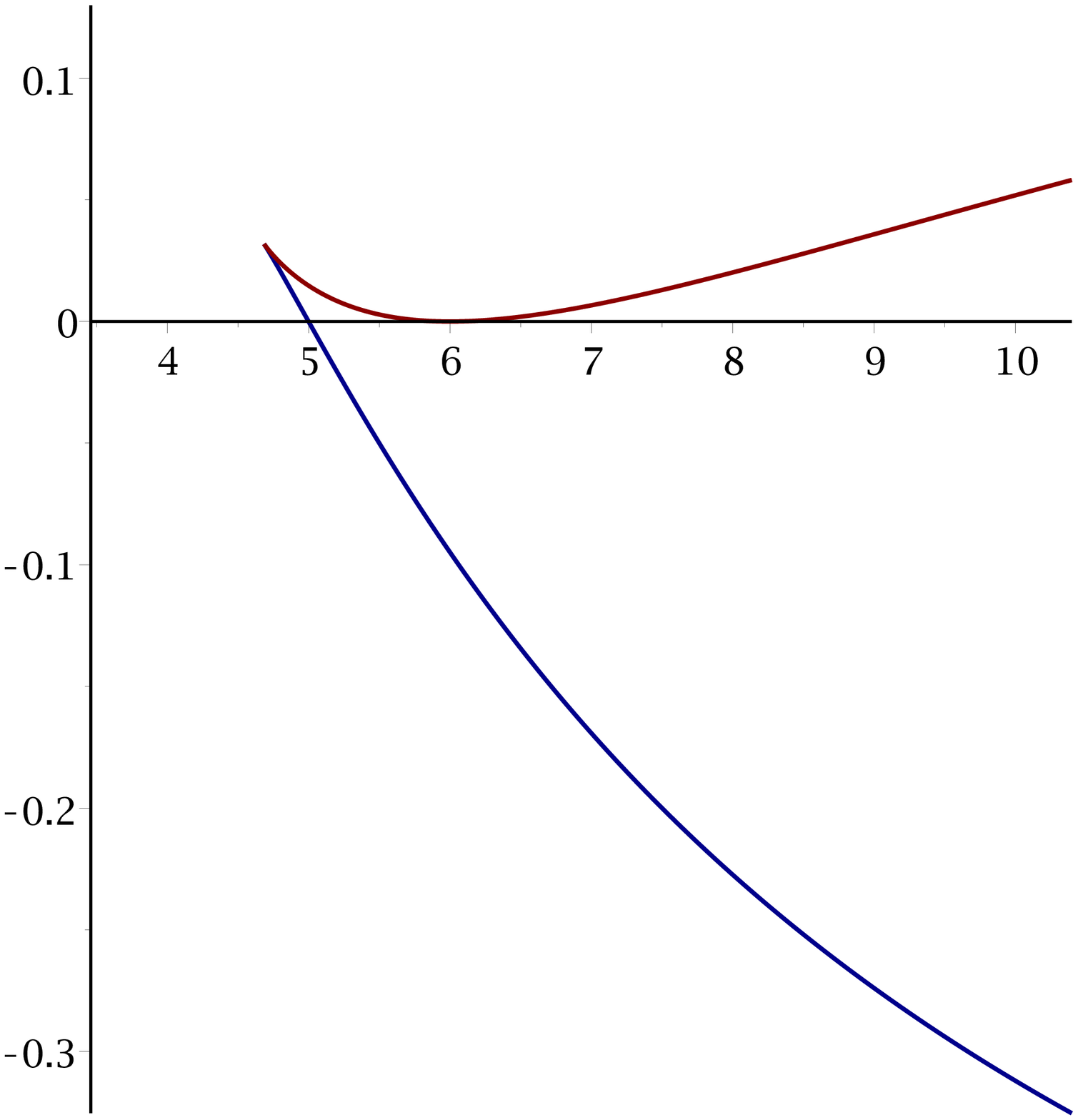}} \label{Fig6b}
\put(-268,136){\mbox{\scriptsize$\alpha_+(\theta)$}}
\put(-268,34){\mbox{\scriptsize$\alpha_-(\theta)$}}
\put(-199,79.5){\mbox{\scriptsize$\theta$}}
\put(-353,177){\mbox{\scriptsize$\alpha$}}
\put(-314,92){\mbox{\scriptsize$C$}}
\put(-70,136){\mbox{\scriptsize$\alpha_+(\theta)$}}
\put(-70,56){\mbox{\scriptsize$\alpha_-(\theta)$}}
\put(-5,114.5){\mbox{\scriptsize$\theta$}}
\put(-163,174){\mbox{\scriptsize$\alpha$}}
\put(-133,137){\mbox{\scriptsize$C$}}

\caption{The graphs of the functions $\alpha_+(\theta)$ and
$\alpha_{-}(\theta)$ (see~\eqref{eq:alpha_pm}): (a) $q=2$, $k=3$;
(b) $q=5$, $k=2$. The coordinates of the cusp point $C$ (see
\eqref{eq:theta'c} \strut{}and~\eqref{eq:alphapm-c}) are given
numerically by: (a) $\theta_{\rm c}=2$, $\alpha_\pm(\theta_{\rm
c})=0$;
  (b) $\theta_{\rm c}\doteq 4.6847$,
$\alpha_\pm(\theta_{\rm c})\doteq0.0319$.
The axis $\alpha=0$ is a tangent line for $\alpha_+(\theta)$ at
\strut{}$\theta^+_0\!=1+q/(k-1)$ (see
Proposition~\ref{pr:alpha-zeros}\mbox{(b-i)}): (a)
$\theta^+_0\!=2\;({=}\:\theta_{\rm c})$; (b) $\theta^+_0\!=6$. In
panel (b), the function $\alpha_{-}(\theta)$ has zero at
$\theta^-_0\! =5$ (see Proposition~\ref{pr:alpha-zeros}\mbox{(b-ii)}
and Example~\ref{ex:k=2_m0}), whereas in panel (a) we have
$\theta_0^-\!=\theta_0^+\!=2$ (see
Proposition~\ref{pr:alpha-pm-limits}). Note that the graphs display
the monotone behaviour as predicted by Conjecture~\ref{rm:conj1}.}
\label{Fig6}
\end{figure}

\begin{conjecture}\label{rm:conj1}
The function $\alpha_{-}(\theta)$ is monotone decreasing for all
$\theta\ge \theta_{\rm c}$, whereas $\alpha_{+}(\theta)$ is
decreasing for $\theta\le\theta^+_0$ and increasing for
$\theta\ge\theta^+_0$, with the unique minimum
$\alpha_{+}(\theta^+_0)=0$ at the critical point
\begin{equation}\label{eq:tilde(theta)_cr}
\theta^+_0=\theta^+_0(k,q):=1+\frac{q}{k-1}.
\end{equation}
In the case $q=2$, we have $\theta_{\rm
c}=\theta_0^+=\frac{k+1}{k-1}$ and, by Lemma~\ref{lm:alpha+alpha},
$\alpha_{+}(\theta)\equiv - \alpha_{-}(\theta)$; hence, the function
$\alpha_{+}(\theta)$ should be monotone increasing for all
$\theta\ge \theta_{\rm c}$.
\end{conjecture}
This conjecture is supported by computer plots (see
Figure~\ref{Fig6}). Towards a proof, we have been able to
characterize the unique zero $\theta_0^{-}$ of $\alpha_{-}(\theta)$
and to show rigorously that $\alpha_{+}(\theta^+_0)=0$ and
$\alpha'_{+}(\theta^+_0)=0$ (see
Proposition~\ref{pr:alpha-zeros}(b)), but the monotonicity
properties are more cumbersome to verify.

\begin{remark}
Note that the value \eqref{eq:tilde(theta)_cr} coincides with a
known critical point in the case $\alpha=0$, above which the
solution $\boldsymbol{z}=\boldsymbol{1}$ is unstable (see
\cite[Section~5.2.2.2, Proposition 5.4]{Ro}). Our
Proposition~\ref{pr:alpha-zeros}(b-i) explains the emergence of this
critical point and its explicit value~\eqref{eq:tilde(theta)_cr}.
\end{remark}

\subsection{Properties of $L_m(v;\theta)$ and $K_m(v;\theta)$}\label{sec:7.2}

Here and below, we assume that $q\ge3$. Recall that
$v_m=v_m(\theta)$ is the unique maximum of the function $v\mapsto
L_m(v;\theta)$ (see Lemma~\ref{lm:vmtheta}(a)), and $\theta_m$ is
defined by the relation \eqref{eq:theta*}, where $1\le m\le q-2$.

\begin{remark}\label{rm:m_cont}
All results in this section hold true for a \emph{continuous}
parameter~$m$ (cf.\ Remark~\ref{rm:m_cont0}), which is evident by
inspection of the proofs.
\end{remark}

The next lemma describes the useful scaling properties of the
functions $L_m(v;\theta)$ and $K_m(v;\theta)$ (see \eqref{eq:L}
and~\eqref{eq:K}) under the conjugation $m\mapsto m'$.
\begin{lemma}\label{lm:LK-scaled}
For each $m=1,\dots,q-2$ and $m'=q-1-m$, the following identities
hold for all $v>0$ and $\theta>\theta_m$,
\begin{align}
\label{eq:L-scaled}
L_{m'}(v;\theta)&=v^{k}L_m(v^{-1};\theta),\\
\label{eq:K-scaled} K_{m'}(v;\theta)&=K_m(v^{-1};\theta).
\end{align}
\end{lemma} \proof Recalling the notation \eqref{eq:p(z)}, note
that
\begin{equation}
p_k(v^{-1})=v^{-(k-1)}+\dots+v^{-1}=v^{-k}p_k(v), \label{eq:pk1}
\end{equation}
and similarly
\begin{equation}
p_k(v^{-1})+1=v^{-(k-1)}+\dots+v^{-1}+1 \label{eq:pk2}
=v^{-k+1}\bigl(p_k(v)+1\bigr).
\end{equation}
Using \eqref{eq:L} and \eqref{eq:pk1}, we can write
\begin{align*}
L_m(v^{-1};\theta)&=(\theta-1)\mypp p_{k}(v^{-1})-m v^{-k} -m'\\
&=v^{-k}\left((\theta-1)\mypp p_{k}(v)-m-m' v^k\right)\\
&=v^{-k}L_{m'}(v;\theta),
\end{align*}
and formula \eqref{eq:L-scaled} follows. Furthermore, using
\eqref{eq:L-scaled} and substituting \eqref{eq:pk1} and
\eqref{eq:pk2}, we obtain
\begin{align*}
K_m(v^{-1};\theta)&=\frac{L_m(v^{-1};\theta)\bigl(p_{k}(v^{-1})+1\bigr)^k}
{\bigl(p_k(v^{-1})+L_m(v^{-1};\theta)\bigr)^k}\\
&=\frac{v^{-k}L_{m'}(v;\theta)\cdot v^{-(k-1)\myp
k}\bigl(p_k(v)+1\bigr)^k}
{\bigl(v^{-k}p_k(v)+v^{-k}L_{m'}(v;\theta)\bigr)^k}\\
&=\frac{L_{m'}(v;\theta) \bigl(p_k(v)+1\bigr)^k}
{\bigl(p_k(v)+L_{m'}(v;\theta)\bigr)^k}\\[.2pc]
&=K_{m'}(v;\theta),
\end{align*}
which proves formula \eqref{eq:K-scaled}.
\endproof

For $\theta>\theta_m$, let $v_m=v_m(\theta)$ be the point where the
function $v\mapsto L_m(v;\theta)$ attains its (positive) maximum
value, that is,
$L_m(v_m;\theta)=\max_{v\in\mathscr{V}_m^{+}(\theta)}L_m(v;\theta)$
(see Lemma~\ref{lm:vmtheta}). The next result provides a strict
lower bound for $v_m$ \strut{}(cf.\ Lemma~\ref{lm:vmtheta}(c) for
$\theta=\theta_m$).

\begin{lemma}\label{lm:v>1}
For all $m$ in the range $1\le m\le\frac12\myp(q-1)$, we have
\begin{equation}\label{eq:v>1}
v_m(\theta)>1,\qquad \theta>\theta_m.
\end{equation}
\end{lemma}

\proof To the contrary, suppose first that $v_m<1$ for some
$\theta>\theta_m$. Then, according to Lemma~\ref{lm:LK-scaled}, we
have
\begin{equation}\label{eq:L->L1}
L_m(v_m^{-1};\theta)=v_m^{-k}L_{m'}(v_m;\theta)>L_{m'}(v_m;\theta),
\end{equation}
and furthermore,
\begin{align}
\notag L_{m'}(v_m;\theta)&=(\theta-1)\mypp p_k(v_m)-m'v_m^k - m\\
&= L_{m}(v_m;\theta)+ (m'-m)(1-v_m^k)\ge L_{m}(v_m;\theta),
\label{eq:L->L2}
\end{align}
since $m'=q-1-m\ge m$ by the hypothesis of the lemma. Combining
\eqref{eq:L->L1} and \eqref{eq:L->L2}, we see that
$L_m(v_m^{-1};\theta)>L_m(v_m;\theta)$, which contradicts the
assumption that $L_m(v_m;\theta)$ is the maximum value of the
function $v\mapsto L_m(v;\theta)$.

Assume now that $v_m=1$ for some $\theta>\theta_m$. Then
$$
\left.\frac{\partial L(v;\theta)}{\partial
v}\right|_{v=v_m=1}\!=(\theta-1)\mypp p'_k(1)-k\myp
m=(\theta-1)\mypp\frac{k\myp(k-1)}{2}-k\myp m=0,
$$
whence $(\theta-1)(k-1)=2\myp m$. Hence,
$$
L_m(1;\theta)=(\theta-1)(k-1)-(q-1)=2\myp m-(q-1)\le 0,
$$
which contradicts the assumption $L_m(v_m;\theta)>0$.

Thus, the inequality \eqref{eq:v>1} is proved.
\endproof

For $\theta>\theta_m$, let $w_m=w_m(\theta)$ be the point where the
function $v\mapsto K_m(v;\theta)$ attains its
\textup{(}positive\textup{)} maximum value, that is,
$K_m(w_m;\theta)=\max_{v\in\mathscr{V}_m^{+}(\theta)}K_m(v;\theta)$.
Note that $w_m(\theta_m)=v_m(\theta_m)=v_m^*$ (see
Lemma~\ref{lm:vmtheta}(c)). The importance of the next technical
lemma is pinpointed by involvement of the expression
$p_k(v)-(k-1)\mypp L_m(v;\theta)$ in the partial derivative
$\partial K_m/\partial v$ (see formula~\eqref{eq:K'} below).
\begin{lemma}\label{lm:p-L>0} Let\/ $1\le m\le \frac12\myp(q-1)$ and $\theta>\theta_m$.
\begin{itemize}
\item[\rm (a)]
\strut{}Let $w\in\mathscr{V}_m^{+}(\theta)$ be a critical point of
the function $v\mapsto K_m(v;\theta)$, that is, any solution of the
equation \strut{}$\partial K_m/\partial v=0$. Assume that either
\textup{(i)} $w<v_m$ and $L_m(w;\theta)<1$, or \strut{}\textup{(ii)}
$w\ge v_m$. Then
\begin{equation}\label{eq:p-L>0-mod}
p_k(w)-(k-1)\mypp L_m(w;\theta)>0.
\end{equation}
\item[\rm (b)]
In particular, the inequality~\eqref{eq:p-L>0-mod} holds for
$w=w_m$, that is,
\begin{equation}\label{eq:p-L>0}
p_k(w_m)-(k-1)\mypp L_m(w_m;\theta)>0.
\end{equation}
\end{itemize}
\end{lemma}

\proof (a) From the definition \eqref{eq:K}, compute the partial
derivative
\begin{equation}\label{eq:K'}
\frac{\partial K_m}{\partial
v}=\frac{(p_k+1)^{k-1}}{(p_k+L_m)^{k+1}}
\left((p_k+1)\bigl(p_k-(k-1)\myp L_m\bigr)\myp\frac{\partial
L_m}{\partial v}-k\myp L_m(1-L_m)\mypp p'_k \right),
\end{equation}
with the shorthand notation $p_k:=p_k(v)$ and $L_m:=L_m(v;\theta)$.
Hence, the condition $\partial K_m/\partial v=0$ is reduced to the
equality
\begin{equation}\label{eq:K'=0}
(p_k+1)\bigl(p_k-(k-1)\myp L_m\bigr)\myp\frac{\partial L_m}{\partial
v}-k\myp L_m(1-L_m)\mypp p'_k = 0.
\end{equation}
If $w<v_m$ then $\partial L_m/\partial v|_{v=w}\myn>0$, and the
required inequality~\eqref{eq:p-L>0-mod} readily follows from
equation~\eqref{eq:K'=0} using that $L_m(w;\theta)<1$. Similarly, if
$w>v_m$ then $\partial L_m/\partial v|_{v=w}\myn<0$ and
equation~\eqref{eq:K'=0} implies the inequality~\eqref{eq:p-L>0-mod}
provided that $L_m(w;\theta)>1$. Alternatively, if $L_m(w;\theta)\le
1$ then, noting that $w>v_m>1$ (by Lemma~\ref{lm:v>1}), we obtain,
in agreement with~\eqref{eq:p-L>0-mod},
\begin{equation}\label{eq:L(w)<}
p_k(w)>p_k(1)=k-1\ge (k-1)\mypp L_m(w;\theta),
\end{equation}
because the function $w\mapsto p_k(w)$ is strictly increasing and
$p_k(1)=k-1$.

Lastly, if $w=v_m$ then $\partial L_m/\partial v|_{v=v_m}\myn=0$ and
equation~\eqref{eq:K'=0} implies $L_m(v_m;\theta)=1$. Again using
Lemma~\ref{lm:v>1}, similarly to \eqref{eq:L(w)<} we get
\begin{equation}\label{eq:L(w)<'}
p_k(w)=p_k(v_m)>p_k(1)=k-1=(k-1)\mypp L_m(w;\theta),
\end{equation}
so the inequality~\eqref{eq:p-L>0-mod} holds in this case as well.

\smallskip
(b) Let $w=w_m$ be the point of maximum of the function $v\mapsto
K_m(v;\theta)$. According to part (a), we only have to consider the
case where $w_m<v_m$ and $L_m(w_m;\theta)\ge 1$.

If $L_m(w_m;\theta)>1$, let $\xbar{w}>v_m>w_m$ be such that
$L_m(\xbar{w};\theta)=L_m(w_m;\theta)$, then
\begin{align}
\notag
K_m(\xbar{w};\theta)&=L_m(\xbar{w};\theta)\left(1-\frac{L_m(\xbar{w};\theta)-1}{p_k(\xbar{w})+L_m(\xbar{w};\theta)}\right)^k\\
\notag
&=L_m(w_m;\theta)\left(1-\frac{L_m(w_m;\theta)-1}{p_k(\xbar{w})+L_m(w_m;\theta)}\right)^k\\
\notag
&>L_m(w_m;\theta)\left(1-\frac{L_m(w_m;\theta)-1}{p_k(w_m)+L_m(w_m;\theta)}\right)^k\\
&=K_m(w_m;\theta). \label{eq:K>K-tilde}
\end{align}
Thus, $K_m(\xbar{w};\theta)>K_m(w_m;\theta)$, which contradicts the
assumption that $v=w_m$ provides the maximum value of the function
$v\mapsto K_m(v;\theta)$.

Lastly, suppose that, for some $\theta>\theta_m$,
\begin{equation}\label{eq:Lw=1}
L_m(w_m;\theta)=1.
\end{equation}
In view of the definition \eqref{eq:K}, condition \eqref{eq:Lw=1}
implies that $K_m(w_m;\theta)=1$. Let us prove that in this case we
must have $w_m>1$, which would then automatically imply the required
inequality~\eqref{eq:p-L>0} (cf.\ \eqref{eq:L(w)<}
and~\eqref{eq:L(w)<'}). To the contrary, assume that $w_m\le1$. If
$w_m=1$ then, using the definition \eqref{eq:L} and recalling that
$m+m'=q-1$, the equation \eqref{eq:Lw=1} is reduced to
\begin{equation*}
L_m(1;\theta)=(\theta-1)(k-1)-(q-1)=1,
\end{equation*}
whence we find
\begin{equation}\label{eq:theta-1}
\theta=1+\frac{q}{k-1}.
\end{equation}
Furthermore, noting that
\begin{equation}\label{eq:p'=}
p_k'(1)=\sum_{i=1}^{k-1} i=\frac{k\myp(k-1)}{2}
\end{equation}
and substituting \eqref{eq:theta-1}, from \eqref{eq:L} we get
\begin{equation*}
\frac{\partial L_m(1;\theta)}{\partial
v}=(\theta-1)\mypp\frac{k\myp(k-1)}{2}-k\myp
m=k\left(\frac{q}{2}-m\right)>0,
\end{equation*}
since $m\le\frac12(q-1)<q/2$. Thus, $w_m=1$ is the \emph{left root}
of the equation $L_m(v;\theta)=1$. Denote by $\bar{v}>1$ the
\emph{right root}, that is, $L_m(\bar{v};\theta)=1$ and $\partial
L_m/\partial v|_{v=\bar{v}}\myn<0$. It follows that
$K_m(\bar{v};\theta)=1$ (see~\eqref{eq:K}), so the maximum value $1$
of the function $v\mapsto K_m(v;\theta)$ is also attained at
$v=\bar{v}>1$. Returning to formula \eqref{eq:K'}, observe that
\begin{align}
\frac{\partial K_m(\bar{v};\theta)}{\partial v}
=\frac{p_k(\bar{v})-(k-1)}{p_k(\bar{v})+1}\times\frac{\partial
L_m(\bar{v};\theta)}{\partial v}<0, \label{eq:K'<0}
\end{align}
because $p_k(\bar{v})>p_k(1)=k-1$ and, as mentioned above, $\partial
L_m/\partial v|_{v=\bar{v}}\myn<0$. But the inequality
\eqref{eq:K'<0} implies that there are points $v<\bar{v}$ such that
$K_m(v;\theta)>K_m(\bar{v};\theta)$, a contradiction. Hence, the
case $w_m=1$ is impossible.

Now, suppose that $w_m<1$. Then $p_k(w_m)<p_k(1)=k-1$ and, in view
of the condition~\eqref{eq:Lw=1}, from equation \eqref{eq:K'=0} it
readily follows that $\partial L_m/\partial v|_{v=w_m}\myn=0$, that
is, $L_m(w_m;\theta)=1$ is the maximum value of the function
$v\mapsto L_m(v;\theta)$. Hence, for all $v<w_m$ we have
\begin{equation}\label{eq:L',L<}
L_m(v;\theta)<1,\qquad \frac{\partial L_m(v;\theta)}{\partial v}>0.
\end{equation}
On the other hand, by \eqref{eq:K'=0} and monotonicity of $p_k(v)$,
\begin{align}
\notag p_k(w_m)-(k-1)\myp
L_m(w_m;\theta)&=p_k(w_m)-(k-1)\\
&<p_k(1)-(k-1)=0. \label{eq:p-<0}
\end{align}
By continuity of the functions $v\mapsto p_k(v)$ and $v\mapsto
L_m(v;\theta)$, the inequality \eqref{eq:p-<0} is preserved for all
$v<w_m$ close enough to $w_m$:
\begin{align}
p_k(v)-(k-1)\myp L_m(v;\theta)<0. \label{eq:p-<0-}
\end{align}
Using \eqref{eq:L',L<} and \eqref{eq:p-<0-}, from \eqref{eq:K'} it
follows that for such $v$ we have $\partial K_m/\partial v<0$. But
this means that the function $v\mapsto K_m(v;\theta)$ is
\emph{decreasing} in the left vicinity of $w_m$, and thus $w_m$
cannot be a maximum, in contradiction with our assumption. Thus, we
have proved that $w_m>1$ as required, which completes the proof of
Lemma~\ref{lm:p-L>0}.
\endproof

The next two lemmas provide useful bounds on $w_m=w_m(\theta)$.
First, there is a simple uniform upper bound.
\begin{lemma}\label{lm:w<theta}
For all $m\in[1,q-2\myp]$,
\begin{equation}\label{eq:w<theta}
w_m<\theta,\qquad \theta\ge\theta_m.
\end{equation}
\end{lemma}
\proof Observe, using the definition \eqref{eq:L}, that
\begin{align*}
L_m(v;\theta)|_{v=\theta}&=(\theta-1)\mypp p_{k}(\theta)-m\myp\theta^{k} -m'\\
&=(\theta-1)(\theta^{k-1}+\dots+\theta)-m\myp\theta^{k} -m'\\
&=-(m-1)\myp \theta^{k}-\theta-m'<0,
\end{align*}
and also (cf.~\eqref{eq:R'v})
\begin{align*}
\left.\frac{\partial L_m(v;\theta)}{\partial v}\right|_{v=\theta}&=(\theta-1)\mypp p'_{k}(\theta)-km\myp \theta^{k-1}\\
&=(\theta-1)\mypp \bigl((k-1)\myp\theta^{k-2}+\dots+2\myp\theta+1\bigr)-km\myp\theta^{k-1}\\
&=-(km-k+1)\myp \theta^{k-1}-\theta^{k-2}-\dots-\theta-1<0.
\end{align*}
Hence, the point $v=\theta$ lies \emph{to the right} of the set
$\mathscr{V}_{m}^+(\theta)=\{v>0\colon L_m(v;\theta)>0\}$
(see~\eqref{eq:V+}). But $w_m\in\mathscr{V}_m^{+}(\theta)$ and
therefore $w_m<\theta$, as claimed in~\eqref{eq:w<theta}.
\endproof

The important lower bound for $w_m=w_m(\theta)$ is established next.
\begin{lemma}\label{lm:w>1}\mbox{}
\begin{itemize}
\item[(a)]
For all $m$ in the range $1\le m<\frac12(q-1)$, we have
\begin{equation}
\label{eq:w>1} w_m>1,\qquad \theta>\theta_m.
\end{equation}
\item[(b)]
If $m=\frac12\myp(q-1)$ then the maximum point $w_m$ \textup{(}which
may not be unique\textup{)} can be chosen so that $w_m\ge1$.
\end{itemize}
\end{lemma}
\proof If $L_m(w_m;\theta)\ge1$ then, by the inequality
\eqref{eq:p-L>0} of Lemma~\ref{lm:p-L>0},
$$
p_k(w_m)>(k-1)\myp L_m(w_m;\theta)\ge k-1=p_k(1),
$$
which implies, due to the monotonicity of $p_k(\cdot)$, that
$w_m>1$, in line with~\eqref{eq:w>1}. Thus, it remains to consider
the case $L_m(w_m;\theta)<1$.

Assume first that $w_m=1$ for some $\theta>\theta_m$. Using the
definition \eqref{eq:L} and the value $p_k(1)=k-1$, we have
\begin{align}\notag
L_m(1;\theta)&=(\theta-1)\mypp p_k(1)-m-m'\\
\label{eq:L|1} &=(\theta-1)(k-1)-(q-1),
\end{align}
and also, recalling formula \eqref{eq:p'=},
\begin{align}\notag
\frac{\partial L_m(1;\theta)}{\partial v} &=(\theta-1)\mypp
p'_k(1)-k\myp m\\
\notag
&=\frac{k}{2}\bigl((\theta-1)(k-1)-2\myp m\bigr)\\
\label{eq:L'|1} &=\frac{k}{2}\bigl(L_m(1;\theta)+m'-m\bigr).
\end{align}
Substituting \eqref{eq:p'=}, \eqref{eq:L|1} and \eqref{eq:L'|1} into
\eqref{eq:K'}, it is easy to check that the condition $\partial
K_m/\partial v|_{v=1}\myn=0$ (see~\eqref{eq:K'=0}) is reduced to
\begin{equation}\label{eq:m'-m=0}
\bigl(1-L_m(1;\theta)\bigr)(m'-m)=0.
\end{equation}
Since $L_m(1;\theta)=L_m(w_m;\theta)<1$ by assumption, the condition
\eqref{eq:m'-m=0} is only satisfied if $m'-m=0$, that is,
$m=\frac12(q-1)$. Conversely, if $m=\frac12(q-1)$ (i.e., $m'=m$)
then, by the scaling formula~\eqref{eq:K-scaled} of
Lemma~\ref{lm:LK-scaled}, we have the identity
\begin{equation}\label{eq:K=K}
K_m(v^{-1};\theta)=K_{m'}(v;\theta)=K_m(v;\theta),\qquad\theta>\theta_m,\
v>0,
\end{equation}
which implies that the maximum point $w_m=w_m(\theta)$ can always be
chosen so as to satisfy the inequality $w_m\ge1$, which proves
part~(b) of the lemma.

Finally, let $m<\frac12(q-1)$ and $w_m<1$ for some
$\theta>\theta_m$. Denote $L_m^*(\theta):= L_m(v_m(\theta);\theta)=
\max_{v\in\mathscr{V}_m^{+}(\theta)} L_m(v;\theta)$ (see
Lemma~\ref{lm:vmtheta} and the definition~\eqref{eq:V+}). We need to
distinguish between two subcases, (i) $L_m^*(\theta)\le 1$ and (ii)
$L_m^*(\theta)>1$, which require a different argumentation.

\smallskip
(i) Assuming first that $L_m^*(\theta)\le1$, we will show that then
\begin{equation}\label{eq:K>K}
K_m(w_m^{-1};\theta)>K_m(w_m;\theta),
\end{equation}
which would contradict the assumption that $K_m(w_m;\theta)$ is the
maximum value. By formula \eqref{eq:K-scaled} of
Lemma~\ref{lm:LK-scaled}, we have
$K_m(w_m^{-1};\theta)=K_{m'}(w_m;\theta)$. Hence, recalling the
definition \eqref{eq:K} of the function $K_m(v;\theta)$, the
inequality~\eqref{eq:K>K} is reduced to
\begin{equation}
\frac{L_{m'}(w_m;\theta)}
{\bigl(p_k(w_m)+L_{m'}(w_m;\theta)\bigr)^k}>\frac{L_{m}(w_m;\theta)}
{\bigl(p_k(w_m)+L_{m}(w_m;\theta)\bigr)^k}. \label{eq:L'>L}
\end{equation}
Note that (cf.~\eqref{eq:L->L2})
\begin{align*}
\notag L_{m'}(w_m;\theta)&=(\theta-1)\mypp p_k(w_m)-m'w_m^k - m\\
&= L_{m}(w_m;\theta)+ (m'-m)(1-w_m^k)\\
&> L_{m}(w_m;\theta).
\end{align*}
Hence, for the proof of the inequality \eqref{eq:L'>L}, it suffices
to show that the function $L\mapsto L\myp(p_k(w_m)+L)^{-k}$ is
strictly increasing on the interval
$L\in[L_{m}(w_m;\theta),L_{m'}(w_m;\theta)]$. Computing the
derivative of this function, we see that the claim holds provided
that
$$
\frac{p_k(w_m)}{k-1}> L,\qquad L_{m}(w_m;\theta)\le L\le
L_{m'}(w_m;\theta),
$$
or simply if
\begin{equation}\label{eq:p>}
\frac{p_k(w_m)}{k-1}>L_{m'}(w_m;\theta).
\end{equation}

The inequality~\eqref{eq:p>} is easy to prove. Indeed, using the
assumption $w_m<1$, observe from the definition \eqref{eq:p(z)} that
\begin{equation}\label{eq:p>'}
\frac{p_k(w_m)}{k-1}> w_m^{k-1}>w_m^k.
\end{equation}
On the other hand, according to the scaling formula
\eqref{eq:L-scaled} of Lemma~\ref{lm:LK-scaled}, we have
\begin{equation}\label{eq:L'<}
L_{m'}(w_m;\theta)=w_m^k\mypp L_m(w_m^{-1};\theta)\le w_m^k\mypp
L_m^*(\theta)\le w_m^k,
\end{equation}
by virtue of the assumption $L^*_m(\theta)\le1$. Now, the required
inequality \eqref{eq:p>} readily follows from the estimates
\eqref{eq:p>'} and~\eqref{eq:L'<}.

Thus, the inequality \eqref{eq:K>K} is proved, and therefore the
assumptions $w_m<1$ and $L_m^*(\theta)\le 1$ are incompatible.

\smallskip
(ii) Assume now that $L_m^*(\theta)=L_m(v_m;\theta)>1$ and, as
before, $w_m=w_m(\theta)<1$. Denote
$$
W_m= W_m(\theta):=\left\{w>1\colon \left.\frac{\partial
K_m(v;\theta)}{\partial v}\right|_{v=w}\!=0\right\},
$$
that is, the set of all critical points of the function $v\mapsto
K_m(v;\theta)$ (i.e., satisfying the equation~\eqref{eq:K'=0}) that
lie to the right of point $v=1$. By assumption,
\begin{equation}\label{eq:K<K}
K_m(w_m;\theta)>\max_{w\in W_m} K_m(w;\theta),
\end{equation}
and our aim is to show that this leads to a contradiction.

Since $\partial L_m/\partial v|_{v=v_m}\myn=0$ and
$L_m(v_m;\theta)>1$, formula~\eqref{eq:K'} implies $\partial
K_m/\partial v|_{v=v_m}\myn>0$ and, therefore, there is at least one
critical point $w>v_m$, which then automatically belongs to the set
$W_m$, because $v_m>1$ by Lemma~\ref{lm:v>1}. There may also be
critical points $w\in W_m$ such that $1<w<v_m$; for these we may
assume, without loss of generality, that $L_m(w;\theta)<1$, for
otherwise we consider the point $\xbar{w}>v_m>w$ such that
$L_m(\xbar{w};\theta)=L_m(w;\theta)$, and it follows (similarly to
the derivation of inequality~\eqref{eq:K>K-tilde}) that
$K_m(\xbar{w};\theta)\ge K_m(w;\theta)$, which means that such $w$
can be removed from the set~$W_m$ without affecting the maximum
in~\eqref{eq:K<K}.

Now, the idea is to increase the index $m$. Namely, treating $m$ as
a continuous parameter (see Remark~\ref{rm:m_cont}), differentiate
the function $m\mapsto K_m(w_m(\theta);\theta)$ to obtain
\begin{align}\notag
\frac{\partial K_m(w_m;\theta)}{\partial m}&=\frac{\partial
K_m(v;\theta)}{\partial v}\biggr|_{v=w_m}\!\times \frac{\partial
w_m(\theta)}{\partial m}\\
\notag &\quad+\frac{\partial K_m(w_m;\theta)}{\partial
L}\biggr|_{L=L_m(w_m;\theta)}\!\times \frac{\partial
L_m(v;\theta)}{\partial m}\biggr|_{v=w_m}\\
&=\frac{\bigl(p_k(w_m)+1\bigr)^k\bigl(p_k(w_m)-(k-1)\myp
L_m(w_m;\theta)\bigr)}{\bigl(p_k(w_m)+L_m(w_m;\theta)\bigr)^{k+1}}\times
\bigl(1-w_m^k\bigr), \label{eq:K'_m>0}
\end{align}
where we used the condition $\partial K_m/\partial v|_{v=w_m}\myn=0$
and the definitions \eqref{eq:L} and~\eqref{eq:K}. Owing to
Lemma~\ref{lm:p-L>0}(b), the right-hand side of \eqref{eq:K'_m>0} is
positive and, therefore, the function $m\mapsto K_m(w_m;\theta)$ is
monotone \emph{increasing} as long as $w_m<1$ and
$m<m_0:=\frac12\myp(q-1)$. Likewise, every critical point
$w=w^{(i)}$ from the original (finite) set $W_m$ generates a
continuously differentiable branch $m\mapsto w_m^{(i)}$ as a
function of the increasing variable~$m$, and an argument similar to
\eqref{eq:K'_m>0}, now based on Lemma~\ref{lm:p-L>0}(a), yields that
the corresponding function $m\mapsto \max_{w\in W_m} K_m(w;\theta)$
is monotone \emph{decreasing} up to $m=m_0$.

If for some $\widetilde{m}\in (m,m_0)$ it occurs that
$w_{\widetilde{m}}=1$ then, by continuity, $\partial
K_{\widetilde{m}}/\partial v|_{v=1}\myn=0$, which implies, as was
shown before (see~\eqref{eq:m'-m=0}), that
$L_{\widetilde{m}}(w_{\widetilde{m}};\theta)=1$ and therefore
$K_{\widetilde{m}}(w_{\widetilde{m}};\theta)=1$ is the maximum value
of the function $v\mapsto K_{\widetilde{m}}(v;\theta)$. Moreover,
combining the monotonicity properties established above with the
hypothetical inequality~\eqref{eq:K<K}, this implies
$$
1=K_{\widetilde{m}}(w_{\widetilde{m}};\theta)>
K_{m}(w_{m};\theta)>\max_{w\in W_{m}}K_m(w;\theta)>\max_{w\in
W_{\widetilde{m}}}\!K_{\widetilde{m}}(w;\theta),
$$
that is,
\begin{equation}\label{eq:maxK<1}
\max_{w\in W_{\widetilde{m}}}\!K_{\widetilde{m}}(w;\theta)<1.
\end{equation}
But this cannot be true, because there is
$\xbar{w}>v_{\widetilde{m}}$ where
$L_{\widetilde{m}}(\xbar{w};\theta)=1$, so that
$K_{\widetilde{m}}(\xbar{w};\theta)=1$ is another maximum and,
hence, $\xbar{w}\in W_{\widetilde{m}}$, thus
contradicting~\eqref{eq:maxK<1}.

This shows that we can exploit the monotonicity properties with
respect to variable $m$ up to the final value $m=m_0=\frac12(q-1)$,
so that
\begin{equation}\label{eq:maxK_m}
K_m(w_m;\theta)<K_{m_0}\myn(w_{m_0};\theta)
\end{equation}
and also
\begin{equation}\label{eq:maxK_m-tilde}
\max_{w\in W_{m_0}}\!K_{m_0}(w;\theta)<\max_{w\in
W_{m}}K_m(w;\theta).
\end{equation}
Combining \eqref{eq:maxK_m} and \eqref{eq:maxK_m-tilde}
with~\eqref{eq:K<K}, it follows that
\begin{equation}\label{eq:max<max}
\max_{w\in W_{m_0}}\! K_{m_0}\myn(w;\theta)
<K_{m_0}(w_{m_0};\theta).
\end{equation}
But this is impossible, since $m_0=m_0'$ and, by the scaling
relation \eqref{eq:K-scaled} of Lemma~\ref{lm:LK-scaled},
$K_{m_0}(v;\theta)\equiv K_{m_0}(v^{-1};\theta)$ (see
\eqref{eq:K=K}), which implies that the maximum values of the
function $v\mapsto K_{m_0}(v;\theta)$ over $v<1$ and $v>1$ must be
the same, in contradiction with the inequality~\eqref{eq:max<max}.

Thus, the hypothesis \eqref{eq:K<K} is false, together with the
assumption $w_m<1$ under case (ii) (i.e., with $L^*_m(\theta)>1$).
This completes the proof of Lemma~\ref{lm:w>1}.
\endproof

\subsection{Properties of $\theta_m$ and $\alpha_m(\theta)$}\label{sec:7.3}
\begin{proposition}\label{pr:theta-symmetry}
For each $m=1,\dots,q-2$ and $m'=q-1-m$, we have
\begin{equation}\label{eq:thetamm'}
\theta_m=\theta_{m'}.
\end{equation}
Moreover, the functions $\theta\mapsto\alpha_m(\theta)$
\textup{(}see~\textup{\eqref{eq:alpham}}\textup{)} satisfy the
symmetry relation
\begin{equation}\label{eq:alphamm'}
\alpha_{m'}(\theta)\equiv \alpha_{m}(\theta),\qquad \theta>\theta_m.
\end{equation}
\end{proposition}
\proof Like in Lemma~\ref{lm:vmtheta}(c), denote
$v_m^*\myn:=v_m(\theta_m)$. Observe that $v_m^*$ satisfies the
conjugation property
\begin{equation}\label{eq:v*m'}
v_{m'}^*=\frac{1}{v_m^*},\qquad m=1,\dots,q-2,
\end{equation}
where $m'=q-1-m$. Indeed, computing the left-hand side of \eqref{vy}
for $v=1/v_m^*$ and with $m$ replaced by $m'$, we get, due to
Lemma~\ref{lm:vmtheta}(c),
\begin{align*}
m'\sum_{i=1}^{k-1}i\left(\frac{1}{v_m^*}\right)^{k-i}\!-m\sum_{i=1}^{k-1}
i\left(\frac{1}{v_m^*}\right)^{i-k}\!=-\left(m\sum_{i=1}^{k-1}
i\mypp (v_m^*)^{k-i}-m'\sum_{i=1}^{k-1}i\mypp(v_m^*)^{i-k}\right)=0,
\end{align*}
whence \eqref{eq:v*m'} follows due to the uniqueness of solution.

Now, using \eqref{eq:v*m'} and the scaling property
\eqref{eq:L-scaled}, we have
\begin{align*}
L_{m'}(v_{m'}^*;\theta_m)&=L_{m'}\bigl((v_{m}^*)^{-1};\theta_m\bigr)\\
&=(v_{m}^*)^{-k}L_{m}(v_{m}^*;\theta_m)=0,
\end{align*}
according to \eqref{eq:theta*}, and by the uniqueness of solution to
the equation $L_{m'}(v_{m'}(\theta);\theta)=0$ (see
Lemma~\ref{lm:vmtheta}(b)), the equality \eqref{eq:thetamm'}
follows.

Finally, the identity \eqref{eq:alphamm'} is valid due to the
definition \eqref{eq:alpham} and formula~\eqref{eq:K-scaled}.
\endproof

\begin{proposition}\label{pr:theta-monotone}
Let $q\ge5$, and set $m_0:=\lfloor \frac12\myp(q-1)\rfloor$. Then
for $m=1,\dots,m_0-1$
\begin{align}\label{eq:theta<theta}
\theta_{m}&<\theta_{m+1},\\
\label{eq:alpha<alpha} \alpha_{m}(\theta)&>\alpha_{m+1}(\theta),
\qquad \theta>\theta_{m+1}.
\end{align}
\end{proposition}
\proof Treating $m$ as a continuous parameter (see
Remark~\ref{rm:m_cont0}), differentiate the identity
\eqref{eq:theta*} to obtain
\begin{align*}
\frac{\dif L_m(v^*_m;\theta_m)}{\dif m}&=\frac{\partial
L_m(v;\theta_m)}{\partial v}\biggr|_{v=v_m^*}\!\times \frac{\dif
v_m^*}{\dif m}\\
&\quad+\frac{\partial L_m(v_m^*;\theta)}{\partial
\theta}\biggr|_{\theta=\theta_m}\!\times \frac{\dif\theta_m}{\dif
m}+\frac{\partial L_m(v;\theta)}{\partial
m}\biggr|_{v=v_m^*,\,\theta=\theta_m}\!\equiv 0.
\end{align*}
Using \eqref{eq:L} and \eqref{eq:R'v*}, the last identity is reduced
to
\begin{align*}
p_k(v_m^*)\,\frac{\dif\theta_m}{\dif m}+1-(v_m^*)^k=0,
\end{align*}
which yields
\begin{equation}\label{eq:theta'}
\frac{\dif\theta_m}{\dif m}=\frac{(v_m^*)^k-1}{p_k(v_m^*)}.
\end{equation}
Recalling that $v_m^*>1$ for all $m<\frac12\myp(q-1)$ (see
Lemma~\ref{lm:vmtheta}(c)), from \eqref{eq:theta'} it follows that
$\dif\theta_m/\dif m>0$ for $m<\frac12\myp(q-1)$. For integer
$m=1,\dots,m_0$, this transcribes as the
inequality~\eqref{eq:theta<theta}.

Turning to the proof of \eqref{eq:alpha<alpha}, for a given
$\theta\ge\theta_1$ let $m^*\ge1$ be the root of the equation
$\theta_m=\theta$. We shall prove a (stronger) \emph{continuous
version} of the inequality \eqref{eq:alpha<alpha}, namely, that the
function $m\mapsto \alpha_m(\theta)$ (defined for $m\ge m^*$) is
monotone decreasing. As before, denote by $w_m=w_m(\theta)$ the
point where the function $v\mapsto K_m(v;\theta)$ attains its
maximum value, and set $K_m^*(\theta):=K_m(w_m(\theta);\theta)$.
Differentiating the function $m\mapsto K_m^*(\theta)$, we obtain
(see~\eqref{eq:K'_m>0})
\begin{align}
\frac{\partial K_m^*}{\partial m}
&=\frac{\bigl(p_k(w_m)+1\bigr)^k\bigl(p_k(w_m)-(k-1)\myp
L_m(w_m;\theta)\bigr)}{\bigl(p_k(w_m)+L_m(w_m;\theta)\bigr)^{k+1}}\times
\bigl(1-w_m^k\bigr). \label{eq:K'_m<0}
\end{align}
Now, owing to Lemmas \ref{lm:p-L>0} and~\ref{lm:w>1} (see also
Remark~\ref{rm:m_cont}), the right-hand side of \eqref{eq:K'_m<0} is
negative \strut{}for all $m\in\bigl[m^*,\frac12\myp(q-1)\bigr)$ and,
therefore, the function $m\mapsto K_m^*(\theta)$ is monotone
\strut{}decreasing in the closed interval $[m^*,\frac12\myp(q-1)]$.
\strut{}By the definition \eqref{eq:alpham}, the same holds for the
function $m\mapsto\alpha_m(\theta)$, \strut{}as claimed.
\endproof

\begin{proposition}\label{pr:alpham-limits}
For all $m=1,\dots,q-2$, the functions
$\theta\mapsto\alpha_m(\theta)$ defined by formula \eqref{eq:alpham}
satisfy the upper bound
\begin{equation}\label{eq:alpham<k-1}
\alpha_m(\theta)<k-1,\qquad \theta>\theta_m.
\end{equation}
Moreover, they have the following ``boundary'' values,
\begin{equation}\label{eq:alpha-limits}
\lim_{\theta\downarrow\myp \theta_m}\alpha_m(\theta)=-\infty,\qquad
\lim_{\theta\uparrow\infty}\alpha_m(\theta)=k-1.
\end{equation}
\end{proposition}

\proof Let $w_m=w_m(\theta)$ be the point of maximum of the function
$v\mapsto K_m(v;\theta)$, so that $K_m^*(\theta)=K_m(w_m;\theta)$.
Treating the term $L_m=L_m(v;\theta)$ in the expression \eqref{eq:K}
as an independent parameter $L\ge 0$, we can write
\begin{equation}\label{eq:max_w}
K_m^*(\theta)\le (p_k(w_m)+1)^k\max_{L\ge0}\frac{L}{(p_k(w_m)+L)^k}.
\end{equation}
By differentiation, it easy to verify that the maximum on the
right-hand side of \eqref{eq:max_w} is attained at
$L_0:=p_k(w_m)/(k-1)$, hence
\begin{align}
\notag
K_m^*(\theta)&\le (p_k(w_m)+1)^k\frac{L}{(p_k(w_m)+L)^k}\biggr|_{L=L_0}\\
\notag
&=\left(\frac{k-1}{k}\right)^k\left(1+\frac{1}{p_k(w_m)}\right)^k\frac{p_k(w_m)}{k-1}\\
&\le \frac{p_k(w_m)}{k-1}. \label{eq:k<(1)}
\end{align}
Furthermore, $w_m<\theta$ by Lemma~\ref{lm:w<theta}, so that
$$
p_m(w_m)<p_m(\theta)<(k-1)\mypp\theta^{k-1}.
$$
Substituting this estimate into the right-hand side of
\eqref{eq:k<(1)}, we obtain
$$
K_m^*(\theta)< \theta^{k-1},
$$
and therefore (see~\eqref{eq:alpham})
$$
\alpha_m(\theta)=\frac{\ln K_m^*(\theta)}{\ln\theta}<k-1,
$$
which proves the bound \eqref{eq:alpham<k-1}. In particular, this
implies that
\begin{equation}\label{eq:K-limsup}
\limsup_{\theta\to\infty}\alpha_m(\theta)\le k-1.
\end{equation}

To obtain a matching lower bound, take a specific value
$$
v=v_0:=\frac{t}{m}\left(1-\frac{\ln t}{t}\right),\qquad t:=\theta-1,
$$
then, as $t\to\infty$,
\begin{align*}
p_k(v_0)&=\frac{t^{k-1}}{m^{k-1}}\left(1-\frac{\ln
t}{t}\right)^{k-1}\!+O(t^{k-2})\\
&=\frac{t^{k-1}}{m^{k-1}}\left(1-\frac{(k-1)\ln
t}{t}\right)+O(t^{k-2})
\end{align*}
and
\begin{align*}
L_m(v_0;\theta)&=t\myp p_k(v_0)-m\myp v_0^{k}-m'\\
&=\frac{t^{k}}{m^{k-1}}\left(1-\frac{(k-1)\ln
t}{t}\right)-\frac{t^k}{m^{k-1}}\left(1-\frac{k\ln
t}{t}\right)+O(t^{k-1})\\
&=\frac{t^{k-1}\ln t}{m^{k-1}}+O(t^{k-1}).
\end{align*}
Hence,
$$
p_k(v_0)+L_m(v_0;\theta) \sim \frac{t^{k-1}\ln t}{m^{k-1}}
$$
and
\begin{align*}
K_m(v_0;\theta)&=\frac{L_m(v_0;\theta)\myp
\bigl(p_k(v_0)+1\bigr)^k}{\bigl(p_k(v_0)+L_m(v_0;\theta)\bigr)^k}
\sim\left(\frac{t}{m\ln t}\right)^{k-1}.
\end{align*}
Therefore,
$$
\ln K_m(v_0;\theta)\sim (k-1)\ln t\sim (k-1)\ln \theta,\qquad
\theta\to\infty,
$$
so that
\begin{equation}\label{eq:K-liminf}
\liminf_{\theta\to\infty}\alpha_m(\theta) \ge
\lim_{\theta\to\infty}\frac{ \ln K_m(v_0;\theta)}{\ln\theta}=k-1.
\end{equation}
Thus, combining \eqref{eq:K-limsup} and \eqref{eq:K-liminf}, we
obtain the second limit in~\eqref{eq:alpha-limits}.

Finally, we turn to the proof of the first limit
in~\eqref{eq:alpha-limits}. By virtue of
Proposition~\ref{pr:theta-symmetry}, we may assume that
$m\le\frac12\myp(q-1)$. Then, by Lemma~\ref{lm:w>1}, $w_m\ge1$ and
therefore $p_k(w_m)\ge k-1$. Hence, from the definition \eqref{eq:K}
we get
\begin{align} \notag 0< K_m^*(\theta)&\le
L_m(w_m;\theta)\left(1+\frac{1}{p_k(w_m)}\right)^k\\
&\le L_m^*(\theta)\left(\frac{k}{k-1}\right)^k, \label{eq:K<L-max}
\end{align}
where
$L_m^*(\theta)=L_m(v_m(\theta);\theta)=\max_{v>0}L_m(v;\theta)$. By
continuity,
$$
\lim_{\theta\downarrow\theta_m} L_m^*(\theta)= L_m^*(\theta_m)=0,
$$
and it follows from the bound \eqref{eq:K<L-max} that
$\lim_{\theta\downarrow\theta_m} K_m^*(\theta)=0$, which implies the
first limit in~\eqref{eq:alpha-limits}. Thus, the proof of
Proposition~\ref{pr:alpham-limits} is complete.
\endproof

\begin{proposition}\label{pr:K-monotone}
For each $m=1,\dots,q-2$, the function $\theta\mapsto K_m^*(\theta)$
is monotone increasing for $\theta>\theta_m$.
\end{proposition}
\proof By virtue of Lemma~\ref{lm:LK-scaled},
$K_m^*(\theta)=K_{m'}^*(\theta)$, where $m'=q-1-m$; hence, it
suffices to prove the claim for $m$ in the range $1\le m\le
\frac12(q-1)$. Using the definitions \eqref{eq:L} and~\eqref{eq:K},
differentiate with respect to $\theta$ to obtain
\begin{align}
\notag \frac{\dif K_m^*}{\dif \theta}&=\frac{\partial K_m}{\partial
v}\biggr|_{v=w_m}\!\times \frac{\dif w_m}{\dif \theta}
+\frac{\partial K_m}{\partial
L}\biggr|_{v=w_m,\,L=L_m(w_m;\myp\theta)}\!\times \frac{\partial
L_m}{\partial \theta}\biggr|_{v=w_m}\\
&=\frac{\bigl(p_k(w_m)+1\bigr)^k\bigl(p_k(w_m)-(k-1)\myp
L_m(w_m;\theta)\bigr)}{\bigl(p_k(w_m)+L_m(w_m;\theta)\bigr)^{k+1}}\times
p_k(w_m), \label{eq:K'_m<0+}
\end{align}
on account of the identity $\partial K_m/\partial
v|_{v=w_m}\myn\equiv 0$. To complete the proof, it remains to notice
that the right-hand side of \eqref{eq:K'_m<0+} is positive due to
Lemma~\ref{lm:p-L>0}(b).
\endproof

\begin{remark}
The result of Proposition~\ref{pr:K-monotone} is not trivial (unlike
the similar statement for $L^*_m(\theta)$, see the footnote in the
proof of Lemma~\ref{lm:vmtheta}(b)), because, for each $v>0$, we
have that $L_m(v;\theta)\to\infty$ and, therefore, $K_m(v;\theta)\to
0$ as $\theta\to\infty$ (see formula~\eqref{eq:K}).
\end{remark}

\begin{figure}
\begin{center}
\includegraphics[width=7.3cm]{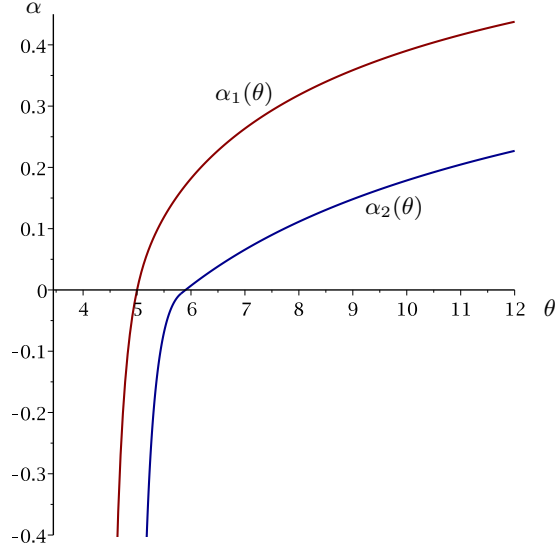}
\put(4,89.7){\mbox{\scriptsize$\theta$}}
\put(-190,204){\mbox{\scriptsize$\alpha$}}
\put(-119,171){\mbox{\scriptsize$\alpha_1(\theta) $}}
\put(-63,128){\mbox{\scriptsize$\alpha_2(\theta)$}}\end{center}
\caption{The graphs of the functions $\alpha_m(\theta)$
($\theta>\theta_m$) for $k=2$, $q=5$ and $m=1,2$ (see the
definition~\eqref{eq:alpham}). According to formula
\eqref{eq:theta|k=2} (with $q=5$), $\theta_1=1+2\sqrt{3}\doteq
4.4641<\theta_2=5$.
Note that $\theta_{1}^0=5$ and $\theta_{2}^0=1+2\myp\sqrt{6}\doteq
5.8990$ are the zeros of $\alpha_1(\theta)$ and $\alpha_2(\theta)$,
respectively (see Example~\ref{ex:k=2_m0}). Note that the graphs are
monotone increasing in line with
Conjecture~\ref{rm:conj2}.}\label{Fig7}
\end{figure}

\begin{conjecture}\label{rm:conj2}
For each $m=1,\dots,q-2$, the function $\theta\mapsto
\alpha_m(\theta)=\ln K_m^*(\theta)/\ln\theta$ is monotone
increasing.
\end{conjecture}
This conjecture is confirmed by computer plots (see
Figure~\ref{Fig7}) and is easy to prove at least for $\theta\le
\theta_m^{\myp0}$, where $\theta_m^{\myp0}$ is the root of the
equation $K_m^*(\theta)=1$ (cf.\
Proposition~\ref{pr:alpha-zeros}(a)); that is, $\alpha_m(\theta)\le
0$ for $\theta\le \theta_m^{\myp0}$. Indeed, for any
$\theta\in(\theta_m,\theta_m^{\myp0}]$, we have
\begin{align*}
\frac{\dif \alpha_m(\theta)}{\dif\theta}&=\frac{\dif
K_m^*(\theta)}{\dif\theta}\times \frac{1}{\ln\theta\cdot
K_m^*(\theta)}-\frac{\ln K_m^*(\theta)}{\theta\ln^2\theta}>0,
\end{align*}
because $\dif K_m^*/\dif\theta>0$ (Proposition~\ref{pr:K-monotone}),
whereas $\ln K_m^*(\theta)\le \ln K_m^*(\theta_m^{\myp0})=0$.

\subsection{Zeros of $\alpha_{\pm}(\theta)$ and $\alpha_m(\theta)$}
Recall that the functions $\alpha_{\pm}(\theta)$ and
$\alpha_m(\theta)$ are defined in \eqref{eq:alpha_pm} and
\eqref{eq:alpham}, respectively. As was observed in numerical
examples (see Figure \ref{Fig3} and also Figures \ref{Fig6}
and~\ref{Fig7}), the functions $\alpha_{-}(\theta)$ and
$\alpha_1(\theta)$ have the same zero, $\theta_0^-=\theta_1^{0}$,
whereas $\alpha_{+}(\theta)=0$ at $\theta^+_{0}=1+q/(k-1)$. In this
subsection, we give a proof of these observations.

Let us first state and prove a lemma. Recall the notation
$L_m^*(\theta)=\max_{v\in\mathscr{V}_m^+(\theta)}L_m(v;\theta)$ and
$K_m^*(\theta)=\max_{v\in\mathscr{V}_m^+(\theta)}K_m(v;\theta)$.

\begin{lemma}\label{lm:K=L=1}Let $q\ge3$ and $k\ge2$.
\begin{itemize}
\item[\rm (a)]
For any $m\in[1,q-2\myp]$, if $L_m(v;\theta)<1$ for some $v\ge1$ and
$\theta>\theta_m$ then $K_m(v;\theta)<1$.

\smallskip
\item[\rm (b)] Let $1\le m\le \frac12(q-1)$. If $L_m^*(\theta)<1$ for some
$\theta>\theta_m$ then $K_m^*(\theta)<1$. In particular,
\strut{}$L_m^*(\theta)=1$ if and only if $K_m^*(\theta)=1$.
\end{itemize}
\end{lemma}
\proof (a) Denoting $s:= (L_m(v;\theta))^{1/k}\!<1$ and using the
definition~\eqref{eq:K}, the required inequality $K_m(v;\theta)<1$
can be rewritten as
$$
p_k(v)>\frac{\,s-s^k}{\!1-s}=p_k(s),
$$
and the last inequality holds by monotonicity of $p_k(v)$, since
$v\ge 1>s$.

\smallskip
(b) Let $v=w_m$ be such that $K_m(w_m;\theta)=K_m^*(\theta)<1$; by
Lemma~\ref{lm:w>1}, $w_m\ge 1$. On the other hand,
$L_m(w_m;\theta)\le L_m^*(\theta)<1$, and by part (a) it follows
that $K_m(w_m;\theta)<1$. The last claim in part (b) then follows by
continuity and monotonicity of both $L_m^*(\theta)$ and
$K_m^*(\theta)$ (see Lemma~\ref{lm:vmtheta}(b) and
Proposition~\ref{pr:K-monotone}, respectively), also recalling that
$L_m(v;\theta)=1$ implies $K_m(v;\theta)=1$ (see~\eqref{eq:K}).
\endproof

\begin{proposition}\label{pr:alpha-zeros} Let $q\ge 3$, and set
$m_0:=\bigl\lfloor\tfrac12(q-1)\bigr\rfloor$.
\begin{itemize}
\item[\rm (a)]
For each $m$ in the range $1\le m\le m_0$, the function
$\alpha_m(\theta)$ has a unique zero given by
\begin{equation}\label{eq:theta_m0}
\theta_m^{\myp0}=\frac{m\myp(v_m^{\myp0})^k+m'+1}{p_k(v_m^{\myp0})},
\end{equation}
where $v_m^{\myp0}>1$ is a sole positive root of the equation
\begin{equation}\label{vy1}
m\sum_{i=1}^{k-1}i\myp v^{k-i}-(m'+1)\sum_{i=1}^{k-1}i\myp
v^{i-k}=0.
\end{equation}

\item[\rm (b)]\begin{itemize}
\item[\rm(i)]\label{item1}
The function $\alpha_{+}(\theta)$ has a unique zero given by
$\theta^{+}_{0}=1+\frac{q}{k-1}$. Moreover,
$\alpha'_+(\theta^{+}_{0})=0$.

\smallskip
\item[\rm(ii)]
The function $\alpha_{-}(\theta)$ has a unique zero
$\theta^{-}_{0}$, which coincides with the zero $\theta_1^{\myp0}$
of the function $\alpha_1(\theta)$.
\end{itemize}

\smallskip
\item[\rm (c)] The zeros
$\theta_1^{\myp0},\dots,\theta_{m_0}^{\myp0}$ follow in ascending
order and are strictly below $\theta_0^+$,
\begin{equation}\label{eq:1<...<m0}
\theta_1^{\myp0}<\dots<\theta_{m_0}^{\myp0}<\theta_0^+=1+\frac{q}{k-1}.
\end{equation}
\end{itemize}
\end{proposition}

\proof (a) By the definition~\eqref{eq:alpham}, the condition
$\alpha_m(\theta)=0$ means that $K_m^*(\theta)=1$ and hence, by
Lemma \ref{lm:K=L=1}(b), $L_m^*(\theta)=1$. Eliminating $\theta$
from the system of equations $L_m(v;\theta)=1$, $\partial L_m(v;\myp
\theta)/\partial v=0$ gives for the root $v=v_m^{\myp0}$ the
equation (cf.~\eqref{eq:L(v)=0})
\begin{equation*}
mk\myp v^{k-1}p_k(v)-\bigl(m\myp v^{k}+m'+1\bigr)\mypp p'_k(v)=0,
\end{equation*}
which can be rearranged to the form~\eqref{vy1}. Uniqueness of
positive solution $v_m^{\myp0}$ of the equation \eqref{vy1} is
obvious, noting that the left-hand side of \eqref{vy1} is a
continuous, increasing function in $v>0$, with the range from
$-\infty$ to $+\infty$. To show that $v_m^{\myp0}>1$, it suffices to
check that the left-hand side of \eqref{vy1} at $v=1$ is negative,
which is indeed true since $2m\le q-1<q$. Expressing $\theta$ from
the equation $L_m(v^{\myp0}_m;\theta)=1$, we obtain
formula~\eqref{eq:theta_m0}.

\smallskip
(b) In the limit $\alpha\to0$, the equation \eqref{rm=1} always has
root $u=1$, while for $u\ne1$, by virtue of the identity
\eqref{eq:p(z)}, it is reduced to equation \eqref{eq:1=} with $m=1$.
Using the notation \eqref{eq:L}, the latter equation can be
rewritten as $L_1(u;\theta)=1$, which in turn has up to two
(positive) roots (see Lemma~\ref{lm:vmtheta}). In total, there are
three positive roots, and for this number to reduce to \emph{two}
(which is the condition of belonging to the curves
$y=\alpha_\pm(\theta)$), either (i) one zero of the function
$u\mapsto L_1(u;\theta)-1$ must coincide with $u=1$, or (ii) the
equation $L_1(u;\theta)=1$ must have a double root, thus also
satisfying the condition $\partial L_1(u;\theta)/\partial u =0$.

In case (i), the condition $L_1(u;\theta)|_{u=1}\myn=1$ transcribes
as $(\theta-1)(k-1)-(q-1)=1$, which immediately yields the root
$\theta_0^+=1+q/(k-1)$. According to the substitution \eqref{eq:x}
(with $\alpha=0$), the corresponding root of the quadratic equation
\eqref{eq:quadratic-intro} is given by $x=\theta_0^+/(q-1)$, which
appears to be the \emph{smaller} of the two roots, $x=x_{-}$.
Therefore, in view of formulas \eqref{eq:a_pm} and
\eqref{eq:alpha_pm}, the value $\theta_0^+$ is a zero of the
function  $\alpha_+(\theta)$. Indeed, using the definition
\eqref{eq:b} of $b=b(\theta)$, the second root of
\eqref{eq:quadratic-intro} is found to be
$$
x_{+}=\frac{b}{x_{-}}=\frac{\theta_0^+\mypp(\theta_0^+
+q-2)}{q-1}\cdot\frac{q-1}{\theta_0^+}=\theta_0^++q-2
>\frac{\theta_0^+}{q-1}=x_{-},
$$
as claimed.

In case (ii), according to the proof of part (a), the unique
solution of the system $L_1(u;\theta)=1$, $\partial
L_1(u;\theta)/\partial u=0$ is given by
$(u,\theta)=(v_1^0,\theta_1^{\myp0})$, where
\begin{equation}\label{eq:theta0a}
\theta_1^{\myp0}=1+\frac{(v_1^0)^k+2}{p_k(v_1^0)}
\end{equation}
and $v_1^0>1$ is a sole root of the equation \eqref{vy1} with $m=1$,
that is,
\begin{equation}\label{vy1'}
\sum_{i=1}^{k-1}i\myp v^{k-i}-(q-1)\sum_{i=1}^{k-1}i\myp v^{i-k}=0.
\end{equation}
Again by the substitution \eqref{eq:x} with $\alpha=0$, the
corresponding root of the quadratic equation
\eqref{eq:quadratic-intro} is given by
$$
x=\frac{\theta_1^{\myp0}\mypp (v_1^0)^k }{q-1},
$$ and we wish to
prove that this is the \emph{bigger} of the two roots, $x=x_+$,
which would imply like before that $\theta_1^{\myp 0}$ is a zero of
the function $\alpha_-(\theta)$. Since the other root of
\eqref{eq:quadratic-intro} equals $b/x$, with $b=b(\theta)$ defined
in \eqref{eq:b}, our claim is expressed as $x>b/x$, that is,
\begin{equation}\label{eq:v1^2>}
(v_1^0)^{2k}>(q-1)\left(1+\frac{q-2}{\theta_1^{\myp0}}\right).
\end{equation}
Furthermore, recalling that $\theta_1^{\myp0}>1$
(see~\eqref{eq:theta0a}), we have
$$
1+\frac{q-2}{\theta_1^{\myp0}}>q-1,
$$
so for the proof of \eqref{eq:v1^2>} it suffices to show that
\begin{equation}\label{eq:v1^2>'}
v_1^0\ge(q-1)^{1/k}.
\end{equation}
Since the function on the left-hand side of \eqref{vy1'} is monotone
increasing, we only need to check that its value at $v=(q-1)^{1/k}$
is non-positive, that is,
\begin{equation*}
\sum_{i=1}^{k-1}i\myp v^{k-i}-\sum_{i=1}^{k-1}i\myp v^{i-k}\le 0.
\end{equation*}
which can be rewritten as
\begin{equation*}
\sum_{i=1}^{k-1}(2\myp i-k)\mypp v^{i-k/2}\ge0.
\end{equation*}
Now, the latter inequality holds because the left-hand side is
obviously monotone increasing as a function of $v\ge 1$, being equal
to $0$ at $v=1$. Hence, \eqref{eq:v1^2>'} follows.

Thus, we have proved that $\theta_0^+=1+q/(k-1)$ and
$\theta_0^-=\theta_1^{\myp0}$ are the sole roots of the functions
$\alpha_+(\theta)$ and $\alpha_-(\theta)$, respectively.

Finally, since the function $\alpha_+(\theta)$ is known to be
positive both at $\theta_{\rm c}<\theta_0^+$ and at infinity (see
Proposition~\ref{pr:alpha-pm-limits}), it readily follows that it
has a minimum value $0$ at $\theta=\theta_0^+$, hence
$\alpha'_+(\theta)=0$, as claimed in part~(b)(i).

\smallskip
(c) The ordering inequalities between $(\theta_m^0)$ in
\eqref{eq:1<...<m0} readily follow from the monotonicity property
\eqref{eq:alpha<alpha} proved in Proposition~\ref{eq:theta<theta}.
Thus, it remains to show that $\theta_m^{\myp0}<\theta_0^+$. Observe
that the value $L_m(v;\theta)|_{v=1}\myn=(\theta-1)\myp(k-1)-(q-1)$
does not depend on $m$. Recalling that $L_1(1;\theta_0^+)=1$ (see
the proof of part~(a)), we get that $L_m(1;\theta_0^+)=1$, but
\begin{align*}
\left.\frac{\partial L_m (v;\theta_0^+}{\partial
v}\right|_{v=1}\!&=(\theta_0^+-1)\mypp
p_k'(1)-mk\\
&=\frac{q}{k-1}\cdot\frac{k\myp(k-1)}{2}-mk\\
&=k\left(\frac{q}{2}-m\right)>0,
\end{align*}
since $m\le \frac12(q-1)<q/2$. Hence, $L_m^*(\theta_0^+)>L_m
(1;\theta_0^+)=1$, and by monotonicity of the function
$\theta\mapsto L_m(v;\theta)$ it follows, according to the proof in
part~(a), that $\theta_m^{\myp0}<\theta_0^+$.

An alternative simple argument is that, as shown in part~(a), the
maximum value of $v\mapsto L_m(v;\theta_m^{\myp0})=1$ is attained at
$v=v_m^0>1$, hence $L_m(1;\theta_m^{\myp0})<1$ and, again by
monotonicity, it follows that $\theta_0^+>\theta_m^{\myp0}$.
\endproof

A version of Proposition~\ref{pr:alpha-zeros}(b) for $q=2$ is easy
to obtain.
\begin{proposition}\label{pr:alpha-pm_q=2}
In the case $q=2$, the functions $\alpha_{\pm}(\theta)$ have a
unique zero at $\theta_{\rm c}=\frac{k+1}{k-1}$, which coincides
with $\theta_0^+=1+2/(k-1)$. Moreover, $\alpha'_{\pm}(\theta_{\rm
c}+)=0$.
\end{proposition}
\proof By Lemma~\ref{lm:alpha+alpha} with $q=2$, we have
$\alpha_{-}(\theta)=-\alpha_{+}(\theta)$ for all $\theta\ge
\theta_{\rm c}=\frac{k+1}{k-1}$. Thus, it suffices to consider the
function $\alpha_{+}(\theta)$. Treating the index $q\ge2$ as a
continuous variable and taking the limit from the domain $q>2$ as
$q\to 2+$, we see that the unique zero of $\alpha_{+}(\theta)$,
given by $\theta_0^+(k,q)=1+q/(k-1)$, converges to
$1+2/(k-1)=\frac{k+1}{k-1}=\theta_{\rm c}$. On the other hand, the
derivative $\alpha'_{+}$ vanishes at $\theta_0^+(k,q)$ for each
$q>2$, hence its limiting value at $\theta_{\rm c}$ is also zero,
that is, $\alpha'_{+}(\theta_{\rm c}+)=0$, as claimed.

This result can also be obtained by a direct calculation. Namely,
for $q=2$ the definition \eqref{eq:b} is reduced to
$b(\theta)=\theta^2$. From equation \eqref{eq:quadratic-intro} with
$q=2$, it is easy to see that $x_\pm(\theta)\to\theta_{\rm c}$ as
$\theta\to\theta_{\rm c}+$. Moreover, a simple asymptotic analysis
shows that
$$
x_{\pm}(\theta)=\theta_{\rm c}\pm\sqrt{2\myp k\myp\theta_{\rm
c}\myp(\theta-\theta_{\rm c})}+(k+1)\myp(\theta-\theta_{\rm
c})+o(\theta-\theta_{\rm c}),\qquad \theta\to\theta_{\rm c}+.
$$
Hence, from \eqref{eq:a_pm} it follows
$$
\ln a_{\pm}(\theta)=-(k+1)\ln \theta_{\rm
c}-(k-1)\myp(\theta-\theta_{\rm c})+o(\theta-\theta_{\rm c}),\qquad
\theta\to\theta_{\rm c}+.
$$
Finally, substituting this into \eqref{eq:alpha_pm} we obtain
\begin{align*}
\alpha_{\pm}(\theta)&=-(k+1)+\frac{(k+1)\ln
\theta_{\rm c}+(k-1)\myp(\theta-\theta_{\rm c})+o(\theta-\theta_{\rm c})}{\ln\theta_{\rm c}+\theta_{\rm c}^{-1}
(\theta-\theta_{\rm c})+o(\theta-\theta_{\rm c})}\\
&=\frac{\theta-\theta_{\rm c}}{\ln\theta_{\rm c}}\left((k-1)-\frac{k+1}{\theta_{\rm c}}\right)+o(\theta-\theta_{\rm c})\\
&=o(\theta-\theta_{\rm c}),\qquad \theta\to\theta_{\rm c}+,
\end{align*}
which implies that $\alpha'\pm{-}(\theta_{\rm c}+)=0$, as claimed.
\endproof

\begin{example}\label{ex:k=2_m0}
Consider the case $k=2$. Then the zero of the function
$\alpha_{+}(\theta)=0$ specializes to \strut{}$\theta^{+}_0=1+q$.
Furthermore, the equation \eqref{vy1} is easily solved to yield
$v_m^0=\sqrt{(q-m)/m}$, and from \eqref{eq:theta_m0} we readily find
\begin{equation}\label{eq:theta=0-}
\theta_{m}^{\myp0}=1+2\mypp\sqrt{m\myp(q-m)},\qquad 1\le m\le
\tfrac12(q-1).
\end{equation}
It is of interest to note, by comparing \eqref{eq:theta=0-} with
\eqref{eq:theta|k=2}, that $\theta_m^{\myp0}=\theta_{m+1}$ (cf.\
\cite[equation~(2.1), page~192]{KRK}). Finally, taking $m=1$ in
\eqref{eq:theta=0-} gives
$\theta^{-}_0=\theta_1^{\myp0}=1+2\mypp\sqrt{q-1}$.
\end{example}

\subsection*{Acknowledgements}

This research was partially supported by a Royal Society
International Exchange Grant IES\textbackslash R2\textbackslash
181078 (December 2018 -- March 2019). U.A.R.\ acknowledges support
by the Institute for Advanced Study of Aix-Marseille University
(IM\'eRA) under a research residency scheme (February\allowbreak
--July 2015), by the London Mathematical Society (LMS Scheme 2
visiting grant, Leeds, February 2016), and by the School of
Mathematics, University of Leeds (visiting grant, February--March
2016). Research of L.V.B.\ was partially supported by a visiting
grant from SFB\,701 (Bielefeld University, June 2015).

\begin{appendix}

\section{Proof of Proposition \ref{pr:TI}}\label{sec:A}

\emph{Necessity.} Assume that the measure $\mu^h$ is translation
invariant, that is, the condition \eqref{tim} holds. Pick the set
$\varLambda\subset V$ to be the unit ball $V_1\equiv
V_1(x_\circ)=\{x_\circ, x_1,\dots,x_{k+1}\}$ centred at the root
$x_\circ$, with $\partial\{x_\circ\}=\{x_1,\dots,x_{k+1}\}$, where
the numbering in $(x_j)$ is consistent with the bijection
$\mathfrak{b}\colon V\to\mathscr{A}_k$, that is,
$x_j=\mathfrak{b}^{-1}(a_j)$ (see Section~\ref{sec:2.3.1}). For
$z\in V$, the shifted set
$V_1(z):=\widetilde{T}_z(V_1)=\{z,z_1,\dots,z_{k+1}\}$
\strut{}(where $z_j=\widetilde{T}_z(x_j)$, $j=1,\dots,k+1$) is the
unit ball centred at $z=\widetilde{T}_z(x_\circ)$, so that
$\partial\{z\}=\{z_1,\dots,z_{k+1}\}$. Consider an arbitrary
configuration $\varsigma\in\varPhi^{V_1}$, \strut{}with spin values
$$
\varsigma(x_\circ)\mynn=i_0,\qquad  \varsigma(x_j)=i_j, \qquad
j=1,\dots,k+1.
$$
Note that if $y=\widetilde{T}_z(x)$ then, according to
\eqref{eq:Tsigma},
\begin{equation}\label{eq:Tsigma=}
(\widetilde{T}_z\varsigma)(y)=\varsigma\bigl(\widetilde{T}^{-1}_z(y)\bigr)=\varsigma(x).
\end{equation}
Hence, the property \eqref{tim} of translation invariance of $\mu^h$
specializes as follows,
\begin{multline}
\mu^h\myn\bigl(\sigma\in\varPhi^V\!\colon \sigma(z)=i_0,\,
\sigma(z_1)=i_1,\dots,\,\sigma(z_{k+1})=i_{k+1}\bigr)\\
=\mu^h\myn\bigl(\sigma\in\varPhi^V\!\colon\sigma(x_\circ)=i_0,\,
\sigma(x_1)=i_1,\dots,\,\sigma(x_{k+1})=i_{k+1}\bigr).
\label{tim_V1}
\end{multline}

Using formulas \eqref{eq:H-Lambda} and \eqref{eq:mu-h-ex-Lambda},
and cancelling the common term $\beta\sum_{j=1}^{k+1}
J\myp\delta_{i_0,i_j}$, the equality \eqref{tim_V1} is reduced to
\begin{equation}\label{eq:A2}
\sum_{j=1}^{k+1}
\bigl[\xi_{i_j}(z_j)+h^\dag_{i_j}(z_j,z)\bigr]+\xi_{i_0}(z)=\sum_{j=1}^{k+1}
\bigl[\xi_{i_j}(x_j)+h^\dag_{i_j}(x_j,x_\circ)\bigr]
+\xi_{i_0}(x_\circ)+\frac{1}{\beta}\ln\frac{Z_{V_1(z)}}{Z_{V_1(x_\circ)}}.
\end{equation}
Varying $i_0\in\varPhi$ in \eqref{eq:A2} (whilst keeping all other
$i_j$ fixed) shows that the difference
$\xi_{i_0}(z)-\xi_{i_0}(x_\circ)$ does not depend on $i_0$. More
precisely, on subtracting from \eqref{eq:A2} the same equality with
$i_0=q$, we obtain
\begin{equation}\label{eq:A4}
\xi_{i}(z)-\xi_{q}(z)=\xi_{i}(x_\circ)-\xi_{q}(x_\circ),\qquad
i=1,\dots,q-1.
\end{equation}
Since \eqref{eq:A4} holds for any $z\in V$, this
proves~\eqref{eq:xih} in view of the notation \eqref{hxi}.

Furthermore, by virtue of \eqref{eq:xih} the equality \eqref{eq:A2}
becomes
\begin{equation}\label{eq:A5}
\sum_{j=1}^{k+1} h^\dag_{i_j}(z_j,z)=\sum_{j=1}^{k+1}
h^\dag_{i_j}(x_j,x_\circ)+\!\sum_{x\in
V_1(x_\circ)}\!\xi_{q}(x)-\!\sum_{y\in
V_1(z)}\!\xi_{q}(y)+\frac{1}{\beta}\ln\frac{Z_{V_1(z)}}{Z_{V_1(x_\circ)}}.
\end{equation}
Similarly, varying the values $i_1,\dots,i_{k+1}$ in \eqref{eq:A5}
(one at a time) yields, for each $j=1,\dots,k+1$,
\begin{equation}\label{eq:A6}
\check{h}^\dag_{i}(z_j,z)=\check{h}^\dag_{i}(x_j,x_\circ),\qquad
i=1,\dots,q-1.
\end{equation}
On the other hand, fix $j\in\{1,\dots,k+1\}$ and consider the shift
$\widetilde{T}_{z_j}$, resulting in
$$
\widetilde{T}_{z_j}(x_\circ)=z_j,\qquad \widetilde{T}_{z_j}(x_j)=z.
$$
The latter equality follows by recalling the definition of conjugate
translations (see~\eqref{eq:Tconj}) and noting that
$z_j=\widetilde{T}_{z}(x_j)=\mathfrak{b}^{-1}(\mathfrak{b}(z)\myp\mathfrak{b}(x_j))
=\mathfrak{b}^{-1}(\mathfrak{b}(z)\myp a_j)$ and, therefore,
$$
\widetilde{T}_{z_j}(x_j)=\mathfrak{b}^{-1}(\mathfrak{b}(z_j)\myp
\mathfrak{b}(x_j))=\mathfrak{b}^{-1}\myn(\mathfrak{b}(z)\myp
a_j^2)=\mathfrak{b}^{-1}\myn(\mathfrak{b}(z))=z,
$$
because $a_j^2=e$ (see Section~\ref{sec:2.3.1}). Hence, for
$\widetilde{T}_{z_j}$ the result \eqref{eq:A6} transforms into
\begin{equation}\label{eq:A7}
\check{h}^\dag_{i}(z,z_j)=\check{h}^\dag_{i}(x_j,x_\circ),\qquad
i=1,\dots,q-1.
\end{equation}
Comparing \eqref{eq:A6} and \eqref{eq:A7}, we conclude that
$\check{h}^\dag_{i}(z_j,z)=\check{h}^\dag_{i}(z,z_j)$, and the claim
\eqref{eq:h-dag} follows. Finally, let $y=\widetilde{T}_v(z)$ and
$y_j=\widetilde{T}_v(z_j)$, for some $v\in V$. Observe that
\begin{equation}\label{eq:y-by-v}
y=\widetilde{T}_y(x_\circ),\qquad y_j=\widetilde{T}_y(x_j).
\end{equation}
The first equality in \eqref{eq:y-by-v} is automatic; to check the
second one, note that
\begin{alignat*}{3}
\mathfrak{b}(y_j)&=\mathfrak{b}(v)\mypp
\mathfrak{b}(z_j)&\qquad& [y_j=\widetilde{T}_v(z_j)]\\
&=\mathfrak{b}(v)\mypp( \mathfrak{b}(z)\myp \mathfrak{b}(x_j))&&
[z_j=\widetilde{T}_z(x_j)]\\
&=(\mathfrak{b}(v)\myp\mathfrak{b}(z))\mypp
\mathfrak{b}(x_j)\\
&=\mathfrak{b}(y)\myp \mathfrak{b}(x_j)&& [y=\widetilde{T}_v(z)].
\end{alignat*}
That is, $\mathfrak{b}(y_j)=\mathfrak{b}(y)\myp \mathfrak{b}(x_j)$
and \eqref{eq:y-by-v} follows. Thus, formula \eqref{eq:A6} applied
to the edge $\langle y_j,y\rangle$ gives
\begin{equation*}
\check{h}^\dag_{i}(y_j,y)=\check{h}^\dag_{i}(x_j,x_\circ),\qquad
i=1,\dots,q-1.
\end{equation*}
Combined with \eqref{eq:A6}, this implies
$$
\check{h}^\dag_{i}(z_j,z)=\check{h}^\dag_{i}(y_j,y)=\check{h}^\dag_{i}\bigl(\widetilde{T}_v(z_j),\widetilde{T}_v(z)\bigr),
$$
and \eqref{eq:h-h} follows. This completes the ``only if'' part of
the proof.

\smallskip
\emph{Sufficiency.} Suppose that the conditions \eqref{eq:xih},
\eqref{eq:h-dag} and \eqref{eq:h-h} are satisfied. It suffices to
verify formula \eqref{tim} for the balls $V_n$ ($n\ge 1$). For $z\in
V$, denote $\varLambda:=\widetilde{T}_z(V_{n-1})$, then
$\bar{\varLambda}=\widetilde{T}_z(V_{n})$ and
$\partial\varLambda=\widetilde{T}_z(W_{n})$. Furthermore, for
$\varsigma\in\varPhi^{V_n}$ set
$\varsigma_z:=\widetilde{T}_z(\varsigma)\in\varPhi^{\bar{\varLambda}}$
(see~\eqref{eq:Tsigma}). Observe that if $y=\widetilde{T}_z(x)$
then, according to \eqref{eq:Tsigma=},
\begin{equation}\label{eq:A8}
\varsigma_z(y)=\varsigma(\widetilde{T}^{-1}_z(y))=\varsigma(x).
\end{equation}
Hence, recalling \eqref{eq:H-Lambda}, we have
\begin{align*}
H_{\myn\bar{\varLambda}}(\varsigma_z) &=-\!\!\sum_{\langle y,\myp
y'\rangle\in E_{\bar{\varLambda}}}\!\!J
\myp\delta_{\varsigma_z(y),\myp\varsigma_z(y')}-\sum_{y\in
\bar{\varLambda}}
\xi_{\varsigma_z(y)}(y)\\
&=-\!\!\sum_{\langle x,\myp x'\rangle\in E_{n}}\!\!J
\myp\delta_{\varsigma(x),\myp\varsigma(x')}-\sum_{x\in V_n}
\xi_{\varsigma(x)}(\widetilde{T}_z(x))\\
&=H_n(\varsigma)+\sum_{x\in
V_n}\bigl(\xi_{q}(x)-\xi_{q}(\widetilde{T}_z(x))\bigr),
\end{align*}
where at the last step we used the property~\eqref{eq:xih}. Thus,
from formula \eqref{eq:mu-h-ex-Lambda} we obtain, omitting factors
not depending on $\varsigma$,
\begin{equation}\label{eq:A9}
\mu^h(\sigma_{\myn\bar{\varLambda}}\myn=\varsigma_{\myp z})\propto
\mu^h(\sigma_{V_n}\myn=\varsigma)\cdot\exp\myn\Biggl\{\beta\myn\sum_{y\in
\partial\varLambda}\!h^\dag_{\varsigma_{\myp
z}(y)}(y,y_{\myn\varLambda})-\beta\!\sum_{x\in
W_n}\!h_{\varsigma(x)}(x)\Biggr\},
\end{equation}
where $y_{\myn\varLambda}\myn$ is the unique neighbour of
$y\in\partial\varLambda$ in $\varLambda$. Note that if
$y=\widetilde{T}_z(x)$, with $x\in \partial V_{n-1}\mynn=W_n$, then
$y_{\myn\varLambda}\myn=\widetilde{T}_z(x')$, where $x'\in W_{n-1}$
is the unique vertex
such that $x\in S(x')$. Thus, using \eqref{eq:A8}, \eqref{eq:h-h}
and~\eqref{eq:h-h-0}, we can write
\begin{align*}
\exp\myn\Biggl\{\beta\myn\sum_{y\in
\partial\varLambda}\!h^\dag_{\varsigma_{\myp
z}(y)}\myn(y,y_{\myn\varLambda})\Biggl\}&=
\exp\myn\Biggl\{\beta\myn\sum_{x\in
W_n}\!h^\dag_{\varsigma(x)}\myn\bigl(\widetilde{T}_z(x),\widetilde{T}_z(x')\bigr)\Biggl\}\\
&\propto \exp\myn\Biggl\{\beta\myn\sum_{x\in
W_n}\!h^\dag_{\varsigma(x)}(x,x')\Biggl\}\\
&\propto \exp\myn\Biggl\{\beta\myn\sum_{x\in
W_n}\!h_{\varsigma(x)}(x)\Biggl\}.
\end{align*}
Returning to \eqref{eq:A9}, this gives
$\mu^h(\sigma_{\myn\bar{\varLambda}}\myn=\varsigma_z)\propto
\mu^h(\sigma_{V_n}\myn=\varsigma)$, and since
$$
\sum_{\varsigma\in\varPhi^{V_n}}\mu^h(\sigma_{\myn\bar{\varLambda}}\myn=\varsigma_z)=1
=\sum_{\varsigma\in\varPhi^{V_n}}\mu^h(\sigma_{V_n}\myn=\varsigma),
$$
it follows that
$\mu^h(\sigma_{\myn\bar{\varLambda}}\myn=\varsigma_z)=
\mu^h(\sigma_{V_n}\myn=\varsigma)$, and the proof of the ``if'' part
is complete.

\section{Proof of Lemma~\ref{lm:q=3}}\label{sec:B}
\subsection{Proof of part (a)}\label{sec:B1}
Denote $t:=\theta-1$. For $q=3$ (i.e., with $m=m'\myn=1$), the
partial derivative $\partial K_1/\partial v$ (see~\eqref{eq:K'}) can
be represented as
\begin{align}
\notag \frac{\partial K_1}{\partial v}&=
\frac{(p_k+1)^{k-1}}{(p_k+L_m)^{k+1}}\,\biggl\{\bigl(p_k+1+(1-L_1)(k-1)\bigr)L'_1\myp \bigl(p_k-(k-1)\bigr)\\
 &\qquad - (1-L_1)\myp k\myp  L_1 \left(p_k'-\frac{k\myp
(k-1)}{2}\right)+\frac{k\myp(k-1)}{2}\myp(1-L_1)\bigl(2L_1'-k
L_1\bigr)\biggr\}, \label{eq:S1}
\end{align}
where $L_1=t\myp p_k-v^k-1$ and $L_1'=\partial L_1/\partial v=t\myp
p'_k-k\myp v^{k-1}$.

Note that $\partial K_1/\partial v|_{v=1}\myn=0$
(cf.~\eqref{eq:m'-m=0}); this follows without calculations from the
scaling property $K_1(v;t)=K_1(v^{-1};t)$ (see~\eqref{eq:K-scaled}).
More explicitly, using the formulas
\begin{align}\label{eq:S3}
p_k\myp|_{v=1}\myn=k-1,\qquad  p'_k\myp|_{v=1}\myn=\frac{k\myp
(k-1)}{2},
\end{align}
it is easy to see that the terms in \eqref{eq:S1} vanish at $v=1$.
We will also need the formula
\begin{equation}\label{eq:S3a}
p''_k\myp|_{v=1}\myn=\frac{k\myp (k-1)\myp(k-2)}{3}.
\end{equation}
Such expressions can be obtained by successively differentiating (at
$v=1$) the identity
$$
v^k-1\equiv(v-1)(p_k+1).
$$

To compute the second-order derivative $\partial^2 K_1/\partial v^2$
at $v=1$, we need to differentiate in \eqref{eq:S1} only the factors
that vanish at $v=1$, setting $v=1$ elsewhere. Hence,
\begin{align}
\notag \left.\frac{\partial^2 K_1}{\partial
v^2}\right|_{v=1}&=\frac{(p_k+1)^{k-1}}{(p_k+L_1)^{k+1}}\,\biggl\{(p_k+1)\myp
L'_1\myp p'_k\\[-.2pc]
&\quad +(1-L_1)\left[(k-1)\myp L'_1\myp p'_k- k\myp  L_1\myp p_k''
+\frac{k\myp(k-1)}{2}\mypp(2L''_1-k
L'_1)\right]\biggr\}\biggr|_{v=1}. \label{eq:S4}\end{align} Using
\eqref{eq:S3} and \eqref{eq:S3a}, we find
\begin{align}
\label{eq:S5a} L_1|_{v=1}&=t\myp p_k(1)-2=t\myp(k-1)-2,\\
\label{eq:S5b}L_1'|_{v=1}&=t\myp p'_k(1)-k=t\mypp\frac{k\myp(k-1)}{2} -k,\\
\label{eq:S5c}L_1''|_{v=1}&=t\myp p''_k(1)-k\myp(k-1)=t\mypp
\frac{k\myp(k-1)\myp(k-2)}{3} -k\myp(k-1).
\end{align}
Finally, substituting formulas \eqref{eq:S3}, \eqref{eq:S3a} and
\eqref{eq:S5a}--\eqref{eq:S5c} into \eqref{eq:S4}, after simple
manipulations (verified with {\sf Maple}) we obtain
\begin{align}
\notag \left.\frac{\partial^2 K_1}{\partial v^2}\right|_{v=1}
&=\left(\frac{k}{(k-1)(t+1)-2}\right)^{k+1}\,\frac{k-1}{2}\\
&\qquad
\times\left(\frac{(k-1)^2}{2}\,t^2+\frac{(k-1)\myp(7k-11)}{6}\,t-3k+1\right).
\label{eq:R'}
\end{align} The quadratic polynomial in \eqref{eq:R'} has one positive
zero $t=t^*$,
\begin{equation*}
t^*=\frac{11-7k+\sqrt{49k^2+62k+49}}{6\myp(k-1)},
\end{equation*}
which corresponds to $\tilde{\theta}_1=t^*\!+1$ as defined
in~\eqref{eq:theta-cr-q=3}. Hence, $\partial^2 K_1/\partial
v^2|_{v=1}>0$ for $\theta>\tilde{\theta}_1$, which implies that
$v=1$ is a local minimum of the function $v\mapsto K_1(v;\theta)$.
This completes the proof of Lemma~\ref{lm:q=3}(a).

\subsection{Proof of part (b)}\label{sec:B2}
It suffices to show, for $1\le \theta\le \tilde{\theta}_1$ (i.e.,
$0\le t\le t^*$), that $v=1$ is the sole root of the equation
$\partial K_1/\partial v=0$. A plausible general scheme of the proof
of the latter statement may be as follows.

\begin{itemize}
\item[(1)]
First, the condition $\partial K_1/\partial v=0$ (see~\eqref{eq:K'})
is reduced to $P(v;t)=0$, where $P(v;t)$ is a polynomial in $v$ (and
also a quadratic polynomial in~$t$).

\smallskip
\item[(2)]
Since $P(1;t)=0$, the quotient $R(v;t)=P(v;t)/(v-1)$ is a polynomial
in $v$, and we wish to prove that $R(v;t)<0$ for any $t\le t^*\myn$
and all $v\ne1$.

\smallskip
\item[(3)] According to the proof
in Section~\ref{sec:B1}, the condition $\partial^2K_1/\partial
v^2|_{v=1}\le 0$ is satisfied whenever $R(1;t)\le 0$; moreover, the
critical value $t^*\myn$ is determined by the quadratic equation
$R(1;t^*)=0$, which implies that $R(1;t)\le0$ for $t\in[\myp0,t^*]$.

\smallskip
\item[(4)] Bearing in mind the invariance of $K_1(v;t)$
under the map $v\mapsto v^{-1}$, it should be possible to represent
the polynomial $R(v;t)$ in the form
$$
R(v;t)=\chi(v)\cdot \widetilde{R}(y;t),\qquad y=v+v^{-1}\ge2,
$$
where $\chi(v)>0$ and $\widetilde{R}(y;t)$ is a polynomial in~$y$.

\smallskip
\item[(5)]
The crucial step is to show that, for each $t\in[\myp0,t^*]$, the
function $y\mapsto \widetilde{R}(y;t)$ is \emph{monotone
decreasing}.

\smallskip
\item[(6)] Finally, using steps (3) to (5), for any $y>2$ (i.e., $v\ne1$) we have
$$
\widetilde{R}(y;t)<\widetilde{R}(2;t)= \frac{R(1;t)}{\chi(1)}\le
0,\qquad 0\le t\le t^*.
$$
Hence, for all $t\in[\myp0,t^*]$ and $v\ne1$, we get
$$
R(v;t)=\chi(v)\cdot \widetilde{R}(y;t)<0,
$$
as required.
\end{itemize}

\smallskip
In what follows, we implement this scheme in more detail for the
cases $k=2,3,4$. All calculations were done analytically and also
verified using {\sf Maple}.

\subsubsection{Case $k=2$} According to formula \eqref{eq:theta-cr-q=3_k=2},
$t^*\myn=\frac12\bigl(\sqrt{41}-1\bigr)$. The polynomial $P(v;t)$ is
found to be
\begin{align*}
P(v;t)&=t^2v^2-t^2v-tv^3+4tv^2-4tv+t-4v^3+2v^2-2v+4\\
&=(v-1)\bigl\{t^2 v-t\mypp(v^2-3v+1)-(4v^2+2v+4)\bigr\}.
\end{align*}
Hence,
\begin{align*}
R(v;t)&=t^2 v-t\mypp(v^2-3v+1)-(4v^2+2v+4\\
&=v\left(t^2-t\left(v-3+\frac{1}v\right)-4\left(v+\frac{1}v\right)-2\right)\\
&=v\left(t^2-t\myp(y-3)-4y-2\right),
\end{align*}
where $y=v+v^{-1}$. This gives
$$
\widetilde{R}(y;t)=t^2-t\myp(y-3)-4y-2=-y\myp(t+4)+t^2+3t-2,
$$
which is clearly a decreasing function of $y$ for any $t\ge0$.

\subsubsection{Case $k=3$} By \eqref{eq:theta-cr-q=3_k=2}, we have $t^*\!=\frac43$.
The quotient $R(v;t)=P(v;t)/(v-1)$ is explicitly given by
\begin{align*}
R(t,v)&=(v+1)(t+3)\bigl(2tv^3+2tv -2v^4 -2+ 5tv^2 -4v^3-4v\bigr)\\
&=v^2(v+1)(t+3)\left(2t\left(v+\frac{1}{v}\right)-2\left(v^2+\frac{1}{v^2}\right)+5t-4\left(v+\frac{1}{v}\right)\right)\\
&=v^2(v+1)(t+3)\bigl(2ty-2\myp(y^2-2)+5t-4y\bigr).
\end{align*}
Thus,
$$
\widetilde{R}(y;t)=(t+3)\bigl(2ty-2\myp(y^2-2)+5t-4y\bigr),
$$
and it is easy to check that this function is decreasing in $y$ for
any $t\le t^*\!=\frac43$.

\subsubsection{Case  $k=4$} Elementary but tedious
calculations yield
\begin{align*}
R(v;t)&=t^2\left(3v^7+11v^6+25v^5+30v^4+25v^3+11v^2+3v\right)\\
&\quad
-t\left(3v^8+v^7-13v^6-57v^5-72v^4-57v^3-13v^2+v+3\right)\\
&\qquad
-\left(8v^8+24v^7+48v^6+36v^5+32v^4+36v^3+48v^2+24v+8\right).
\end{align*}
Rearranging under the substitution $y=v+v^{-1}$ gives
$R(v;t)=v^4\myp\widetilde{R}(y;t)$ with
\begin{align}
\notag
\widetilde{R}(y;t)&=t^2(3y^3+11y^2+16y+8)\\
\notag
&\quad-t\myp(3y^4+y^3-25y^2-60y-40)\\
&\qquad-(8y^4+24y^3+16y^2-36y-48). \label{eq:R-tilde}
\end{align} In
particular, if $y=2$ then
$$
\widetilde{R}(2;t)=12\myp\bigl(9\myp t^2+17\myp t-22\bigr)=0
$$
for $t=\frac1{18}\bigl(\sqrt{1081}-17\bigr)=t^*$
(cf.~\eqref{eq:theta-cr-q=3_k=2}), as it should be.

Finally, we need to verify that the function $y\mapsto
\widetilde{R}(y;t)$ is decreasing for any $t\in[\myp0,t^*]$.
Unfortunately (but inevitably), technicalities involved in a purely
analytic check become quite substantial; however, using {\sf Maple}
to plot the graph of \eqref{eq:R-tilde}, with parameter $t$ ranging
from $0$ to $t^*$, makes the monotonicity evident.
\end{appendix}


\begin{thebibliography}{99}


\bibitem{Abou-Chacra}
Abou-Chacra, R., Anderson, P.W.\ and Thouless, D.J. A selfconsistent
theory  of localization. \emph{J.~Phys.~C} {\bf 6} (1973),
1734--1752.
(\href{https://doi.org/10.1088/0022-3719/6/10/009}{doi:\allowbreak
10.\allowbreak 1088/\allowbreak 0022-3719/\allowbreak 6/\allowbreak
10/009})

\bibitem{Aizenman}
Aizenman, M.\ and Warzel, S. The canopy graph and level statistics
for random operators on trees. \emph{Math.\ Phys.\ Anal.\ Geom.}
{\bf 9} (2007), 291--333.
(\href{https://doi.org/10.1007/s11040-007-9018-3}{doi:\allowbreak
10.\allowbreak 1007/\allowbreak s11040-007-9018-3}, \MR{2329431})


\bibitem{Athreya}
Athreya, S., Bandyopadhyay, A.\ and Dasgupta, A. Random walks in
i.i.d.\ random environment on Cayley trees. \emph{Statist.\ Probab.\
Lett.} {\bf 92} (2014), 39--44.
(\href{https://doi.org/10.1016/j.spl.2014.04.026}{doi:\allowbreak
10.\allowbreak 1016/\allowbreak j.\allowbreak spl.\allowbreak
2014.\allowbreak 04.026}, \MR{3230470})


\bibitem{Ba} Baxter, R.J. \textit{Exactly Solved Models in Statistical
Mechanics}. Academic Press, London, 1982. (\MR{0690578})


\bibitem{Be1}
Beaudin, L.,  Ellis-Monaghan, J., Pangborn, G.\ and Shrock,~R. A
little statistical mechanics for the graph theorist. \emph{Discrete
Math.}\ {\bf 310} (2010),
2037--2053.
(\href{https://doi.org/10.1016/j.disc.2010.03.011}{doi:\allowbreak
10.\allowbreak 1016/\allowbreak j.\allowbreak disc.\allowbreak
2010.\allowbreak 03.\allowbreak 011}, \MR{2629922})

\bibitem{Bethe}
Bethe, H.A. Statistical theory of superlattices. \emph{Proc.\ Roy.\
Soc.~A} {\bf 150} (1935), 552--575.
(\href{https://doi.org/10.1098/rspa.1935.0122}{doi:\allowbreak
10.\allowbreak 1098/\allowbreak rspa.\allowbreak 1935.\allowbreak
0122})

\bibitem{BRSSZ} Bleher, P., Ruiz, J., Schonmann, R.H.,
Shlosman, S.\ and Zagrebnov, V. Rigidity of the critical phases on a
Cayley tree. \emph{Mosc.\ Math.~J.} {\bf 1} (2001),
345--363.
(\href{https://doi.org/10.17323/1609-4514-2001-1-3-345-363}{doi:\allowbreak
10.\allowbreak 17323/\allowbreak 1609-4514-2001-1-3-345-363},
\MR{1877597})

\bibitem{BRZ} Bleher, P.M., Ruiz, J.\ and Zagrebnov, V.A. On the
phase diagram of the random field Ising model on the Bethe lattice.
\emph{J.\ Stat.\ Phys.} {\bf 93} (1998), 33--78.
(\href{https://doi.org/10.1023/B:JOSS.0000026727.43077.49}{doi:\allowbreak
10.\allowbreak 1023/\allowbreak B:\allowbreak JOSS.\allowbreak
0000026727.43077.49}, \MR{1656364})


\bibitem{Bov}
Bovier, A. \emph{Statistical Mechanics of Disordered Systems: A
Mathematical Perspective}. Cambridge Series in Statistical and
Probabilistic Mathematics, {\bf 18}. Cambridge University Press,
Cambridge, 2006.
(\href{https://doi.org/10.1017/CBO9780511616808}{doi:\allowbreak
10.\allowbreak 1017/\allowbreak CBO9780511616808}, \MR{2252929})

\bibitem{Boy} Boykov, Y.,  Veksler, O.\ and Zabih R. Fast approximate energy minimization via graph cuts.
\emph{IEEE Trans.\ Pattern Anal.\ Machine Intell.}\ {\bf 23}
(2001), 1222--1239.
(\href{https://doi.org/10.1109/34.969114}{doi:\allowbreak
10.1109/34.969114})

\bibitem{Bricmont}
Bricmont, J., Kuroda, K.\ and Lebowitz, J.L. First order phase
transitions in lattice and continuous systems: Extension of
Pirogov--Sinai theory. \emph{Comm.\ Math.\ Phys.}\ {\bf 101} (1985),
501--538. (\href{https://doi.org/10.1007/BF01210743}{doi:\allowbreak
10.\allowbreak 1007/\allowbreak BF01210743}, \MR{0815198})

\bibitem{Br}
Brookings, T., Carlson, J.M.\ and Doyle, J. Three mechanisms for
power laws on the Cayley tree. \emph{Phys.\ Rev.~E} {\bf 72} (2005),
056120, 18 pp.
(\href{https://doi.org/10.1103/PhysRevE.72.056120}{doi:\allowbreak10.1103/\allowbreak
PhysRevE.\allowbreak 72.056120}, \MR{2198305})

\bibitem{Chen}
Chen, M.-S., Onsager, L., Bonner, J.\ and Nagle,~J. Hopping of ions
in ice. \emph{J.~Chem.\ Phys.} {\bf 60} (1974), 405--419.
(\href{https://doi.org/10.1063/1.168105660}{doi:\allowbreak
10.\allowbreak 1063/\allowbreak 1.\allowbreak 168105660})

\bibitem{DMR}
Daskalakis, C., Mossel, E.\ and Roch, S. Evolutionary trees and the
Ising model on the Bethe lattice: a proof of Steel's conjecture.
\emph{Probab.\ Theory Related Fields} {\bf 149} (2011),
149--189.
(\href{https://doi.org/10.1007/s00440-009-0246-2}{doi:\allowbreak
10.\allowbreak 1007/\allowbreak s00440-009-0246-2}, \MR{2773028})


\bibitem{Dembo2}
Dembo, A., Montanari, A., Sly, A.\ and Sun, N. The replica symmetric
solution for Potts models on $d$-regular graphs. \emph{Comm.\ Math.\
Phys.}\ {\bf 327} (2014), 551--575.
(\href{https://doi.org/10.1007/s00220-014-1956-6}{doi:\allowbreak
10.\allowbreak 1007/\allowbreak s00220-014-1956-6}, \MR{3183409})


\bibitem{Domb}
Domb, C. Configurational studies of the Potts models.
\emph{J.~Phys.~A}
{\bf 7} (1974), 1335--1348.
(\href{https://doi.org/10.1088/0305-4470/7/11/013}{doi:\allowbreak
10.\allowbreak 1088/\allowbreak 0305-4470/\allowbreak 7/\allowbreak
11/013})

\bibitem{Fr}
Friedrich, F., Kempe, A., Liebscher, V.\ and Winkler, G.  Complexity
penalized $M$-estimation: fast computation. \emph{J.~Comput.\
Graph.\ Statist.} {\bf 17} (2008),
201--224.
(\href{https://doi.org/10.1198/106186008X285591}{doi:\allowbreak
10.\allowbreak 1198/\allowbreak 106186008X285591},
 \MR{2424802})

\bibitem{Galanis}
Galanis, A., \v{S}tefankovi\v{c}, D., Vigoda, E.\ and Yang, L.
Ferromagnetic Potts model: refined \#BIS-hardness and related
results. \emph{SIAM J.\ Comput.}\ {\bf 45} (2016),
2004--2065.
(\href{https://doi.org/10.1137/140997580}{doi:\allowbreak
10.\allowbreak 1137/\allowbreak 140997580}, \MR{3572375})


\bibitem{GRR}
Gandolfo, D., Rahmatullaev, M.M.\ and Rozikov, U.A. Boundary
conditions for translation-invariant Gibbs measures of the Potts
model on Cayley trees.  \emph{J.~Stat.\ Phys.} {\bf 167} (2017),
1164--1179.
(\href{https://doi.org/10.1007/s10955-017-1771-5}{doi:\allowbreak
10.\allowbreak 1007/\allowbreak s10955-017-1771-5}, \MR{3647056})



\bibitem{Ga0}
Ganikhodjaev, N.N. Group presentations and automorphisms of the
Cayley tree. (Russian) \emph{Dokl.\ Akad.\ Nauk Respub.\
Uzbekistan}, no.~4 (1994), 3--5.

\bibitem{GaRo} Ganikhodjaev, N.\ and Rozikov, U. On disordered
phase in the ferromagnetic Potts model on the Bethe lattice.
\emph{Osaka J.~Math.} {\bf 37}
(2000), 373--383.
(\href{http://projecteuclid.org/euclid.ojm/1200789203}{http://\allowbreak
projecteuclid.\allowbreak org/\allowbreak euclid.\allowbreak
ojm/\allowbreak 1200789203}, \MR{1772838})

\bibitem{Ge}
Georgii, H.-O. \emph{Gibbs Measures and Phase Transitions}, 2nd ed.
De Gruyter Studies in Mathematics, {\bf 9}.
De Gruyter, Berlin, 2011. (\MR{2807681})

\bibitem{Green} Green, P.J.\ and Richardson, S. Hidden Markov
models and disease mapping. \emph{J.~Amer.\ Statist.\ Assoc.} {\bf
97} (2002),
1055--1070.
(\href{https://doi.org/10.1198/016214502388618870}{doi:\allowbreak
10.\allowbreak 1198/\allowbreak 016214502388618870}, \MR{1951259})

\bibitem{Graner}
Graner, F.\ and Glazier, J. Simulation of biological cell sorting
using a two-dimensional extended Potts model. \emph{Phys.\ Rev.\
Lett.}\ {\bf 69} (1992),
2013--2017.
(\href{https://doi.org/10.1103/PhysRevLett.69.2013}{doi:\allowbreak
10.\allowbreak 1103/\allowbreak PhysRevLett.\allowbreak
69.\allowbreak 2013})


\bibitem{Gr1}
Grimmett, G. \emph{The Random-Cluster Model}. Grundlehren der
mathematischen Wissenschaften [A~Series of Comprehensive Studies in
Mathematics], {\bf 333}. Springer, Berlin, 2006.
(\href{https://doi.org/10.1007/978-3-540-32891-9}{doi:\allowbreak
10.\allowbreak 1007/\allowbreak 978-3-540-32891-9}, \MR{2243761})

\bibitem{Hagg}
H\"aggstr\"om, O. The random-cluster model on a
homogeneous tree. \emph{Probab.\ Theory Related Fields} {\bf 104}
(1996), 231--253.
(\href{https://doi.org/10.1007/BF01247839}{doi:\allowbreak
10.\allowbreak 1007/\allowbreak BF01247839}, \MR{1373377})

\bibitem{Kotecky-Shlosman}
Koteck\'y, R.\ and Shlosman, S.B. First-order phase transitions in
large entropy lattice models. \emph{Comm.\ Math.\ Phys.}\ {\bf 83}
(1982), 493--515.
(\href{https://doi.org/10.1007/BF01247839}{doi:\allowbreak
10.\allowbreak 1007/\allowbreak BF01247839}, \MR{1373377})

\bibitem{Kramers-Wannier}
Kramers, H.A.\ and Wannier, G.H. Statistics of the two-dimensional
ferromagnet. Part~I. \emph{Phys.\ Rev.}\ {\bf 60} (1941),
252--262.
(\href{https://doi.org/10.1103/PhysRev.60.252}{doi:\allowbreak
10.\allowbreak 1103/\allowbreak PhysRev.\allowbreak 60.\allowbreak
252}, \MR{0004803})

\bibitem{KR}
K\"ulske,  C.\ and Rozikov, U.A. Fuzzy transformations and
extremality of Gibbs measures for the Potts model on a Cayley tree.
\emph{Random Structures Algorithms} {\bf 50} (2017),
636--678. (\href{https://doi.org/10.1002/rsa.20671}{doi:\allowbreak
10.\allowbreak 1002/\allowbreak rsa.\allowbreak 20671},
\MR{3660523})

\bibitem{KRK}
K\"ulske, C., Rozikov, U.A.\ and Khakimov, R.M. Description of the
translation-invariant splitting Gibbs measures for the Potts model
on a Cayley tree. \emph{J.~Stat.\ Phys.} {\bf 156} (2014), 189--200.
(\href{https://doi.org/10.1007/s10955-014-0986-y}{doi:\allowbreak
10.\allowbreak 1007/\allowbreak s10955-014-0986-y}, \MR{3215122})


\bibitem{Laanait}
Laanait, L., Messager, A.\ and Ruiz, J. Phases coexistence and
surface tensions for the Potts model. \emph{Comm.\ Math.\ Phys.}\
{\bf 105} (1986),
527--545. (\href{https://doi.org/10.1007/BF01238932}{doi:\allowbreak
10.\allowbreak 1007/\allowbreak BF01238932}, \MR{0852089})


\bibitem{Magnus}
Magnus, W., Karrass, A.\ and Solitar, D. \emph{Combinatorial Group
Theory: Presentations of Groups in Terms of Generators and
Relations}, 2nd revised ed. Dover, New York, 1976. (\MR{0422434})


\bibitem{Martin}
Martin, P. \emph{Potts Models and Related Problems in Statistical
Mechanics.} Series on Advances in Statistical Mechanics, {\bf 5}.
World Scientific, Singapore,
1991. (\href{https://doi.org/10.1142/0983}{doi:\allowbreak
10.\allowbreak 1142/\allowbreak 0983}, \MR{1103994})

\bibitem{Martinelli}
Martinelli, F., Sinclair, A.\ and Weitz, D. Glauber dynamics on
trees: Boundary conditions and mixing time. \emph{Comm.\ Math.\
Phys.} {\bf 250} (2004),
301--334.
(\href{https://doi.org/10.1007/s00220-004-1147-y}{doi:\allowbreak
10.\allowbreak 1007/\allowbreak s00220-004-1147-y}, \MR{2094519})

\bibitem{Martirosian}
Martirosian, D.H. Translation invariant Gibbs states in the
$q$-state Potts model. \emph{Comm.\ Math.\ Phys.}\ {\bf 105} (1986),
281--290.
(\href{https://doi.org/10.1007/BF01211103}{doi:10.1007/BF01211103},
\MR{0849209})


\bibitem{Mezard}M\'ezard, M.\ and Montanari, A. Reconstruction on trees and
spin glass transition. \emph{J.~Stat.\ Phys.}\ {\bf 124} (2006),
1317--1350.
(\href{https://doi.org/10.1007/s10955-006-9162-3}{doi:10.1007/s10955-006-9162-3},
\MR{2266446})


\bibitem{Mezard-Parisi}
M\'ezard, M.\ and Parisi, G. The Bethe lattice spin glass revisited.
\emph{Eur.\ Phys.~J.~B} {\bf 20} (2001), 217--233.
(\href{https://doi.org/10.1007/PL00011099}{doi:\allowbreak
10.\allowbreak 1007/\allowbreak PL00011099}, \MR{1832936})

\bibitem{Mos} Mossel, E.
Survey: Information flow on trees. In: \emph{Graphs, Morphisms and
Statistical Physics} (J.~Ne\v{s}et\v{r}il and P.~Winkler, eds.),
DIMACS Series in Discrete Mathematics and Theoretical Computer
Science {\bf 63}, 155--170. Amer.\ Math.\ Soc., Providence, RI,
2004.
(\href{https://doi.org/10.1090/dimacs/063/12}{doi:10.1090/dimacs/063/12},
\MR{2056226})

\bibitem{Murua}
Murua, A.\ and Quintana, F.A. Semiparametric Bayesian regression via
Potts model. \emph{J.~Comput.\ Graph.\ Statist.}\ {\bf 26} (2017),
265--274.
(\href{https://doi.org/10.1080/10618600.2016.1172015}{doi:10.1080/10618600.2016.1172015},
\MR{3640184})

\bibitem{Ost}
Ostilli, M. Cayley trees and Bethe lattices: A concise analysis for
mathematicians and physicists. \emph{Phys.~A} {\bf 391} (2012),
3417--3423.
(\href{https://doi.org/10.1016/j.physa.2012.01.038}{doi:10.1016/j.physa.2012.01.038},
\MR{2904466})


\bibitem{Peruggi1} Peruggi, F., di Liberto, F.\ and Monroy, G.
The Potts model on Bethe lattices. I. General results.
\emph{J.~Phys.~A} {\bf 16} (1983),
811--827.
(\href{https://doi.org/10.1088/0305-4470/16/4/018}{doi:10.1088/0305-4470/16/4/018},
\MR{0706197})

\bibitem{Peruggi2} Peruggi, F., di Liberto, F.\ and Monroy, G.
Phase diagrams of the $q$-state Potts model on Bethe lattices.
\emph{Phys.~A} {\bf 141} (1987),
151--186.
(\href{https://doi.org/10.1016/0378-4371(87)90267-6}{doi:10.1016/0378-4371(87)90267-6},
\MR{0879715})

\bibitem{Potts} Potts, R.B. Some generalized order-disorder
transformations. \emph{Proc.\ Cambridge Philos.\ Soc.} {\bf 48}
(1952), 106--109.
(\href{https://doi.org/10.1017/S0305004100027419}{doi:10.1017/S0305004100027419},
\MR{0047571})


\bibitem{Pr}
Preston, C.J. \emph{Gibbs States on Countable Sets}. Cambridge
Tracts in Mathematics, {\bf 68}. Cambridge University Press, London,
1974.
(\href{https://doi.org/10.1017/CBO9780511897122}{doi:10.1017/CBO9780511897122},
\MR{0474556})


%

\bibitem{Reichardt}
Reichardt, J.\ and Bornholdt, S. Detecting fuzzy community
structures in complex networks with a Potts model. \emph{Phys.\
Rev.\ Lett.}\ {\bf 93} (2004), Art.~218701, 4~pp.
(\href{https://doi.org/10.1103/PhysRevLett.93.218701}{doi:10.1103/PhysRevLett.93.218701})

\bibitem{Riva}
Riva, V.\ and Cardy, J. Holomorphic parafermions in the Potts model
and stochastic Loewner evolution. \emph{J.~Stat.\ Mech.\ Theory
Exp.}\ (2006), no.\,12, P12001, 19~pp.
(\href{https://doi.org/10.1088/1742-5468/2006/12/P12001}{doi:\allowbreak
10.\allowbreak 1088/\allowbreak 1742-\allowbreak  5468/\allowbreak
2006/\allowbreak 12/\allowbreak P12001}, \MR{2280251})



\bibitem{Ro}
Rozikov, U.A. \emph{Gibbs Measures on Cayley Trees}. World
Scientific, Hackensack, NJ, 2013.
(\href{https://doi.org/10.1142/8841}{doi:\allowbreak 10.\allowbreak
1142/\allowbreak 8841}, \MR{3185400})


\bibitem{Roz-DNA}
Rozikov, U.A. Tree-hierarchy of DNA and distribution of Holliday
junctions. \emph{J.~Math.\ Biol.} {\bf 75} (2017), 1715--1733.
(\href{https://doi.org/10.1007/s00285-017-1136-3}{doi:\allowbreak
10.\allowbreak 1007/\allowbreak s00285-017-1136-3}, \MR{3712328})

\bibitem{RS} Rozikov, U.A.\ and Shoyusupov, Sh.A.  Gibbs
measures for the SOS model with four states on a Cayley tree.
(Russian) \emph{Teoret.\ Mat.\ Fiz.} {\bf 149} (2006), no.~1,
18--31; translation in \emph{Theor.\ Math.\ Phys.} \textbf{149}
(2006), no.~1, 1312--1323.
(\href{https://doi.org/10.1007/s11232-006-0120-7}{doi:\allowbreak
10.\allowbreak 1007/\allowbreak s11232-006-0120-7}, \MR{2297108})

\bibitem{RoSu} Rozikov, U.A.\ and Suhov, Y.M. Gibbs measures for SOS models on a Cayley
tree. \emph{Infin.\ Dimens.\ Anal.\ Quantum Probab.\ Relat.\ Top.}
\textbf{9} (2006), 471--488.
(\href{https://doi.org/10.1142/S0219025706002494}{doi:\allowbreak
10.\allowbreak 1142/\allowbreak
S02\allowbreak1\allowbreak9\allowbreak
0\allowbreak2\allowbreak5\allowbreak7\allowbreak0\allowbreak
6002494}, \MR{2256506})

\bibitem{Rudin}
Rudin, W. \emph{Principles of Mathematical Analysis}, 3rd ed.
International Series in Pure and Applied Mathematics. McGraw-Hill,
New York, 1976. (\MR{0385023})

\bibitem{Sa}
Sanyal, S.\ and Glazier, J.A. Viscous instabilities in flowing
foams: A Cellular Potts Model approach. \emph{J.~Stat.\ Mech.\
Theory Exp.} (2006), no.~10, P10008, 9~pp.
(\href{https://doi.org/10.1088/1742-5468/2006/10/P10008}{doi:\allowbreak
10.\allowbreak 1088/\allowbreak 1742-5468/\allowbreak
2006/\allowbreak 10/P10008})

\bibitem{Sc} Schelling, T.C. Dynamic models of segregation. \emph{J.\ Math.\ Sociology} {\bf 1} (1971),
143--186.
(\href{https://doi.org/10.1080/0022250X.1971.9989794}{doi:\allowbreak
10.\allowbreak 1080/\allowbreak 0022250X.\allowbreak
1971.\allowbreak 9989794})

\bibitem{Schramm}
Schramm, O. Conformally invariant scaling limits: an overview and a
collection of problems. In: \emph{Proceedings of the International
Congress of Mathematicians (Madrid, August 22--30, 2006)}, Vol.~I,
513--543. Eur.\ Math.\ Soc., Z\"urich, 2007.
(\href{https://doi.org/10.4171/022-1/20}{doi:10.4171/022-1/20},
\MR{2334202})


\bibitem{Schulze} Schulze, C. Potts-like model for ghetto formation in multi-cultural
societies. \emph{Int.\ J.\ Modern Phys.~C} {\bf 16}
(2005), 351--355.
(\href{https://doi.org/10.1142/S0129183105007169}{doi:10.1142/S0129183105007169})

\bibitem{Shiryaev}
Shiryaev, A.N. \emph{Probability}, 2nd ed. Graduate Texts in
Mathematics, {\bf 95}. Springer, New York, 1996.
(\href{https://doi.org/10.1007/978-1-4757-2539-1}{doi:\allowbreak
10.\allowbreak 1007/\allowbreak 978-1-4757-2539-1}, \MR{1368405})


\bibitem{Sp}
Spitzer, F. Markov random fields on an infinite tree. \emph{Ann.\
Probab.} {\bf 3} (1975),
387--398.
(\href{https://doi.org/10.1214/aop/1176996347}{doi:\allowbreak
10.\allowbreak 1214/\allowbreak aop/\allowbreak 1176996347},
\MR{0378152})

\bibitem{Sun} Sun, L., Chang, Y.F.\ and Cai, X. A discrete simulation of tumor growth
concerning nutrient concentration. \emph{Int.\ J.\ Modern Phys.~B}
{\bf 18} (2004), 2651--2657.
(\href{https://doi.org/10.1142/S0217979204025853}{doi:10.1142/S0217979204025853})

\bibitem{Sz}
Szab\'o, A.\ and Merks, R.M.H. Cellular Potts modeling of tumor
growth, tumor invasion, and tumor evolution. \emph{Front.\ Oncol.}\
{\bf 3} (2013), Art.~87, 12 pp.
(\href{https://doi.org/10.3389/fonc.2013.00087}{doi:10.3389/fonc.2013.00087})

\bibitem{Takaishi} Takaishi, T. Simulations of financial
markets in a Potts-like model. \emph{Int.\ J.\ Modern Phys.~C} {\bf
16} (2005),
1311--1317.
(\href{https://doi.org/10.1142/S0129183105007923}{doi:\allowbreak
10.\allowbreak 1142/\allowbreak S0129183105007923})

\bibitem{Thouless}
Thouless, D.J. Spin-glass on a Bethe lattice. \emph{Phys.\ Rev.\
Lett.} {\bf 56} (1986),
1082--1085.
(\href{https://doi.org/10.1103/PhysRevLett.56.1082}{doi:\allowbreak
10.\allowbreak 1103/\allowbreak PhysRevLett.\allowbreak
56.\allowbreak 1082})

\bibitem{Ti}
Tikare, V.\ and Cawley, J.D. Application of the Potts model to
simulation of Ostwald ripening. \emph{J.~Amer.\ Ceramic Soc.} {\bf
81}
(1998), 485--491.
(\href{https://doi.org/10.1111/j.1151-2916.1998.tb02366.x}{doi:\allowbreak
10.\allowbreak 1111/\allowbreak j.\allowbreak
1151-2916.1998.\allowbreak tb02366.x})

\bibitem{Weiss}
Weiss, P.R. The application of the Bethe--Peierls method to
ferromagnetism. \emph{Phys.\ Rev.} {\bf 74} (1948),
1493--1504.
(\href{https://doi.org/10.1103/PhysRev.74.1493}{doi:\allowbreak
10.\allowbreak 1103/\allowbreak PhysRev.\allowbreak 74.\allowbreak
1493})

\bibitem{Wu} Wu, F.Y. The Potts model. \emph{Rev.\ Modern Phys.} {\bf 54} (1982),  235--268.
(\href{https://doi.org/10.1103/RevModPhys.54.235}{doi:\allowbreak
10.\allowbreak 1103/\allowbreak RevModPhys.\allowbreak 54.235},
\MR{0641370})


\bibitem{Z0} Zachary, S. Countable state space Markov random fields
and Markov chains on trees. \emph{Ann.\ Probab.} {\bf 11} (1983),
894--903.
(\href{https://doi.org/10.1214/aop/1176993439}{doi:10.1214/aop/1176993439},
\MR{0714953})





\end{thebibliography}
\end{document}